\pdfoutput=1
\documentclass[pdflatex,tex,twocolumn,epjc3]{svjour3}

\RequirePackage[T1]{fontenc}
\RequirePackage{mathptmx}

\usepackage{graphicx}
\usepackage{float}
\usepackage{adjustbox}

\usepackage{amssymb}
\usepackage{amsmath}
\usepackage{dcolumn}
\usepackage{slashed}

\usepackage{indentfirst}
\usepackage{enumitem}
\usepackage{setspace}

\usepackage{multirow}
\usepackage{booktabs}

\RequirePackage{flushend}
\smartqed

\RequirePackage[numbers,sort&compress]{natbib}
\RequirePackage[colorlinks,citecolor=blue,urlcolor=blue,linkcolor=blue]{hyperref}

\journalname{Eur. Phys. J. C}

\begin{document}

\title{Reconstructing dark energy with model independent methods after DESI DR2}

\author{
	Jun-Xian Li\thanksref{e1,addr1}
	\and
	Shuang Wang\thanksref{e2,addr1}}
	
\thankstext{e1}{e-mail: lijx389@mail2.sysu.edu.cn}
\thankstext{e2}{e-mail: wangshuang@mail.sysu.edu.cn \text{(corresponding author)}}
\institute{School of Physics and Astronomy, Sun Yat-Sen University, Zhuhai 519082, P.R.China\label{addr1}}
	
\date{Received: date / Accepted: date}
	
\maketitle

\begin{abstract}
In this paper, we employ two model-independent approaches, including redshift binning method and polynomial interpolation method, to reconstruct dark energy (DE) equation of state (EoS) $w(z)$ and DE density function $f(z)$. Our analysis incorporates data from the Dark Energy Spectroscopic Instrument (DESI) Data Release 2 (DR2), Cosmic Microwave Background (CMB) distance priors from Planck 2018 and Atacama Cosmology Telescope (ACT) DR6, and three type Ia supernovae (SN) compilations (PantheonPlus, Union3, and DESY5). To ensure model independence, we adopt three redshift binning schemes (n=3, 4, 5) and three polynomial interpolation schemes with the same number of nodes (n=3, 4, 5).
Our main conclusions are as follows:
1) After incorporating DESI data, there is a trend that DE should evolve with redshift (with deviations from EoS $w=-1$ reaching at least a $2.13\sigma$ confidence level), indicating that current observations favor a dynamical DE.
2) In the redshift range $0.5 < z < 1.5$, the DE EoS $w(z)$ exhibits a decreasing trend and crosses the phantom divide $w=-1$, suggesting quintom-like behavior.
3) The DE density $f(z)$ first increases at low redshift, reaching a hump around $z\approx 0.5$, and then decreases at $0.5 < z < 1.5$, with a rapid decrease at $z>1.0$.
4) For $z > 1.5$, current data are insufficient to place strong constraints on the evolution of DE, resulting in large uncertainties in the DE reconstruction.
It must be emphasized that these four main conclusions are independent of specific reconstruction models and are insensitive to the choice of SN compilations.
\end{abstract}

\section{Introduction}
Since the discovery of cosmic acceleration \cite{SupernovaSearchTeam:1998fmf, SupernovaCosmologyProject:1998vns}, dark energy (DE) has become one of the most profound mysteries in modern cosmology. 
As a standard cosmological model, the $\Lambda$CDM model, which contains the cosmological constant $\Lambda$ together with cold dark matter, has been widely recognized for its success in describing a wide range of observational data. 
However, in recent years, the robustness of $\Lambda$CDM has encountered increasing challenges. 
The most pressing issue is the "Hubble tension", a $\sim 5 \sigma$ discrepancy between the directly measured current cosmic expansion rate and its inferred value from early-universe observations (see \cite{Poulin:2023lkg, Rong-Gen:2023dcz, Efstathiou:2024dvn, DiValentino:2025sru} for recent reviews). 
Moreover, both the DESI DR1 and DR2 disfavor the $\Lambda$CDM model \cite{DESI:2024mwx, DESI:2025zgx}. For example, by using DESI DR2 and CMB data, DESI collaboration found a 3.1$\sigma$ evidence in favor of dynamical DE. In addition, after adding the PantheonPlus \cite{Brout:2022vxf}, Union3 \cite{Rubin:2023ovl} or DESY5 \cite{DES:2024jxu} SN datasets to the combination, the preference for the Chevallier-Polarski-Linder (CPL) model \cite{Chevallier:2000qy, Linder:2002et} over the $\Lambda$CDM model is 2.8$\sigma$, 3.8$\sigma$ or 4.2$\sigma$, respectively.
Because of these observational challenges to the $\Lambda$CDM model, it is necessary to explore dynamical DE models.

Dynamical DE models suggest that the accelerated expansion of the universe is driven by a time-evolving DE, characterized by an EoS $w(z)$ that varies with redshift $z$. 
The popular dynamical DE models include scalar field DE models \cite{Yin:2024hba, Shlivko:2024llw, DESI:2024kob, Notari:2024rti, Wolf:2024eph, Shajib:2025tpd, Anchordoqui:2025fgz, Gialamas:2025pwv}, holographic DE models \cite{Li:2024bwr, Tyagi:2024cqp, Astashenok:2024jje, Li:2024qus, Han:2024sxm}, interacting DE models \cite{Giare:2024smz, Montani:2024pou, Li:2024qso, Li:2025owk, Yang:2025vnm, Zhai:2025hfi, Pan:2025qwy, Yang:2025boq, vanderWesthuizen:2025iam}, early DE models \cite{Chaussidon:2025npr, Qu:2024lpx, Seto:2024cgo}, and some others \cite{Wang:2024rus, RoyChoudhury:2024wri, Odintsov:2025kyw, Scherer:2025esj, RoyChoudhury:2025dhe, Paliathanasis:2025hjw, Paliathanasis:2025xxm, Odintsov:2025jfq}. For more detailed progress, we refer the reader to Refs. \cite{Yoo:2012ug,  Arun:2017uaw, Bahamonde:2017ize, Wang:2016och, Wang:2023gov, Cai:2025mas}.

Another approach of probing DE is the model independent reconstruction of DE \cite{Huterer:2002hy, Shafieloo:2005nd, Holsclaw:2010sk, Clarkson:2010bm, Shafieloo:2012ht, Zhao:2012aw, Nesseris:2012tt, Shafieloo:2012yh, Zhao:2017cud, Wang:2018fng, LHuillier:2019imn, Raveri:2021dbu, DESI:2024aqx}. 
One of the widely used methods is the redshift binning method.
For examples, in Ref. \cite{Pang:2024qyh}, the authors parameterize the DE EoS as three constants in each redshift bin with three bins artificially selected.
In Ref. \cite{Bansal:2025ipo}, the authors employ a piecewise-constant parameterization of the Hubble expansion rate $H(z)$ to investigate potential deviations from the standard $\Lambda$CDM.
In Ref. \cite{Reboucas:2024smm} the authors focus on low redshift data to investigate the EoS with constant values in five and ten redshift bins.
Another widely used method is the polynomial interpolation method.
For examples, in Ref. \cite{Orchard:2024bve}, the authors employ a quadratic polynomial parameterization of the DE density function to study the evidence for DE evolution.
In Ref. \cite{Ormondroyd:2025exu, Ormondroyd:2025iaf}, the authors utilize a free-form “flexknot” parameterization to represent $w(a)$ as a linear spline between free-moving nodes.
In Ref. \cite{Berti:2025phi}, the authors reconstruct the DE density through a third-degree piece-wise polynomial interpolation, allowing for direct constraints on its redshift evolution.
Other model independent reconstruction methods \cite{Luongo:2024fww, Mukherjee:2024ryz, Jiang:2024xnu, Liu:2024yib, Yang:2025kgc, Paliathanasis:2025cuc, Gao:2025ozb, Gu:2025xie, Cheng:2025cmb, Li:2025ula, Gonzalez-Fuentes:2025lei} are also useful in probing the characteristics of DE, enriching our understanding of its nature.

In this work, in order to investigate the properties of DE,  we focus on two mainstream model-independent reconstruction methods, i.e. redshift binning method and polynomial interpolation method.
It should be pointed out that, previous studies of model independent DE reconstruction always employ a single method to reconstruct DE EoS or DE density. In this work, we adpot two methods (both redshift binning method and polynomial interpolation method) to reconstruct both the DE EoS and the DE density.
Based on this cross-validation strategy, one can extract the information of DE properties, which is independent of the specific reconstruction method.

The structure of this paper is as follows: Section \ref{sec:method} introduces the model-independent methodology in this paper. Section \ref{sec:data_analysis} describes the observational datasets and the analysis framework in our investigation. Section \ref{sec:results} presents the main results of cosmology-fits. Finally, Section \ref{sec:conclusion} summarizes our conclusions.

\section{Model Independent Methodology}
\label{sec:method}
In this section, we first review the basics of standard cosmology, and then we introduce two model independent reconstruction methods.

Assuming a flat universe, the evolution of $H(z)$ is governed by the Friedmann equation, which can be written as
\begin{equation}
    \frac{H(z)}{H_0} = \sqrt{ \Omega_r (1+z)^4 + \Omega_m (1+z)^3  + (1-\Omega_{r}-\Omega_{m}) f(z) },
\end{equation}
where $H_0=100h$ km $\text{s}^{-1} \text{Mpc}^{-1}$ with $h$ the dimensionless Hubble constant, $\Omega_r$ and $\Omega_m$ refer to the energy densities in radiation and matter. $f(z)$ is the effective (normalized) DE density, which is related to the EoS parameter $w(z)$ as
\begin{equation}
\label{eq.fz}
    f(z)\equiv \frac{\rho_{de}(z)}{\rho_{de}(0)}=\text{exp} \left[ 3\int_0^z \frac{1+w(z')}{1+z'} dz' \right].
\end{equation}

A common approach used to explore alternative DE models is to parameterize the time evolution of its EoS. As one of the most popular model, the CPL parameterization is widely used in various studies,
\begin{equation}
    w= w_0 + w_a \frac{z}{1+z},
\end{equation}
where $w_0$ and $w_a$ are constants. The corresponding $f(z)$ is given by
\begin{equation}
    f(z) = (1+z)^{3(1+w_0+w_a)} \text{exp}\left( -\frac{3w_a z}{1+z} \right).
\end{equation}
In this paper, we take the CPL model as a fiducial reference.

\subsection{Redshift binning method}
\label{sec:Redshift_Binning_Method}
The key idea of redshift binning method is dividing the redshift range into several bins and assuming one DE quantity (EoS $w(z)$, or energy density $\rho_{de}$) as a piecewise constant.
For the redshift binning method, a key point is the choice of redshift intervals.
One common practice is to choose uniformly spaced redshift intervals \cite{Zhao:2007ew, Serra:2009yp, Zhao:2009ti, AlbertoVazquez:2012ofj, Wang:2015wga, Wang:2025vfb}. 
Alternatively, one can choose redshift nodes arbitrarily \cite{Huterer:2004ch, Qi:2008zk, Gong:2009ye, Cai:2010qp, Lazkoz:2012eh, Liu:2015mkm, Zheng:2017ulu, Dai:2018zwv, Luo:2018yvq}. 
A further sophisticated approach treats the discontinuity points of redshift as free model parameters, which can be determined by observational data \cite{Huang:2009rf, Wang:2010vj, Escamilla:2021uoj}.

In this paper, we divide the redshift range into uniformly spaced intervals.
The reason is as follows.
Generally, the observable $H(z)$ is assumed constant within each redshift slice. 
Therefore, the redshift slices should be selected such that variations in $H(z)$ remain approximately uniform with $z$, ensuring that the assumption of constancy holds well.
Ref. \cite{Wang:2009gt} demonstrated that using bins with a constant width $\Delta z$ for redshift slices is optimal, as it maintains the validity of assuming $H(z)$ as constant.

In our analysis, we evenly divide the total redshift range into several bins. Note that the highest redshift data point in the DESI BAO data is at $z=2.33$, we set an upper limit at $z_{max}=2.4$. Beyond this redshift ($z>2.4$), we impose a cosmological constant-like condition with $w=-1$ (or $f(z)=1$), following the treatment adopted by DESI in Ref. \cite{DESI:2024aqx}. In order to ensure that the conclusions do not depend on the specific number of bins, we divide the redshift range into three, four, and five bins.
Therefore, the detailed redshift intervals are:
\begin{itemize}[label=\textbullet]
    \item n=3: $z_i=$ 0.8/1.6/2.4;
    \item n=4: $z_i=$ 0.6/1.2/1.8/2.4;
    \item n=5: $z_i=$ 0.48/0.96/1.44/1.92/2.4.
\end{itemize}

\subsubsection{Binned EoS $w(z)$}
First, we assume that DE is parameterized in terms of $w(z)$, which is defined to be constant in each redshift bin, with a value $w_i$ in the $i^{\text{th}}$ redshift bin. 
The EoS can be written as
\begin{equation}
    w(z)=
    \begin{cases}
         w_1, & 0 \leq z \leq z_1 \\ 
         w_i, & z_{i-1} \leq z \leq z_i \quad (i>1)
    \end{cases} \quad ,
\end{equation}
Based on the continuity condition, the corresponding $f(z)$ is given by Ref. \cite{Sullivan:2007pd}, which can be written as
\begin{equation}
    f(z_{n-1} < z \leq z_n) = (1+z)^{3(1+w_n)} \prod_{i=0}^{n-1} (1+z_i)^{3(w_i-w_{i+1})}.
\end{equation}

\subsubsection{Binned density $f(z)$}
\label{subsubsec:binnedf}
As seen in Eqs. (\ref{eq.fz}) and (\ref{eq.codist}), the multiple integral relations smear out information about $w$ and its time variation \cite{Maor:2000jy}. It is hard to constrain $w$ without making specific assumptions about it. Therefore, if we constrain the DE density $\rho_{de}(z)$ instead, we minimize the smearing effect by removing one integral. In this case, we consider the piecewise constant $\rho_{de}$, $f(z)$ can be written as
\begin{equation}
    f(z) = 
    \begin{cases} 
        1, & 0 \leq z \leq z_1 \\ 
        f_i, & z_{i-1} \leq z \leq z_i \quad (i>1)
    \end{cases} \quad ,
\end{equation}
where $f_i$ is a piecewise constant that is free model parameter. From the relation $E(0)=1$, one can easily obtain $f_1=1$.

\subsection{Polynomial interpolation method}
Since the redshift binning method provides only the average information of DE within each bin, the reconstructed function describes its behavior in discrete segments. 
As an alternative, interpolation method can model the evolution of DE as a smooth, continuous function.
As a simple and effective method, we take into account the Lagrange polynomial interpolation method. 
This method enables us to construct a global polynomial. 

\begin{figure}[ht]
    \centering
    \includegraphics[width=\linewidth]{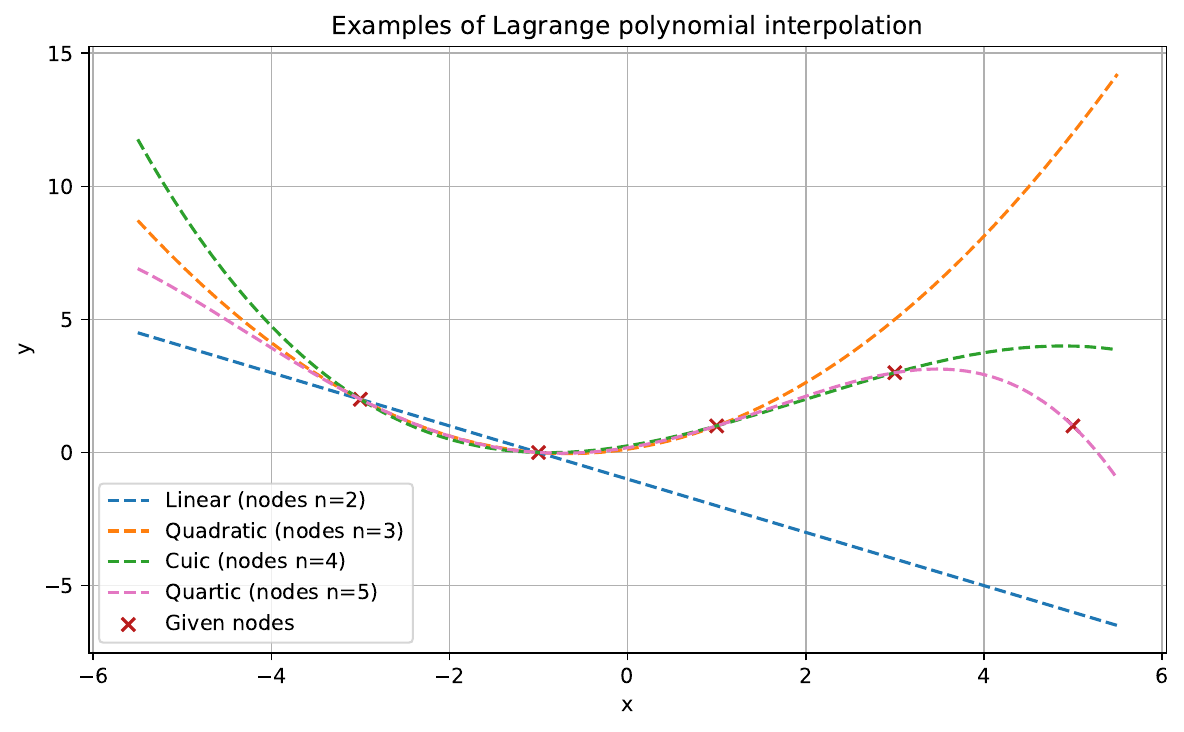}
    \caption{Lagrange polynomial interpolants built from $n=2,3,4,5$ nodes (n is the number of nodes, degree is $n-1$). Red crosses indicate the given nodes. The dashed lines correspond to the linear ($n=2$), quadratic ($n=3$), cubic ($n=4$), and quartic ($n=5$) interpolants. A higher polynomial degree requires more nodes.}
    \label{fig.polynomial}
\end{figure}

When considering the degree of polynomial, linear interpolation (nodes $n=2$) is the simplest case, where the interpolant is a straight line.
However, this form may introduce errors if the underlying function exhibits curvature.
Quadratic interpolation (nodes $n=3$) produces a parabola interpolant but may introduce spurious bulges and local extrema between nodes.
By contrast, cubic and quartic interpolation (nodes $n=4, 5$) provides a better approximation of curvature compared to linear and quadratic cases.
An illustrative figure (Fig. \ref{fig.polynomial}) comparing different polynomial orders is provided to improve clarity.
To ensure cross-validation with section \ref{sec:Redshift_Binning_Method}, we place the interpolation nodes at $z_i=i \times z_{\text{max}}/n$, where $n =3, 4$, and $5$.

Previous studies typically employed this interpolation method to reconstruct the evolution of DE density \cite{Wang:2001ht, Wang:2001da, Cardenas:2014jya, Grandon:2021nls, Bernardo:2021cxi}. 
In our work, we applies this method to both the DE EoS and density. 

\subsubsection{Interpolation of EoS $w(z)$}
In order to constrain the time-variation of DE in a robust manner, it is important that we allow the EoS to be an arbitrary function of redshift $z$. 
This approach avoids assumptions about a specific functional form.
Taking $w=-1$ as a baseline, we interpolate deviations from this value using a Lagrange polynomial \cite{gezerlis2020numerical} of order $n$,
\begin{equation}
    w(z)= -1 +\sum^n_{i=1} \alpha_i \frac{(z-z_1)\cdots (z-z_{i-1}) (z-z_{i+1}) \cdots (z-z_n)}{(z_i-z_1)\cdots (z_i-z_{i-1}) (z_i-z_{i+1}) \cdots (z_i-z_n)},
\end{equation}
where $\alpha_i$ are the independent variables to be estimated from data. Each interpolation node $\alpha_i$ is evaluated at $z_i=i \times z_{\text{max}}/n$. If the data prefer $\alpha_i =0$ for all $i$, then $w(z)$ returns to the cosmological constant case.

\subsubsection{Interpolation of density $f(z)$}
Analogously, we can also assume that the DE density has an $n$-node Lagrange polynomial form as 
\begin{equation}
    f(z)= \sum^n_{i=1} \beta_i \frac{(z-z_1)\cdots (z-z_{i-1}) (z-z_{i+1}) \cdots (z-z_n)}{(z_i-z_1)\cdots (z_i-z_{i-1}) (z_i-z_{i+1}) \cdots (z_i-z_n)},
\end{equation}
The physical normalization $f(0)=1$ imposes one linear constraint on free variables $\beta_i$. In practice we fix $\beta_1$ by other variables $\beta_i$ via the relation $f(0)=1$. Therefore, only the remaining $n-1$ model parameters need to be determined by the data.

\section{Datasets and Analysis Methodology}
\label{sec:data_analysis}
This Section presents the datasets used to explore the DE properties described in Section \ref{sec:method}. We present the individual datasets in Section \ref{subsec:bao} to Section \ref{subsec:sn}. In Section \ref{subsec:chisq}, we detail our analysis methodology.

\subsection{Baryon Acoustic Oscillation}
\label{subsec:bao}
For the BAO data, we adopt the DR2 three-year data released by the DESI collaboration \cite{DESI:2025zgx}.
The quantities of BAO correspond to several key distances: $D_M$, $D_H$, and $D_V$. In a spatially flat FLRW universe, the transverse comoving distance $D_M$ at redshift $z$ is defined as
\begin{equation}
\label{eq.codist}
    D_M(z) = \frac{c}{H_0} \int_0^z \frac{d z'}{H(z') /H_0},
\end{equation}
where $c$ is the speed of light.  The distance variable $D_H$ is related to the Hubble parameter $H(z)$ as $D_H(z) = c/H(z)$. The angle-averaged distance $D_V$ is given by $D_V(z) = [zD_M(z)^2 D_H(z) ]^{1/3}$.

BAO measurements depend on the radius of the sound horizon at the drag epoch $r_d$. This represents the distance that sound can travel between the Big Bang and the drag epoch, which marks the time when baryons decoupled. The sound horizon can be expressed as
\begin{equation}
\label{rs}
    r_s(z) = \int_{z}^\infty \frac{c_s(z')}{H(z')}dz',
\end{equation}
where $c_s(z)$ is the speed of sound. For those models do not deviate too much from $\Lambda$CDM, the drag epoch sound horizon can be approximated by \cite{Brieden:2022heh, Schoneberg:2022ggi}
\begin{equation}
    r_d \simeq |_{\Lambda CDM} \frac{147.05}{\mathrm{Mpc}} \left(\frac{\omega_\mathrm{m}}{0.1432}\right)^{-0.23} \left(\frac{N_{\mathrm{eff}}}{3.04}\right)^{-0.1} \left(\frac{\omega_b}{0.02236}\right)^{-0.13},
\end{equation}
where $\omega_m \equiv \Omega_m h^2$, $\omega_b \equiv \Omega_b h^2$, and $N_{\mathrm{eff}}$ is the effective number of extra relativistic degrees of freedom. We assume standard neutrino content $N_{\mathrm{eff}} = 3.044$ in our analysis.

\subsection{Cosmic Microwave Background}
\label{subsec:cmb}
For the CMB data, we use the latest distance priors from Ref. \cite{Wolf:2024eph} which compresses the CMB data (Planck's TTTEEE \cite{Planck:2018vyg, Planck:2019nip} and ACT DR6 lensing \cite{ACT:2023dou, ACT:2023kun}) into the background quantities. The method of distance priors \cite{Efstathiou:1998xx, Wang:2006ts, Wang:2007mza} serves as a compressed dataset that encapsulates key information from the full CMB data. This approach allows us to substitute the full CMB power spectrum with a more compact representation while retaining key cosmological information.

The distance priors contain two primary features of the CMB power spectrum: the shift parameter $R$ and the acoustic scale $l_a$. The shift parameter $R$ affects the peak heights in the CMB temperature power spectrum along the line of sight, while the acoustic scale $l_a$ influences the spacing of the peaks in the transverse direction. These parameters are defined as
\begin{align}
    R &\equiv \frac{ D_M(z_*)\sqrt{\Omega_m H_0^2}}{c},\\
    l_a &\equiv \frac{\pi D_M(z_*)}{r_s(z_*)},
\end{align}
where $z_*$ is the redshift at the photon decoupling epoch, which can be calculated by an approximate formula \cite{Hu:1995en}:
\begin{equation}
    z_* = 1048[1+0.00124(\Omega_b h^2)^{-0.738}][1+g_1(\Omega_m h^2)^{g_2}],
\end{equation}
where
\begin{align}
    g_1 &= \frac{0.0783(\Omega_b h^2)^{-0.238}}{1+39.5(\Omega_b h^2)^{0.763}},\\
    g_2 &= \frac{0.560}{1+21.1(\Omega_b h^2)^{1.81}}.
\end{align}

\subsection{Type Ia Supernovae}
\label{subsec:sn}
For the SN data, we use three different compilations. The PantheonPlus compilation comprises 1550 supernovae, spanning a redshift range $0.01 \leq z \leq 2.26$ \cite{Brout:2022vxf}, which does not include the calibration sample at lower redshifts.
The Union3 compilation contains 2087 SNe Ia over the redshift range $0.01<z<2.26$, which is processed through the Unity 1.5 pipeline based on Bayesian Hierarchical Modelling \cite{Rubin:2023ovl}. 
The Dark Energy Survey Year 5 (DESY5) compilation consists of 194 low-redshift SNe Ia ($0.025<z<0.1$) and 1635 photometrically classified SNe Ia covering the range $0.1<z<1.3$ \cite{DES:2024jxu}. 
All three likelihoods are implemented in the $\mathbf{Cobaya}$ sampling code and the underlying data are publicly available.\footnote{https://github.com/CobayaSampler/sn\_data.git}

SN data are usually specified in terms of the distance modulus $\mu$. The theoretical distance modulus $\mu_{th}$ in a flat universe is given by
\begin{equation}
    \mu_{th} = 5 \log_{10} \left[ \frac{d_L(z_{hel}, z_{cmb})}{\text{Mpc}} \right] +25,
\end{equation}
where $z_{hel}$ and $z_{cmb}$ are the heliocentric and CMB rest-frame redshifts of SN. The luminosity distance $d_L$ is
\begin{equation}
    d_L(z_{hel}, z_{cmb})= (1+z_{hel}) r(z_{cmb}),
\end{equation}
where $r(z)$ is the comoving distance given by Eq. (\ref{eq.codist}). 

\subsection{Chi-square Statistic}
\label{subsec:chisq}
To quantify the goodness of fit between predicted values from cosmological models and actual measurements from astronomical observations, we use the $\chi^2$ statistic. By minimizing the $\chi^2$ function, one can identify the model parameters that best describe the observed universe.

For independent data points, the $\chi^2$ function is defined as
\begin{equation}
    \chi^2_\xi = \frac{(\xi_{th} - \xi_{obs})^2}{\sigma_\xi ^2},
\end{equation}
where $\xi_{th}$ is the theoretically predicted value, $\xi_{obs}$ is the experimentally measured value, and $\sigma_\xi$ is the standard deviation. For correlated data points, the $\chi^2$ function is given by
\begin{equation}
    \chi^2 = \Delta \xi ^T \text{Cov}^{-1} \Delta \xi,
\end{equation}
where $\Delta \xi \equiv \xi_{th} - \xi_{obs} $, and Cov$^{-1}$ is an inverse covariance matrix that characterizes the errors in the data.

Since we use the BAO data from DESI DR2, the CMB distance priors from Planck 2018 and ACT DR6, and the SN data from PantheonPlus, Union3, and DESY5, respectively, the total $\chi^2$ is
\begin{equation}
    \chi^2 = \chi^2_{BAO} + \chi^2_{CMB} + \chi^2_{SN}.
\end{equation}
Assuming the measurement errors be Gaussian, the likelihood function is given by
\begin{equation}
    \mathcal{L}\propto e^{-\chi^2 /2}.
\end{equation}

In this work, we sample from the dark energy parameter posterior distributions using the MCMC code $\mathbf{Cobaya}$\footnote{https://github.com/CobayaSampler/cobaya} \cite{Torrado:2020dgo} with the $\mathbf{mcmc}$ sampler \cite{Lewis:2013hha, Lewis:2002ah}. 
Convergence of an MCMC run is assessed using the Gelman-Rubin statistic \cite{Gelman:1992zz} with a tolerance of $|R-1| < 0.02$. 
We make use of $\mathbf{Getdist}$\footnote{https://github.com/cmbant/getdist}  \cite{Lewis:2019xzd} to perform statistical analysis of the MCMC samples.
The priors used in the analysis are given in Table \ref{tab:flatprior}.
As pointed out in Ref. \cite{DESI:2024aqx}, unlike the EoS parameter $w(z)$, the direct reconstruction of $f(z)$ allows for the effective energy density to become negative, which can happen in modified gravity and various other dark energy models.
Thus we impose negative priors on $f_i$.

\begin{table}
\begin{center}
\renewcommand{\arraystretch}{1.2}
\begin{tabular}{lll}
\hline
$ $ & $\textbf{Parameters}$ & $\textbf{Priors}$ \\
\hline
\multirow{3}{*}{background}
  &  $\Omega_m$      & $\mathcal{U}[0.01, 0.99]$ \\
  &  $\Omega_{b}h^2$ & $\mathcal{U}[0.005, 0.1]$ \\
  &  $h$         & $\mathcal{U}[0.2, 1.0]$ \\
\hline
\multirow{2}{*}{Binning method}
  &  $w_i$   & $\mathcal{U}[-3.0, 1.0]$ \\
  &  $f_i$     & $\mathcal{U}[-1.0, 2.0]$ \\
\hline
\multirow{2}{*}{Interpolation method}
  &  $\alpha_i$       & $\mathcal{U}[-3.0, 1.0]$ \\
  &  $\beta_i$         & $\mathcal{U}[-1.0, 2.0]$ \\
\hline
\end{tabular}
\end{center}
\caption{Parameters and priors used in the analysis. $\mathcal{U}[a, b]$ represents an uniform distribution from a to b.}
\label{tab:flatprior}
\end{table}

\section{Results and Discussion}
\label{sec:results}
In this section, we present the results of cosmology-fits based on different combinations of observational data and reconstruction methods. 
Subsection \ref{subsec:results_RB} gives the results of redshift binning method, while subsection \ref{subsec:results_PI} gives the results of polynomial interpolation method.

\subsection{Redshift binning method}
\label{subsec:results_RB}
\subsubsection{Binned EoS $w(z)$}
\label{result_binnedw}
\begin{figure*}[htbp]
\small
    \centering
        \includegraphics[width=0.33\textwidth]{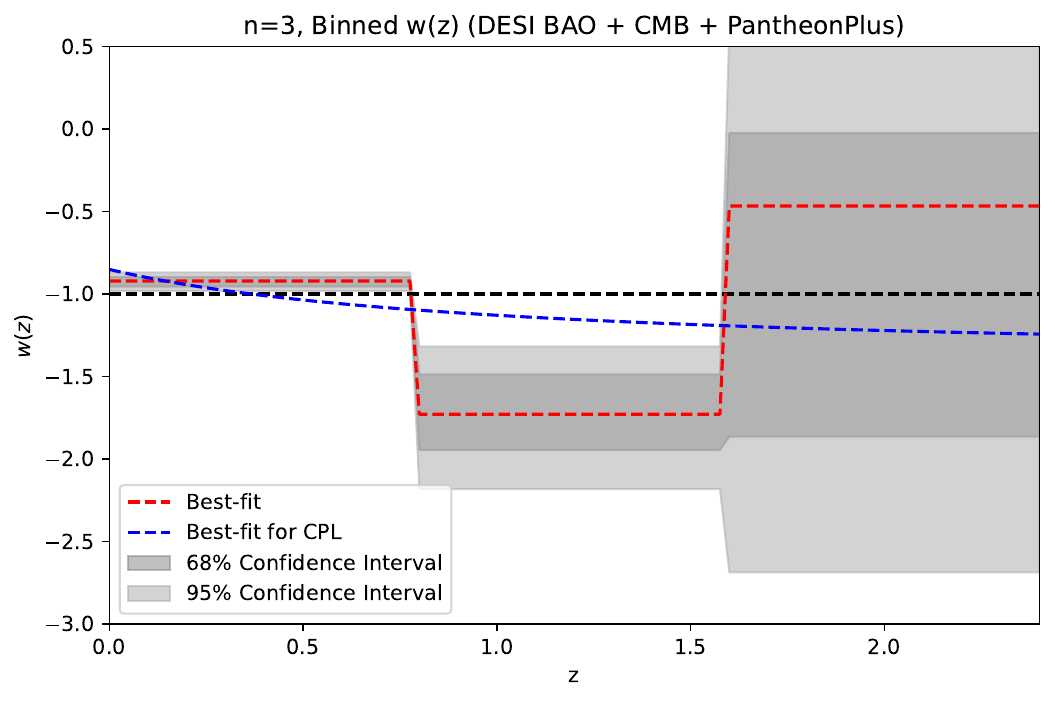}%
        \includegraphics[width=0.33\textwidth]{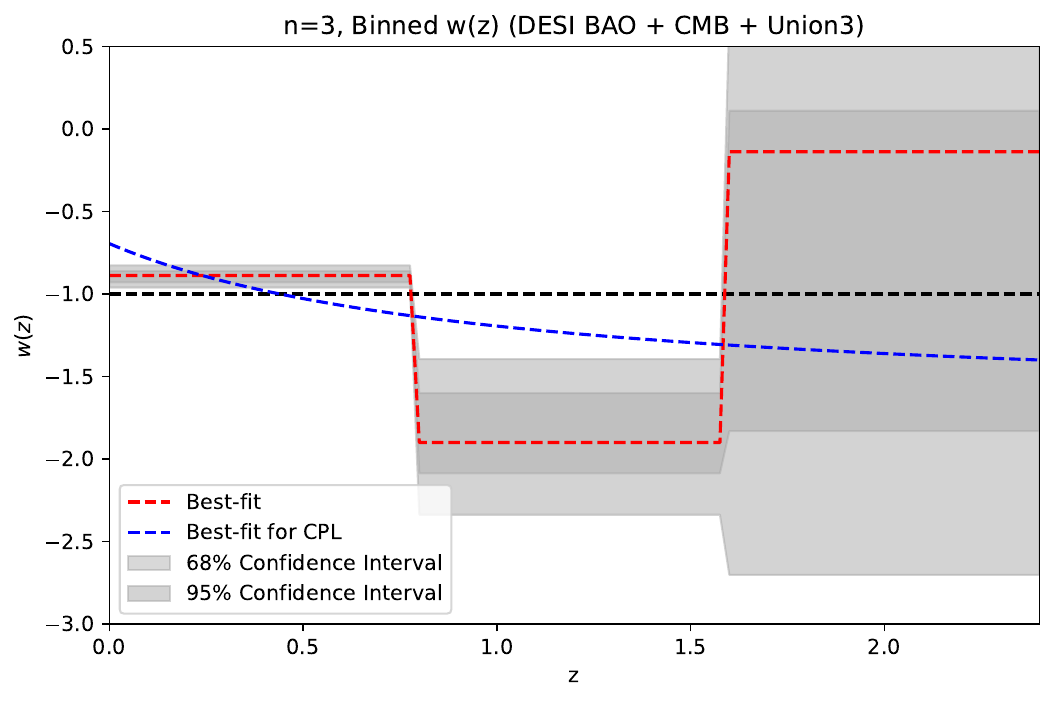}%
        \includegraphics[width=0.33\textwidth]{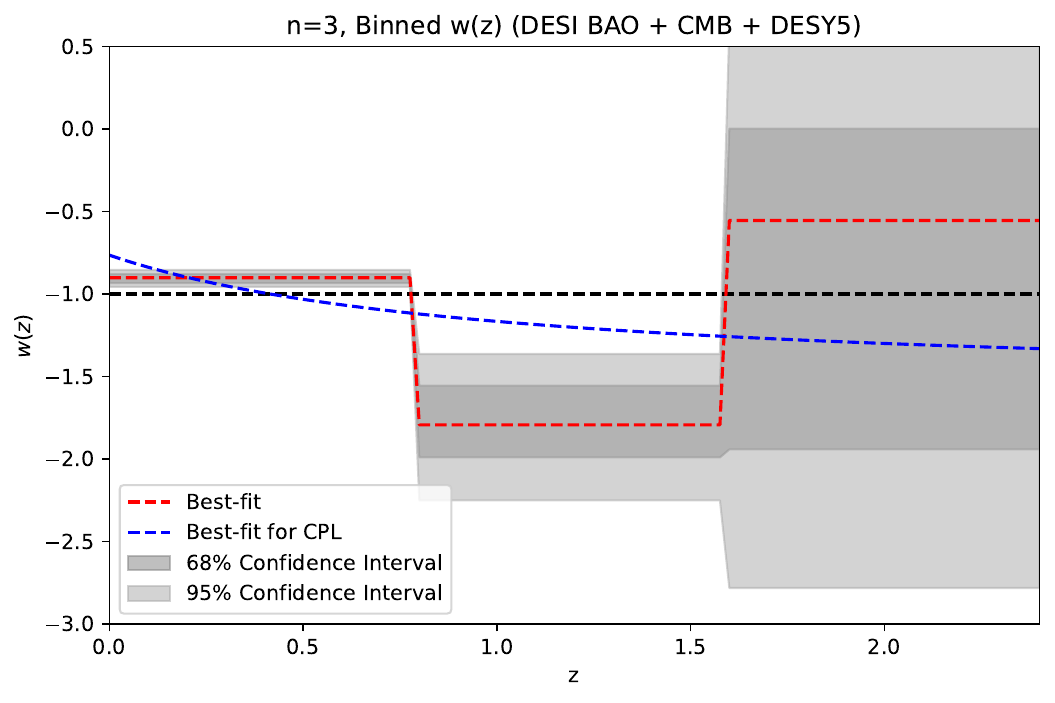}%
    
        \includegraphics[width=0.33\textwidth]{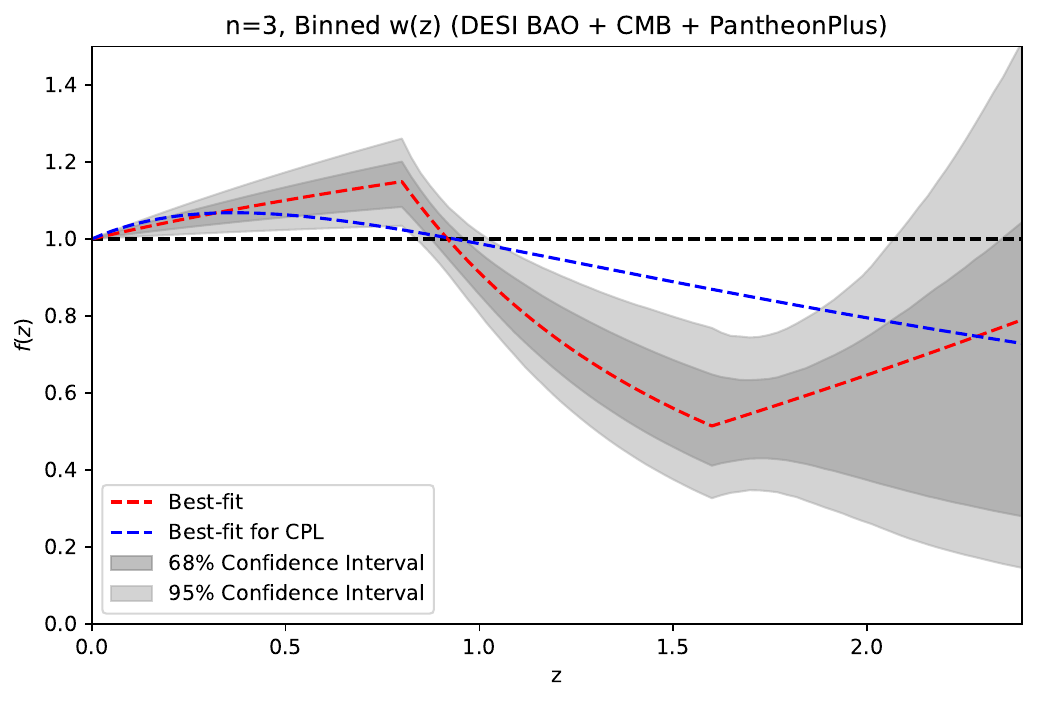}%
        \includegraphics[width=0.33\textwidth]{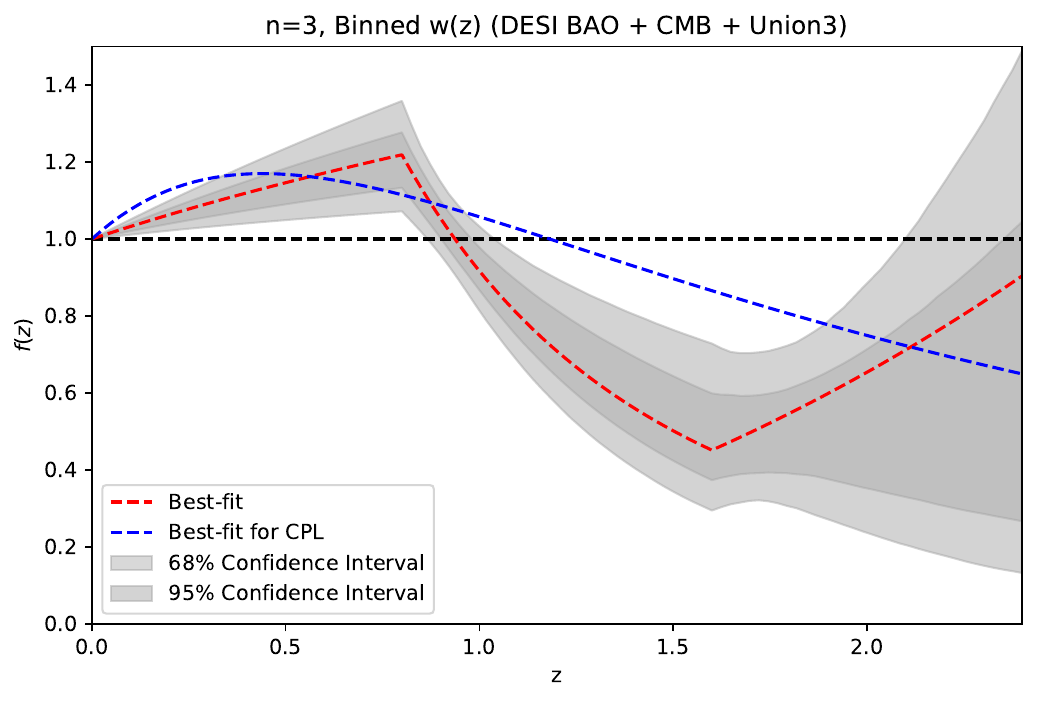}%
        \includegraphics[width=0.33\textwidth]{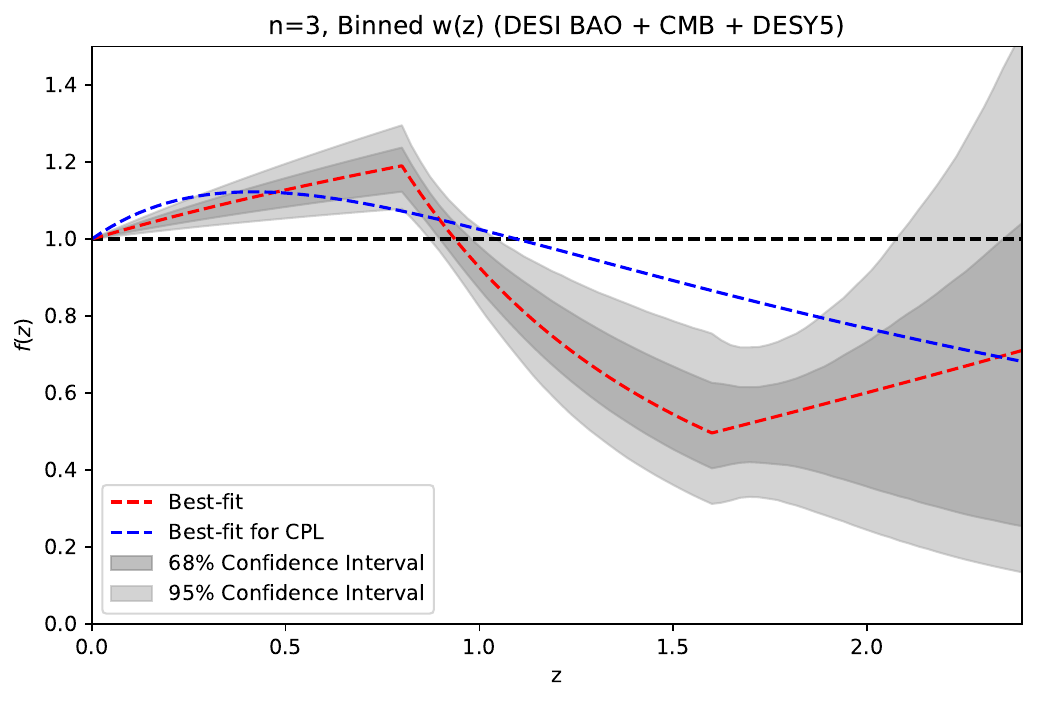}%

    \caption{Reconstruction of the DE EoS $w(z)$ (upper panels) and the corresponding DE density $f(z)$ using three redshift bins. Red dashed lines denote the best-fit values within each bin, while the blue dashed lines correspond to the best-fit CPL model for comparison. Shaded regions represent the 68\% (dark gray) and 95\% (light gray) confidence intervals. Results presented from left to right columns combine DESI BAO + CMB data with PantheonPlus, Union3, and DESY5 supernovae datasets, respectively.}
    \label{fig.3binnedwz}
\end{figure*}

\begin{figure*}[htbp]
\small
    \centering
        \includegraphics[width=0.33\textwidth]{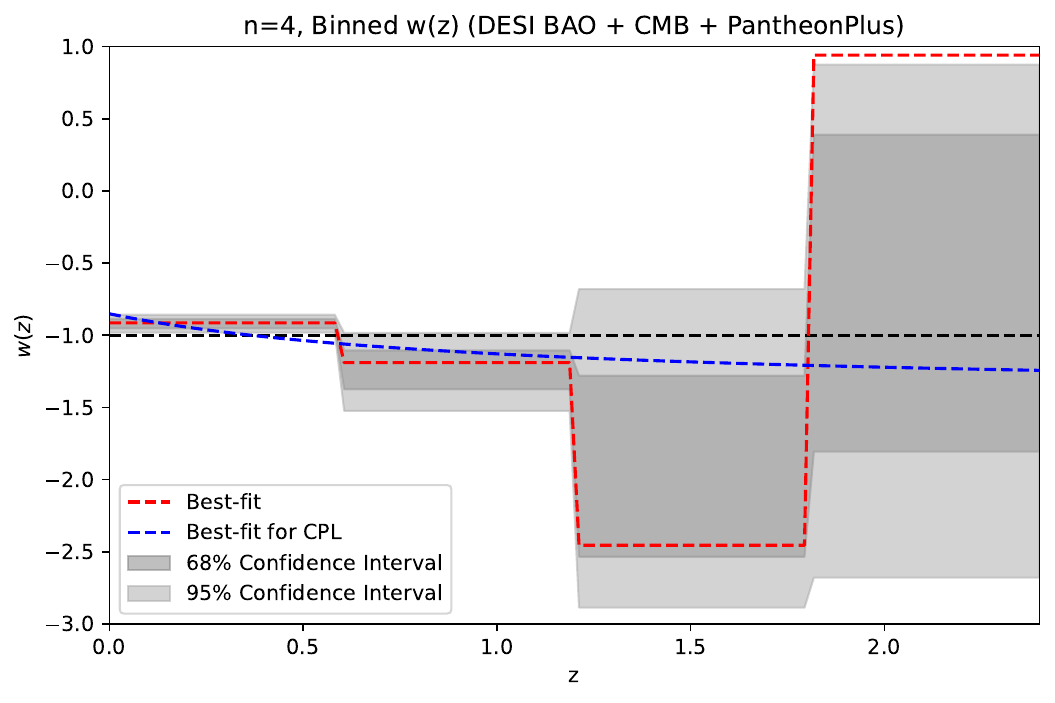}
        \includegraphics[width=0.33\textwidth]{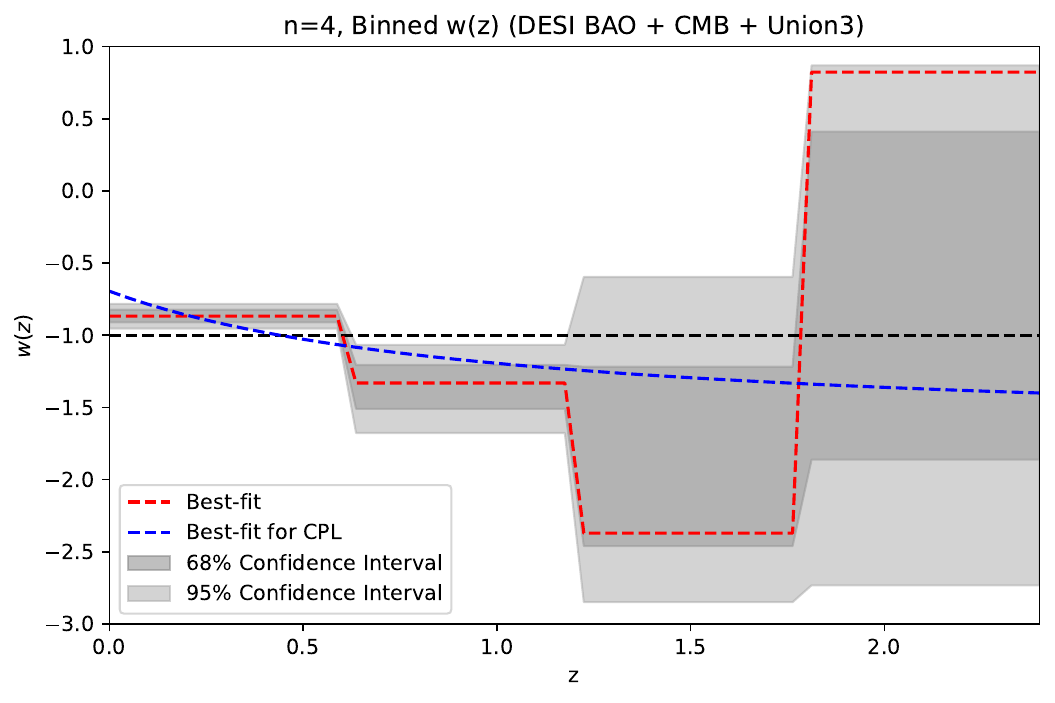}
        \includegraphics[width=0.33\textwidth]{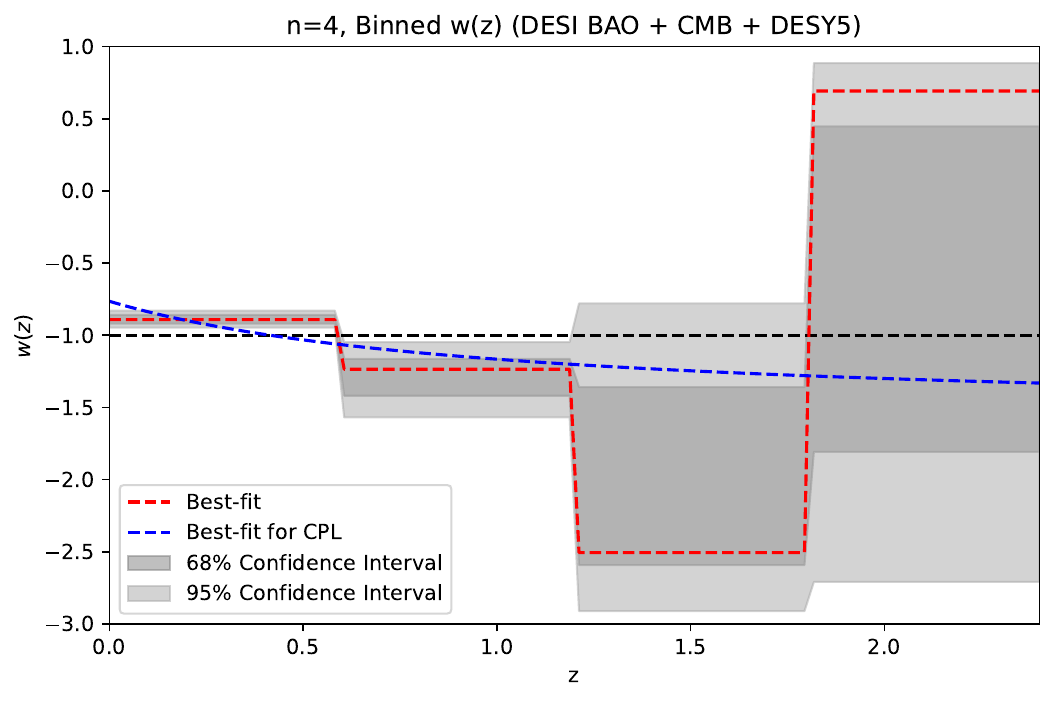}
    
        \includegraphics[width=0.33\textwidth]{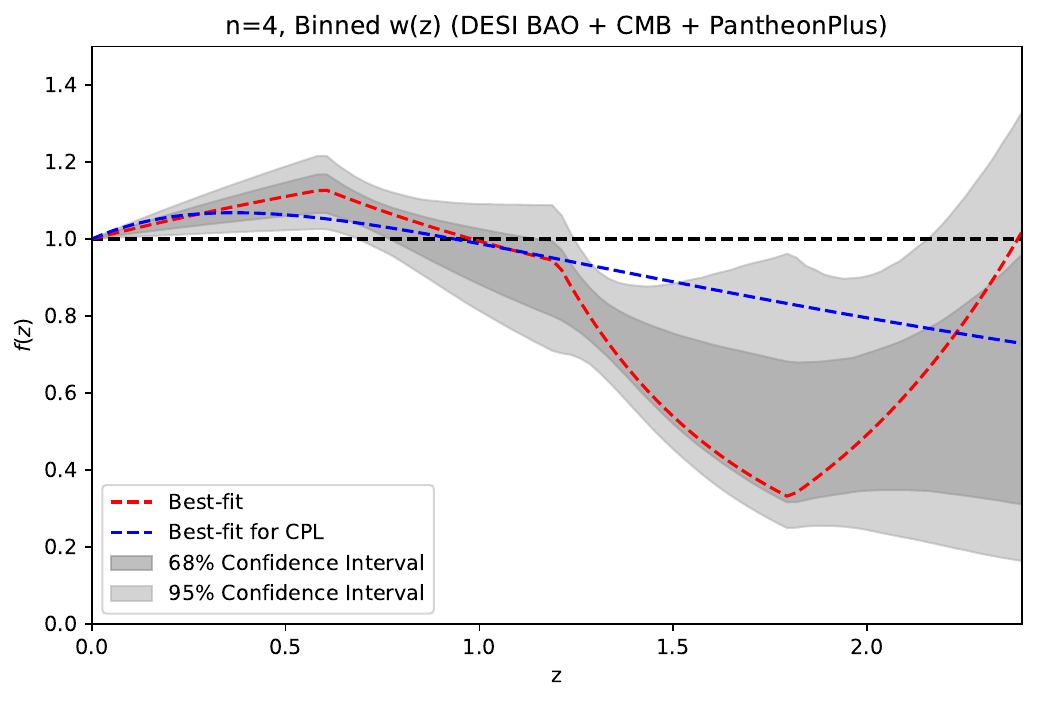}
        \includegraphics[width=0.33\textwidth]{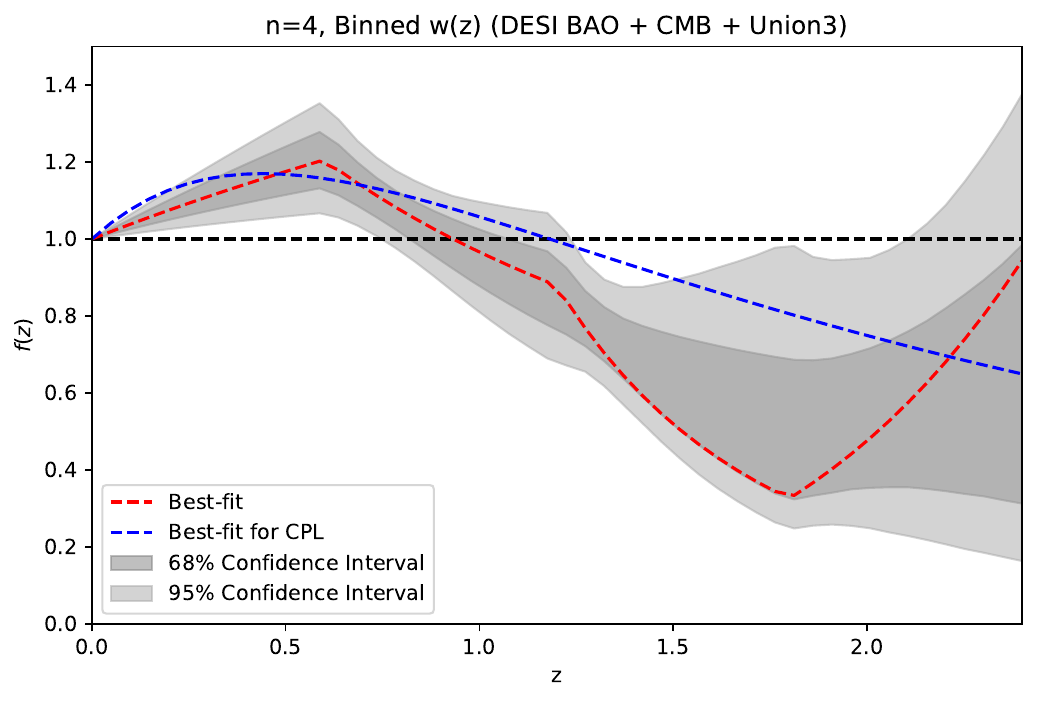}
        \includegraphics[width=0.33\textwidth]{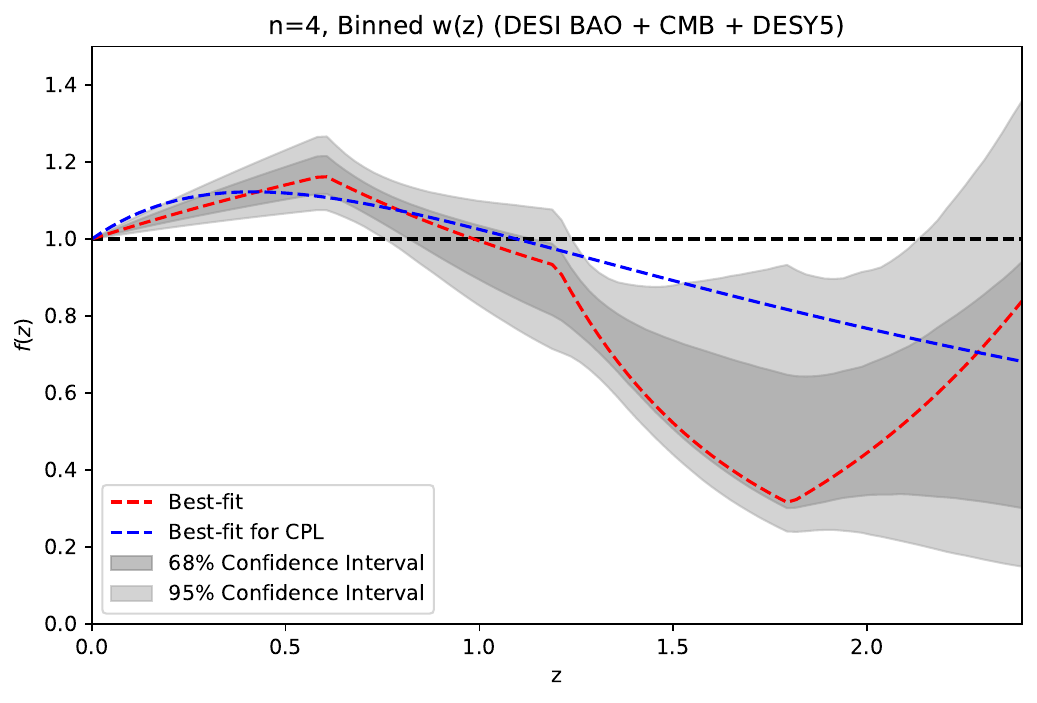}

    \caption{Reconstruction of the DE EoS $w(z)$ (upper panels) and the corresponding DE density $f(z)$ using four redshift bins. Red dashed lines denote the best-fit values within each bin, while the blue dashed lines correspond to the best-fit CPL model for comparison. Shaded regions represent the 68\% (dark gray) and 95\% (light gray) confidence intervals. Results presented from left to right columns combine DESI BAO + CMB data with PantheonPlus, Union3, and DESY5 supernovae datasets, respectively.}
    \label{fig.4binnedwz}
\end{figure*}

\begin{figure*}[htbp]
\small
    \centering
        \includegraphics[width=0.33\textwidth]{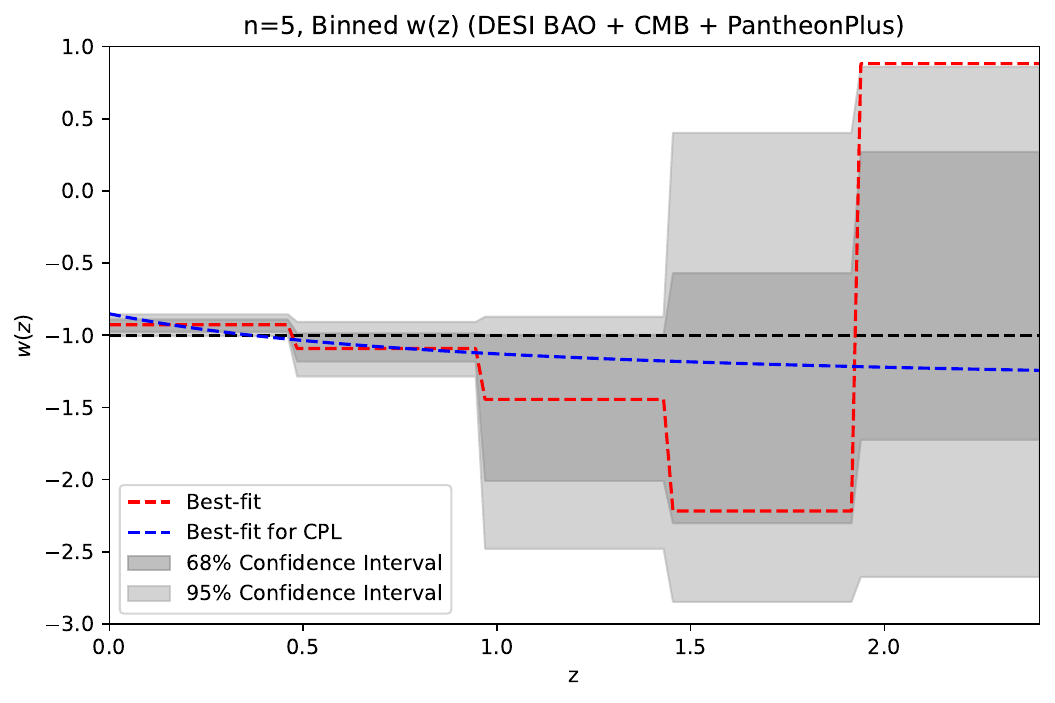}
        \includegraphics[width=0.33\textwidth]{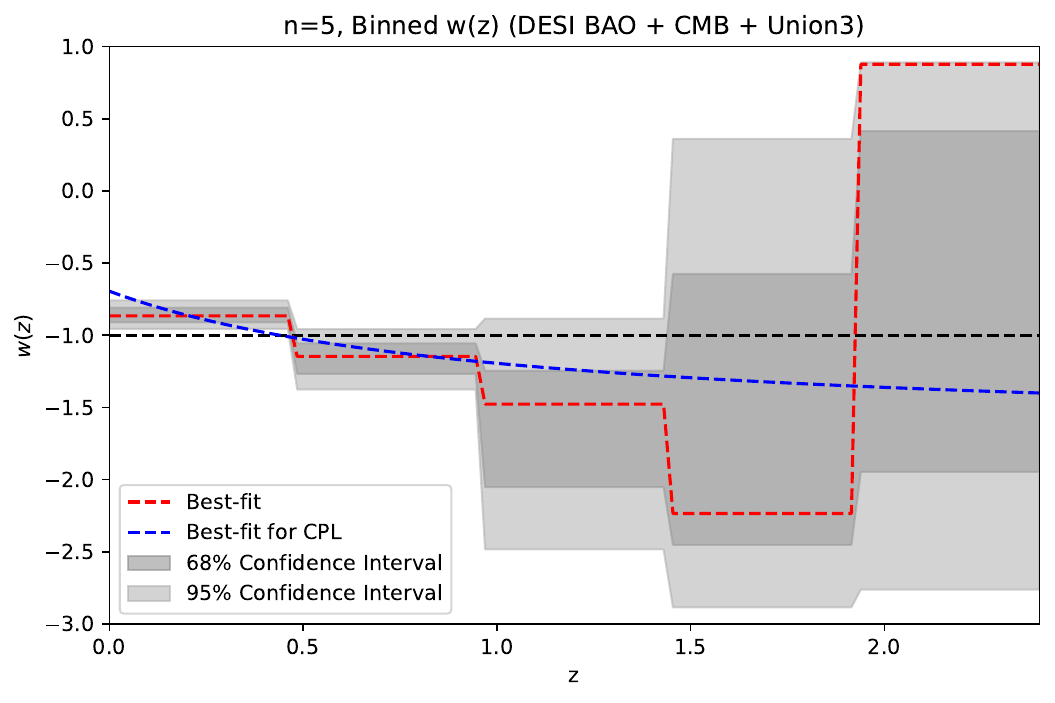}
        \includegraphics[width=0.33\textwidth]{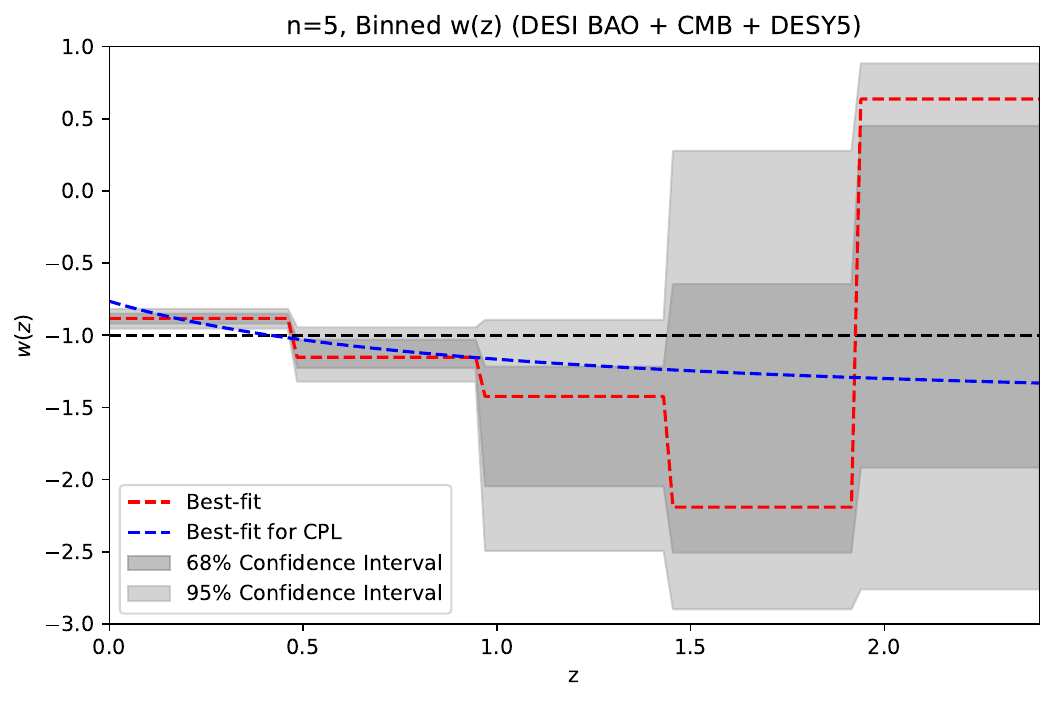}

        \includegraphics[width=0.33\textwidth]{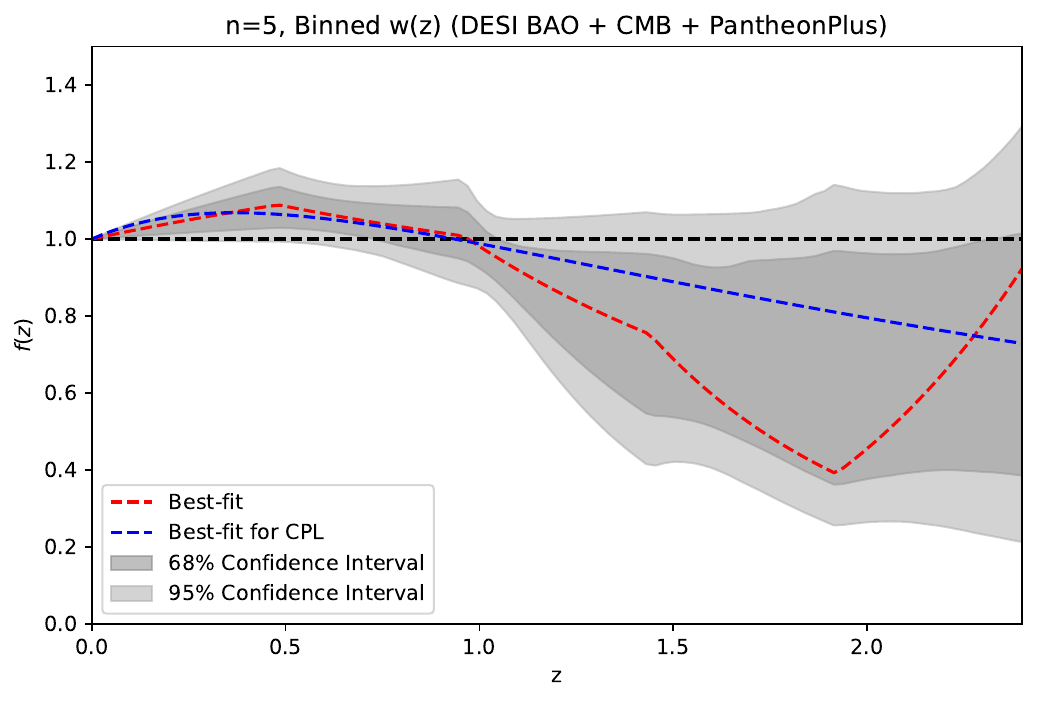}
        \includegraphics[width=0.33\textwidth]{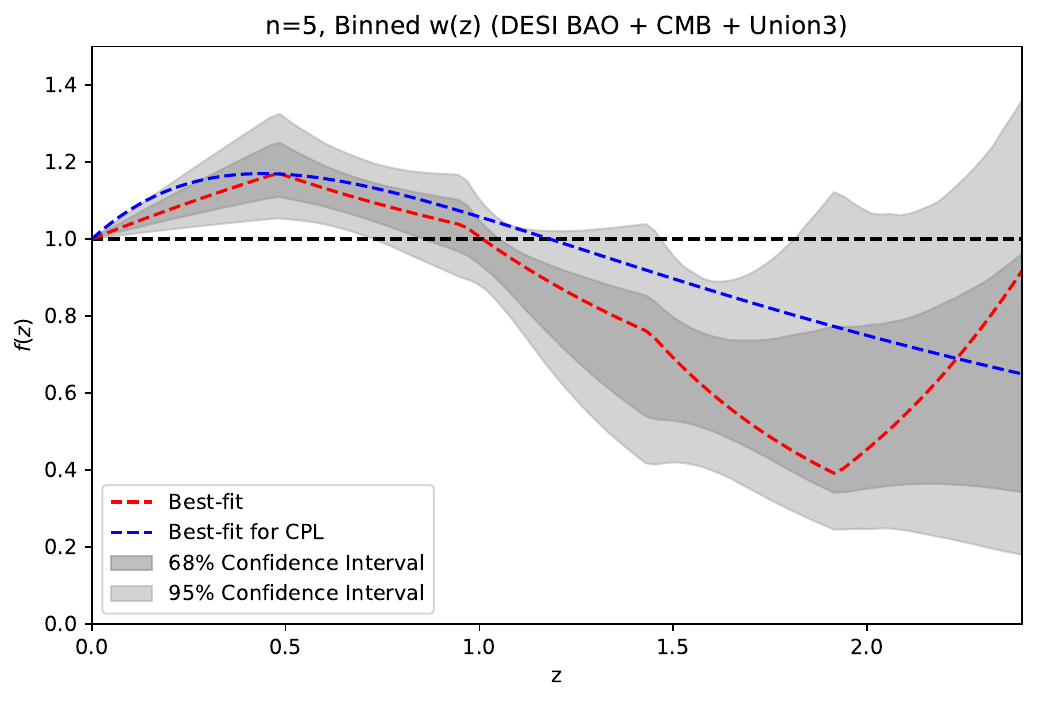}
        \includegraphics[width=0.33\textwidth]{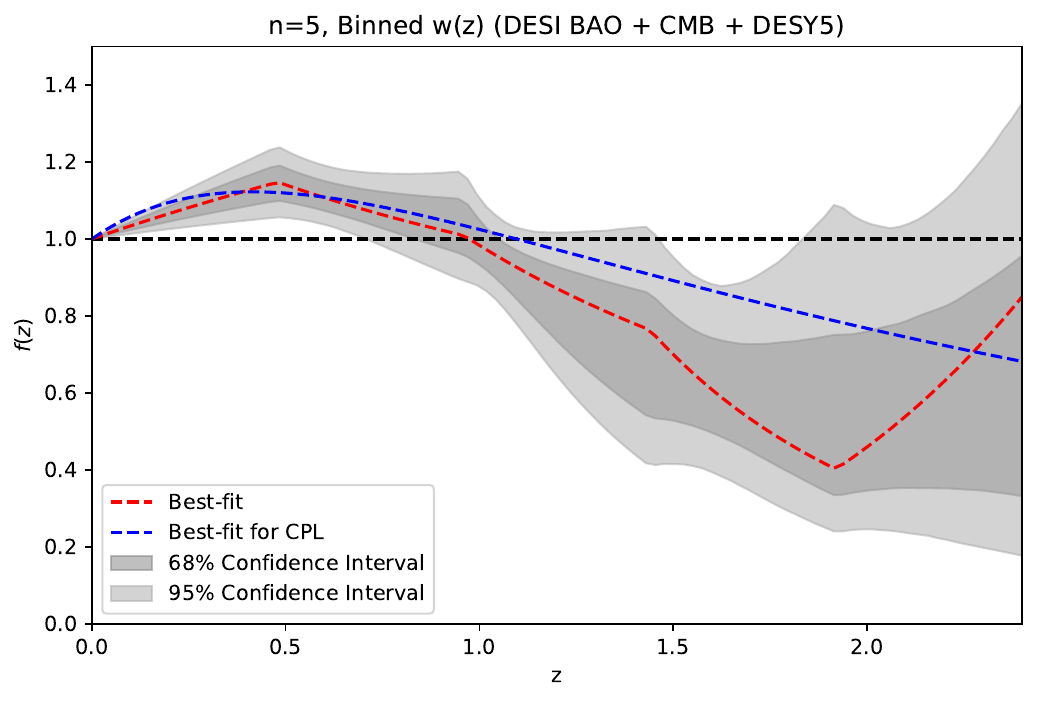}

    \caption{Reconstruction of the DE EoS $w(z)$ (upper panels) and the corresponding DE density $f(z)$ using five redshift bins. Red dashed lines denote the best-fit values within each bin, while the blue dashed lines correspond to the best-fit CPL model for comparison. Shaded regions represent the 68\% (dark gray) and 95\% (light gray) confidence intervals. Results presented from left to right columns combine DESI BAO + CMB data with PantheonPlus, Union3, and DESY5 supernovae datasets, respectively.}
    \label{fig.5binnedwz}
\end{figure*}

For the binned $w(z)$ method, $w(z)$ is characterized by a piecewise constant in each redshift bin. 
The reconstruction results are shown in Figs. \ref{fig.3binnedwz}, \ref{fig.4binnedwz}, \ref{fig.5binnedwz}, where redshift range is divided into three, four, and five bins, respectively.
For simplicity, we refer to these cases as "3bins-w", "4bins-w", and "5bins-w", respectively.
In these figures, the reconstructions of DE EoS $w(z)$ are shown in the upper rows. In order to make a direct comparison between the results of the binned $w(z)$ and the binned $f(z)$, we also present the corresponding DE density results in the lower rows.
Note that three kinds of data combinations (i.e. DESI BAO+CMB+PantheonPlus, DESI BAO+CMB+Union3, and DESI BAO+CMB+DESY5) are considered in this work, so Figs. \ref{fig.3binnedwz}, \ref{fig.4binnedwz}, \ref{fig.5binnedwz} have three columns. The red dashed lines denote the best-fit results of redshift binning models. For comparison, the CPL best-fit results are given in the form of the blue dashed lines. The mean values with standard deviations within each redshift bin are summarized in Table \ref{tab:binningw}. 

\begin{table*}
\small
\begin{center}
\resizebox{\textwidth}{!}{
\renewcommand{\arraystretch}{1.5}
\begin{tabular}{llllll}
\hline
$\text{Model/Dataset}$ & $w_1$/CL & $w_2$/CL & $w_3$/CL & $w_4$/CL & $w_5$/CL \\
\hline
\multirow{4}{*}{}
\textbf{3bins-w} & & & & & \\
DESI+CMB+PantheonPlus & $-0.926\pm 0.028$/ $2.64\sigma$ & $-1.712 \pm 0.221$/ $3.22\sigma$ & $-0.921 \pm 0.897$/ $0.08\sigma$ & - & - \\
DESI+CMB+Unino3  &  $-0.895 \pm 0.034$/ $3.08\sigma$ & $-1.842 \pm 0.239$/ $3.52\sigma$ & $-0.802 \pm 0.922$/ $0.21\sigma$ & - & - \\
DESI+CMB+DESY5  &  $-0.906 \pm 0.027$/ $3.48\sigma$ & $-1.769 \pm 0.220$/ $3.49\sigma$ & $-0.916 \pm 0.932$/ $0.09\sigma$ & - & - \\
\hline
\multirow{4}{*}{}
\textbf{4bins-w} & & & & & \\
DESI+CMB+PantheonPlus & $-0.921 \pm 0.032$/ $2.46\sigma$ & $-1.235 \pm 0.135$/ $1.74\sigma$ & $-1.934 \pm 0.601$/ $1.55\sigma$ & $-0.617 \pm 1.023$/ $0.37\sigma$ & - \\
DESI+CMB+Unino3  &  $-0.865 \pm 0.043$/ $3.14\sigma$ & $-1.358 \pm 0.149$/ $2.40\sigma$ & $-1.843 \pm 0.599$/ $1.40\sigma$ & $-0.587 \pm 1.053$/ $0.39\sigma$ & - \\
DESI+CMB+DESY5  &  $-0.890 \pm 0.030$/ $3.66\sigma$ & $-1.287 \pm 0.131$/ $2.19\sigma$ & $-2.004 \pm 0.590$/ $1.70\sigma$ & $-0.571 \pm 1.055$/ $0.40\sigma$ & - \\
\hline
\multirow{4}{*}{}
\textbf{5bins-w} & & & & & \\
DESI+CMB+PantheonPlus & $-0.923 \pm 0.036$/ $2.13\sigma$ & $-1.084 \pm 0.096$/ $0.87\sigma$ & $-1.646 \pm 0.412$/ $1.56\sigma$ & $-1.484 \pm 0.924$/ $0.52\sigma$ & $-0.700 \pm 1.059$/ $0.28\sigma$ \\
DESI+CMB+Unino3  &  $-0.859 \pm 0.050$/ $2.82\sigma$ & $-1.161 \pm 0.104$/ $1.54\sigma$ & $-1.630 \pm 0.400$/ $1.57\sigma$ & $-1.563 \pm 0.891$/ $0.63\sigma$ & $-0.702 \pm 1.091$/ $0.27\sigma$ \\
DESI+CMB+DESY5  &  $-0.885 \pm 0.034$/ $3.38\sigma$ & $-1.128 \pm 0.096$/ $1.33\sigma$ & $-1.617 \pm 0.412$/ $1.49\sigma$ & $-1.645 \pm 0.887$/ $0.72\sigma$ & $-0.624 \pm 1.086$/ $0.34\sigma$ \\
\hline
\end{tabular}%
}
\end{center}
\caption{Constraints on $w_i$ in each redshift bin, derived from different datasets using the redshift binning method. The confidence level (CL) indicates the deviation from $w(z)=-1$.}
\label{tab:binningw}
\end{table*}

For the 3bins-w case, the results of all three data combinations yield similar trends: the EoS parameter $w_1 > -1$ deviates from $w=-1$ at a confidence level of at least $2.64\sigma$, while $w_2 < -1$ deviates at a level of at least $3.22\sigma$. 
These results reveal a significant deviation from the $\Lambda$CDM model. 
This deviation grows when PantheonPlus is replaced by Union3 or DESY5. 
In addition, $w(z)$ has a behavior of crossing $w=-1$, which corresponds to a quintom type DE. 
As for the second bin, since the binning method provides only the average information of DE within each bin, it is the limitation of binning method that leads to three results deviating from the CPL best-fit.
For the EoS parameter $w_3$, the error bar is very large, this is because the data at this redshift region is rare.
Furthermore, based on the lower row of Fig. \ref{fig.3binnedwz}, one can see a common trend for the DE density. It slowly increases in the first redshift bin and obviously decreases in the second bin.

For the 4bins-w case, the results are broadly consistent with those from the 3bins-w case. 
The EoS parameter $w_1 > -1$ deviates from $w=-1$ at a confidence level of at least $2.4\sigma$, while $w_2 < -1$ and $w_3 < -1$ deviate at levels of at least $1.7\sigma$ and $1.4\sigma$, respectively. 
The decreasing trend of $w(z)$ persists, which is consistent with the CPL best-fit behavior. 
For the EoS parameter $w_4$, the large error bar is due to the limited data points in this redshift range. 
Therefore, the poor constraint on this parameter leads to the best-fit value of $w_4$ lying outside the $1\sigma$ confidence region. 
As shown in the lower row of Fig. \ref{fig.4binnedwz}, the corresponding DE density increases in the first bin and then decreases across the second and third bins.

For the 5bins-w case, the overall trends remain consistent with the previous two cases. 
The EoS parameter $w_1 > -1$ deviates from $w=-1$ at a level of at least $2.13\sigma$, while $w_2 < -1$, $w_3 < -1$, and $w_4 < -1$ deviate at levels of at least $0.87\sigma$, $1.49\sigma$, and $0.52\sigma$, respectively. 
The decreasing trend of $w(z)$ persists, which is consistent with the CPL best-fit behavior. 
For the EoS parameter $w_5$, the best-fit values of $w_5$ lies outside the $1\sigma$ confidence region. 
This is attributed to the limited data points in this redshift range and the poor constraints of this parameter $w_5$.
For the lower row of Fig. \ref{fig.5binnedwz}, the corresponding DE density increases in the first bin and then decreases across the second, third, and fourth bins.

In summary, Figs. \ref{fig.3binnedwz}, \ref{fig.4binnedwz}, \ref{fig.5binnedwz} reveal a consistent trend for the evolution behavior of DE. 
These reconstruction results of $w(z)$ are generally in agreement with Fig.7 of the DESI results \cite{DESI:2025fii}.
The EoS $w(z)$ is greater than $-1$ at low redshift and smaller than $-1$ at high redshift, which corresponds to a quintom type behavior. 
This means that binned $w(z)$ indicates a significant deviation from the $\Lambda$CDM model. 
It is worth mentioning that as the binning number increases, the number of data points in each bin decreases.
Considering the limited data at high redshift, increasing binning number will further reduce the data points in higher redshift bins.
Consequently, the constraints in those bins become weaker, the error bar grows, and the deviations from $\Lambda$CDM lose statistical significance. 
With additional high-redshift data in the future, tighter constraints will enable a more definite test of whether dark energy at high redshift deviates from $\Lambda$CDM or not.
Focusing on the bin with the largest deviation in each case, we find that in the 3bins-w case, $w_2$ deviates at least $3.22\sigma$; in the 4bins-w case, $w_1$ deviates at least $2.46\sigma$; and in the 5bins-w case, $w_1$ deviates at least $2.13\sigma$. 
Therefore, after incorporating DESI data, there is a clear trend that DE EoS should evolve with redshift (at least $2.13\sigma$).

\subsubsection{Binned density $f(z)$}
\begin{figure*}[htbp]
\small
    \centering
        \includegraphics[width=0.33\textwidth]{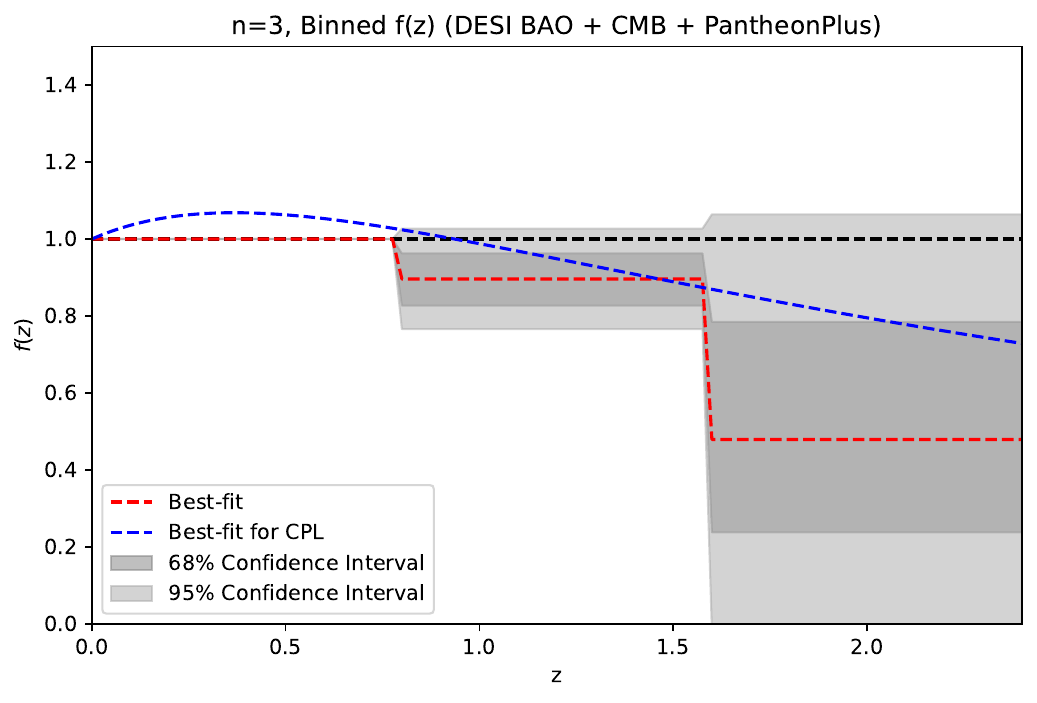}
        \includegraphics[width=0.33\textwidth]{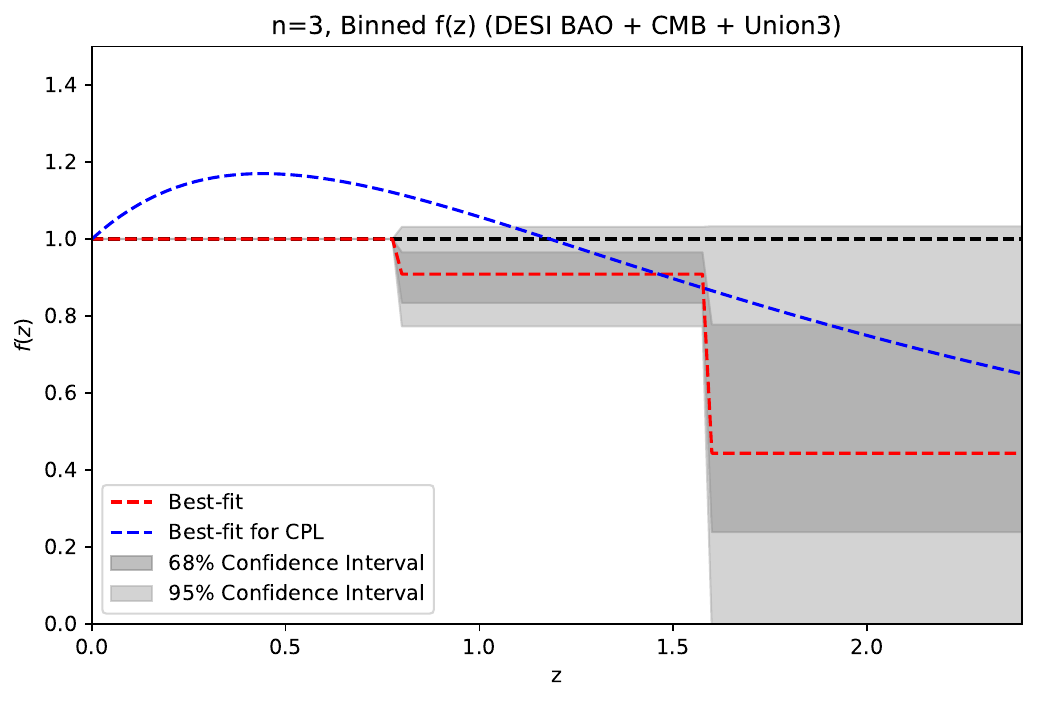}
        \includegraphics[width=0.33\textwidth]{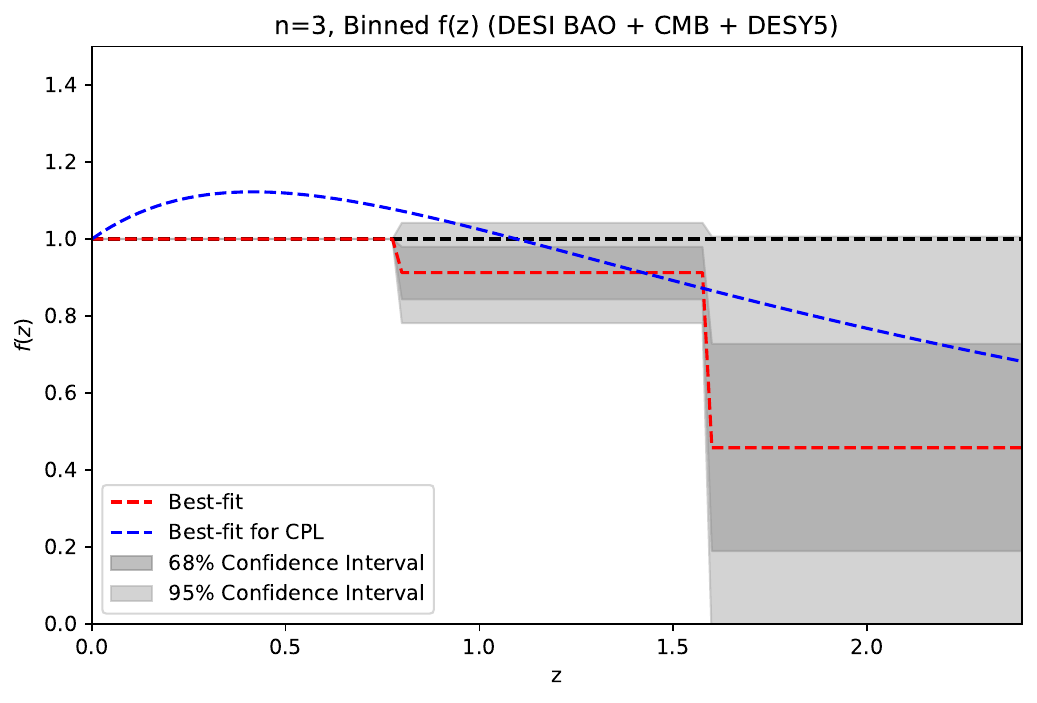}
    
        \includegraphics[width=0.33\textwidth]{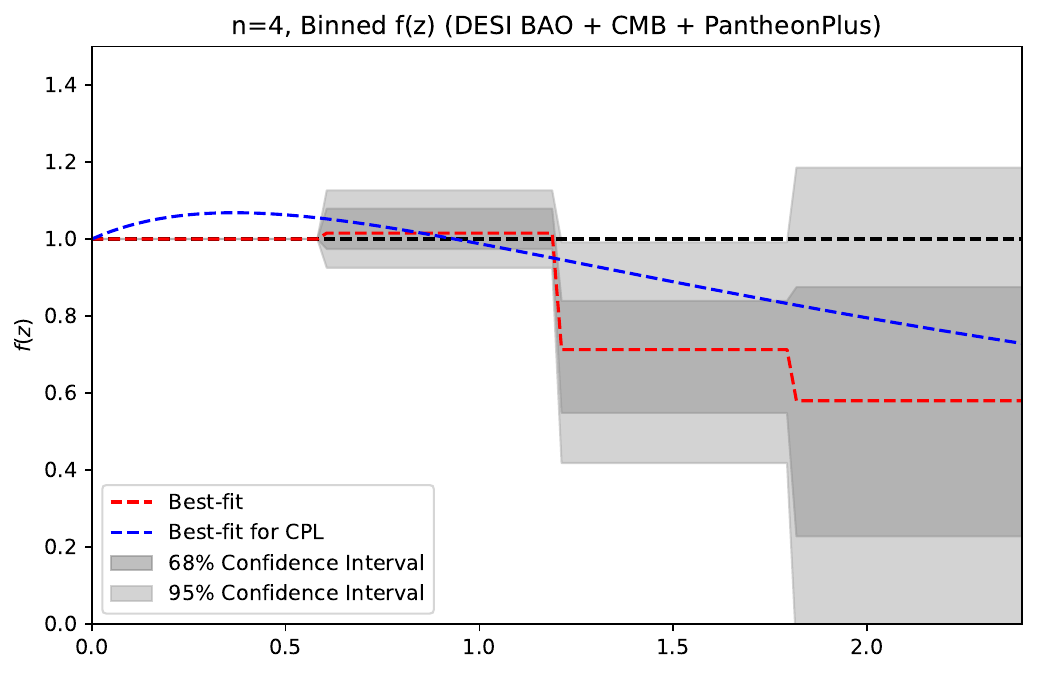}
        \includegraphics[width=0.33\textwidth]{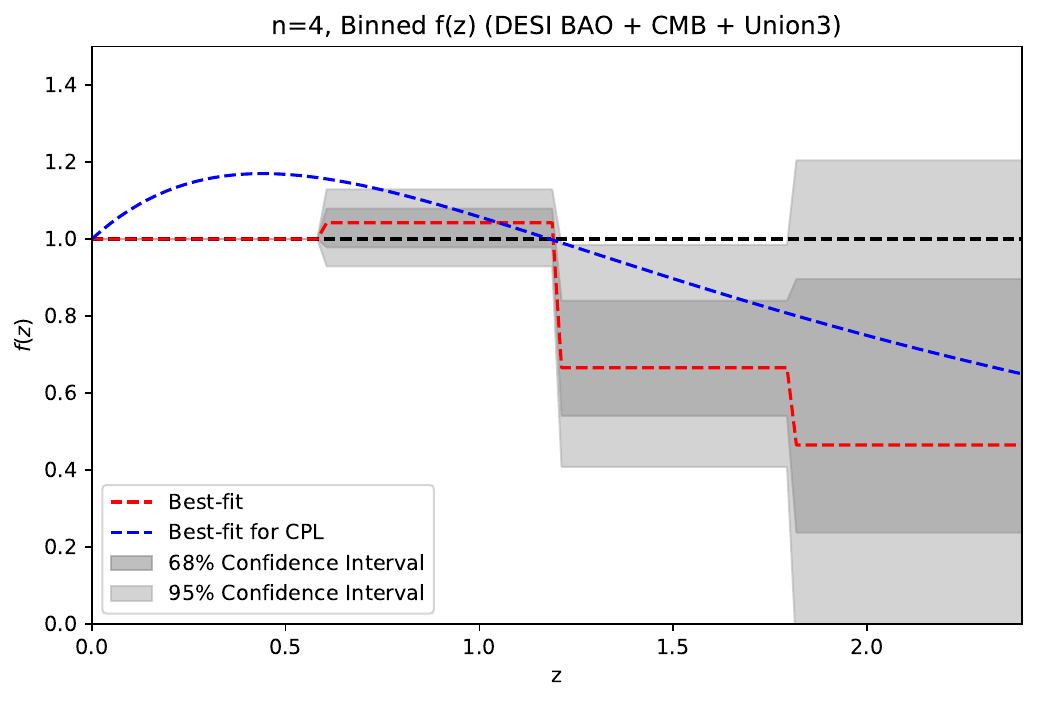}
        \includegraphics[width=0.33\textwidth]{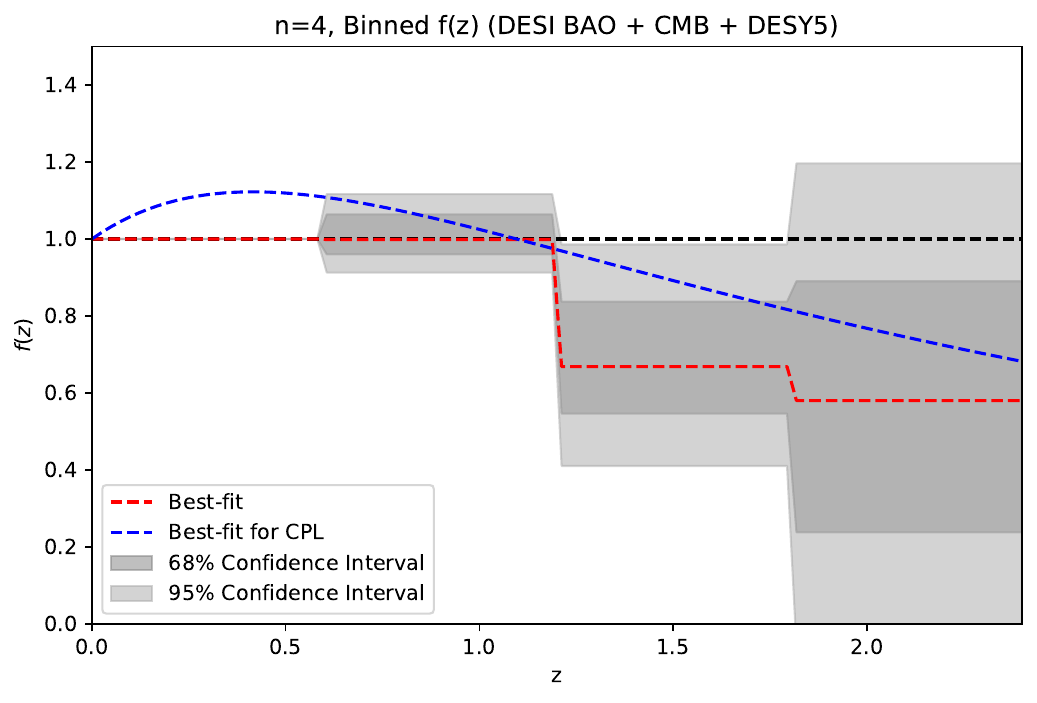}
    
        \includegraphics[width=0.33\textwidth]{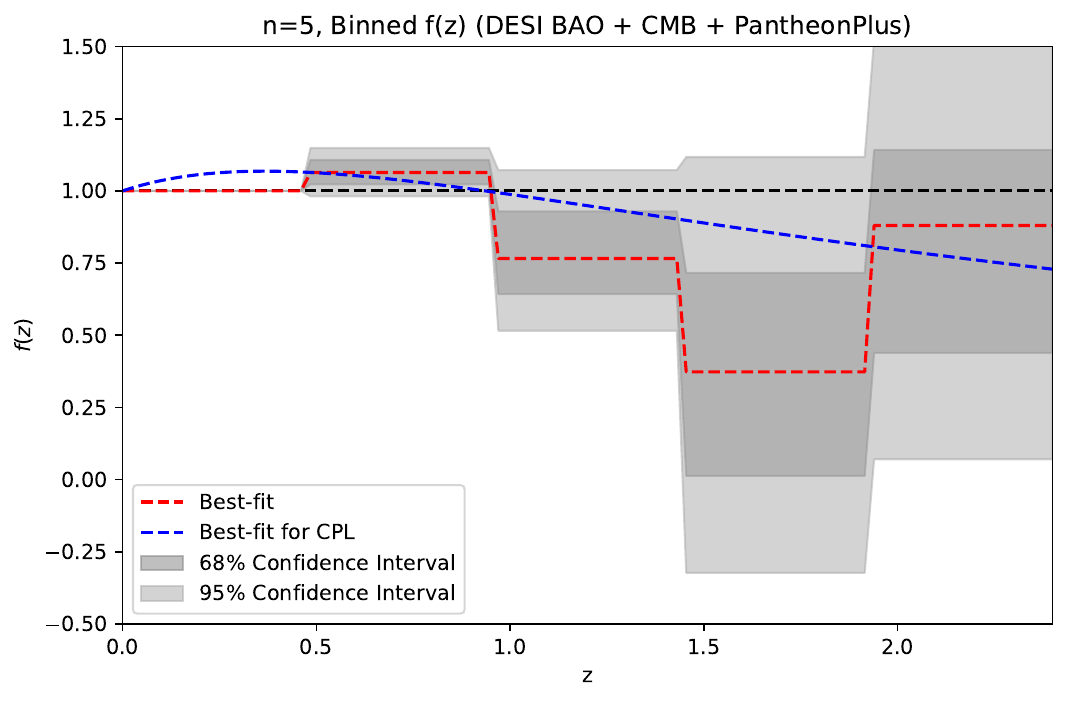}
        \includegraphics[width=0.33\textwidth]{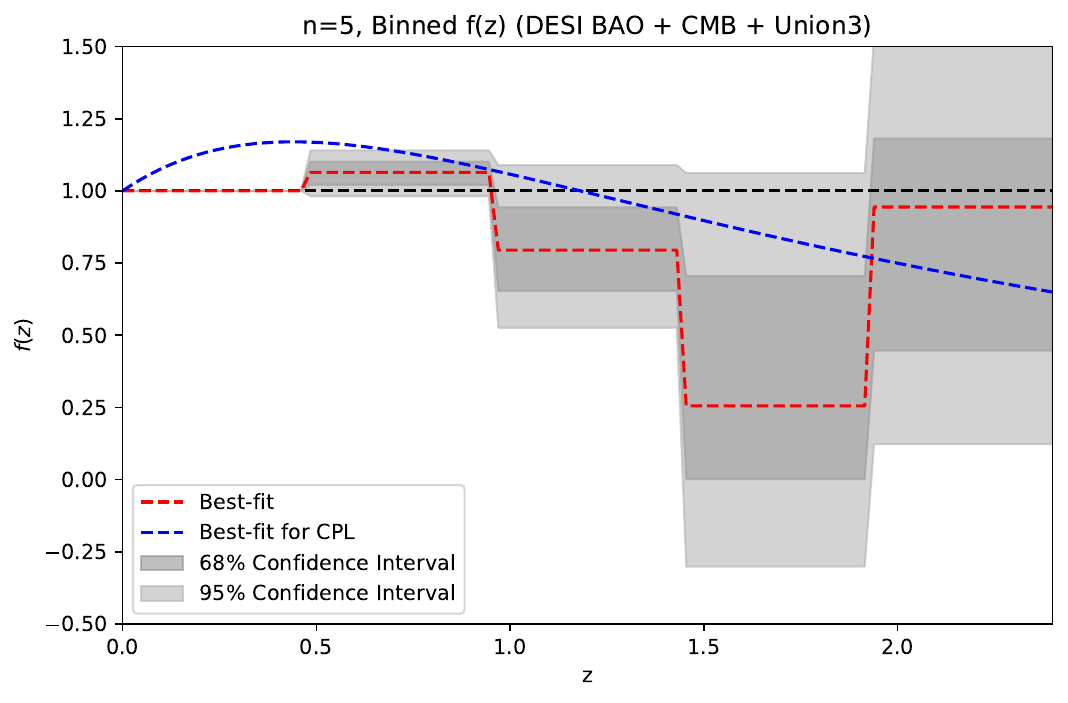}
        \includegraphics[width=0.33\textwidth]{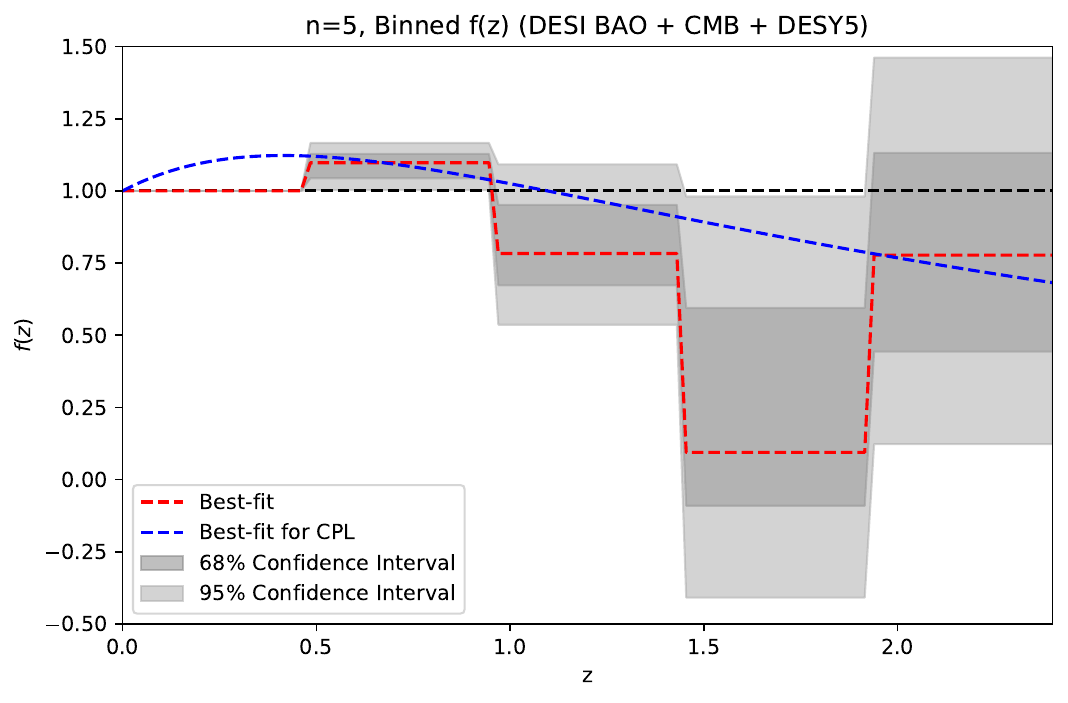}

    \caption{Reconstruction of the DE density $f(z)$ using three, four, and five redshift bins (upper, middle, and lower rows, respectively). Red dashed lines denote the best-fit values within each bin, while the blue dashed lines correspond to the best-fit CPL model for comparison. Shaded regions represent the 68\% (dark gray) and 95\% (light gray) confidence intervals. Results presented from left to right columns combine DESI BAO + CMB data with PantheonPlus, Union3, and DESY5 supernovae datasets, respectively.}
    \label{fig.binnedfz}
\end{figure*}

For the binned $f(z)$ method, the density function $f(z)$ is characterized by a piecewise constant in each redshift bin.
As described in subsubsection \ref{subsubsec:binnedf}, we set $f_1=1$ from the relation $E(0) = 1$.
The reconstruction results are presented in Fig. \ref{fig.binnedfz}, where the upper, middle, and lower rows correspond to cases with three, four, and five bins, respectively.
For simplicity, we refer to these cases as "3bins-f", "4bins-f", and "5bins-f", respectively.
The three columns represent the results obtained by combining DESI BAO and CMB with PantheonPlus, Union3, and DESY5, respectively.
As a comparison, the CPL best-fit under the same dataset is included in each panel.
The red dashed lines denote the best-fit results of redshift binning models, while the blue dashed lines represent the CPL best-fit curves. 
The corresponding mean values and standard deviations within each redshift bin are summarized in Table \ref{tab:binnedf}. 

\begin{table*}
\small
\begin{center}
\resizebox{\textwidth}{!}{%
\renewcommand{\arraystretch}{1.5}
\begin{tabular}{lllll}
\hline
$\text{Model/Dataset}$ & $f_2$/CL & $f_3$/CL & $f_4$/CL & $f_5$/CL \\
\hline
\multirow{4}{*}{}
\textbf{3bins-f} & & & & \\
DESI+CMB+PantheonPlus & $0.895 \pm 0.066$/ $1.59\sigma$ & $0.505 \pm 0.270$/ $1.83\sigma$ & - & - \\
DESI+CMB+Unino3  &  $0.901 \pm 0.066$/ $1.50\sigma$ & $0.502 \pm 0.267$/ $1.86\sigma$ & - & - \\
DESI+CMB+DESY5  &  $0.908 \pm 0.066$/ $1.39\sigma$ & $0.464 \pm 0.260$/ $2.06\sigma$ & - & - \\
\hline
\multirow{4}{*}{}
\textbf{4bins-f} & & & & \\
DESI+CMB+PantheonPlus & $1.027 \pm 0.052$/ $0.51\sigma$ & $0.692 \pm 0.143$/ $2.15\sigma$ & $0.556 \pm 0.324$/ $1.37\sigma$ & - \\
DESI+CMB+Unino3  &  $1.030 \pm 0.050$/ $0.60\sigma$ & $0.688 \pm 0.146$/ $2.13\sigma$ & $0.557 \pm 0.325$/ $1.36\sigma$ & - \\
DESI+CMB+DESY5  &  $1.010 \pm 0.051$/ $0.19\sigma$ & $0.695 \pm 0.141$/ $2.16\sigma$ & $0.558 \pm 0.320$/ $1.38\sigma$ & - \\
\hline
\multirow{4}{*}{}
\textbf{5bins-f} & & & & \\
DESI+CMB+PantheonPlus & $1.065 \pm 0.041$/ $1.58\sigma$ & $0.783 \pm 0.140$/ $1.55\sigma$ & $0.370 \pm 0.349$/ $1.80\sigma$ & $0.793 \pm 0.351$/ $0.59\sigma$ \\
DESI+CMB+Unino3  & $1.061 \pm 0.040$/ $1.52\sigma$ & $0.802 \pm 0.144$/ $1.37\sigma$ & $0.352 \pm 0.348$/ $1.86\sigma$ & $0.796 \pm 0.358$/ $0.57\sigma$ \\
DESI+CMB+DESY5  &  $1.088 \pm 0.040$/ $2.20\sigma$ & $0.810 \pm 0.138$/ $1.37\sigma$ & $0.249 \pm 0.341$/ $2.20\sigma$ & $0.787 \pm 0.346$/ $0.61\sigma$ \\
\hline
\end{tabular}%
}
\end{center}
\caption{Constraints on $f_i$ in each redshift bin, derived from different datasets using the redshift binning method. The confidence level (CL) indicates the deviation from $f(z)=1$.}
\label{tab:binnedf}
\end{table*}

For the 3bins-f case, the results of three data combinations exhibit a common trend.
In the second bin, the density parameter $f_2 <1$ deviates from $f=1$ at a confidence level of at least $1.39\sigma$. 
While in the third bin, $f_3 <1$ deviates at a level of at least $1.83\sigma$. 
These results are very similar to those of the 3bins-w case (see lower row of Fig. \ref{fig.3binnedwz}), suggesting a deviation from the $\Lambda$CDM model.

For the 4bins-f case, the results are broadly consistent with those from the 3bins-f case. 
In the second bin, the density $f_2 >1$ deviates from $f=1$ at a confidence level of at least $0.19\sigma$. 
While in the third and fourth bins, $f_3 <1$ and $f_4 <1$ deviate at levels of at least $2.13\sigma$ and $1.36\sigma$, respectively. 
This behavior is close to the 4bins-w case (see lower row of Fig. \ref{fig.4binnedwz}), implying a decreasing trend in DE density at high redshift.

For the 5bins-f case, the results of three data combinations also exhibit a consistent trend.
In the second bin, the density $f_2 >1$ deviates from $f=1$ at a confidence level of at least $1.52\sigma$. 
While in the third, fourth, and fifth bins, $f_3 <1$, $f_4 <1$, and $f_5 <1$ deviate at levels of at least $1.37\sigma$, $1.80\sigma$, and $0.57\sigma$, respectively. 
The DE density is greater than $f=1$ in the second bin, and smaller than $f=1$ in the following bins. 
This trend is similar to the case of 5bins-w (see lower row of Fig. \ref{fig.5binnedwz}), implying that DE density would first increase and then decrease with redshift.

In summary, as shown in Fig. \ref{fig.binnedfz}, we adopt three redshift binning schemes and three data combinations to probe the evolution of DE density. All the cases give a common trend. The DE density $f(z)$ would significantly decrease in the redshift range $1.0<z<2.0$. It is clear that this conclusion is independent of both the redshift binning scheme and data combination. 

\subsection{Polynomial interpolation method}
\label{subsec:results_PI}
\subsubsection{Interpolation of EoS $w(z)$}
\label{subsubsec:result_piw}
\begin{figure*}[htbp]
\small
    \centering
        \includegraphics[width=0.33\textwidth]{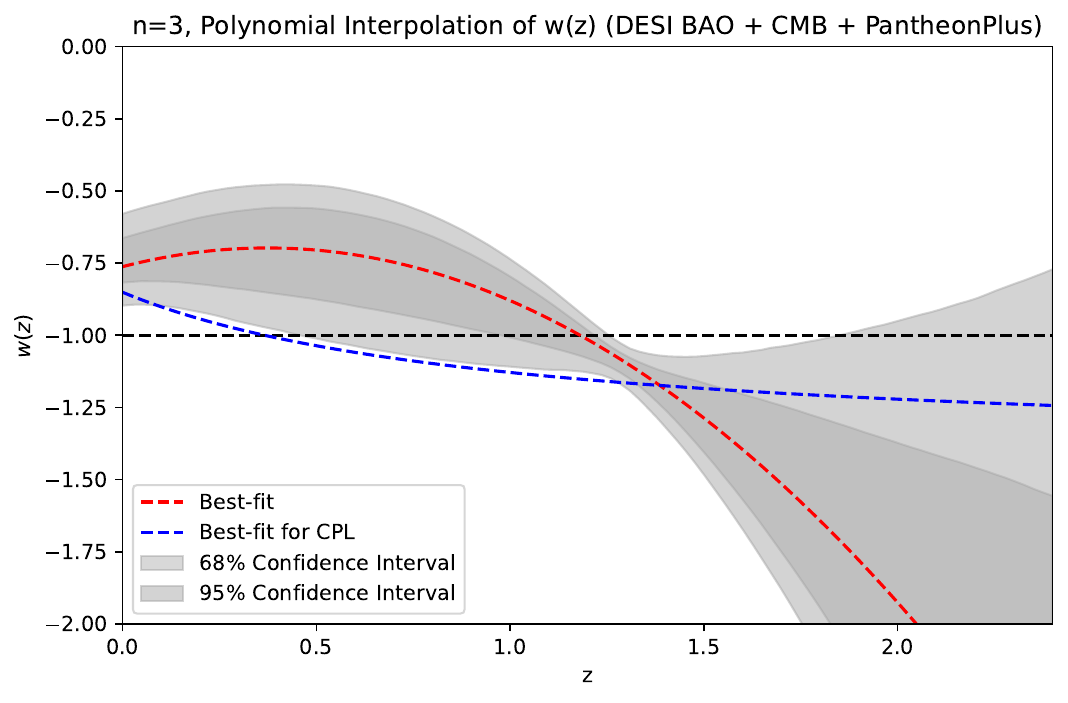}
        \includegraphics[width=0.33\textwidth]{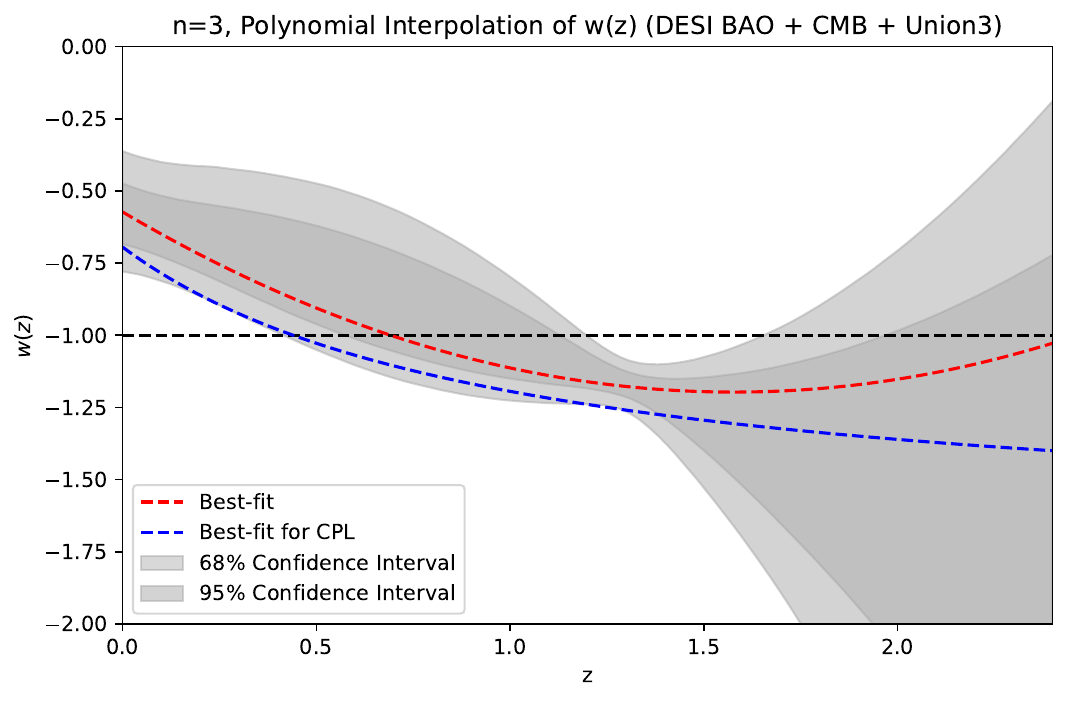}
        \includegraphics[width=0.33\textwidth]{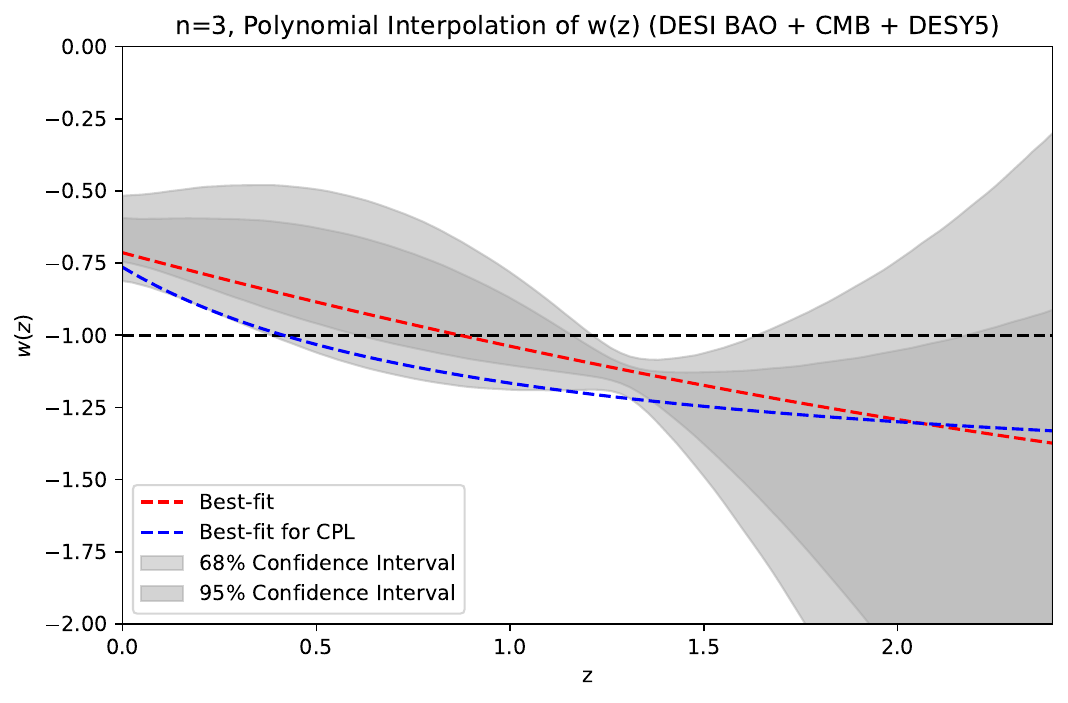}
    
        \includegraphics[width=0.33\textwidth]{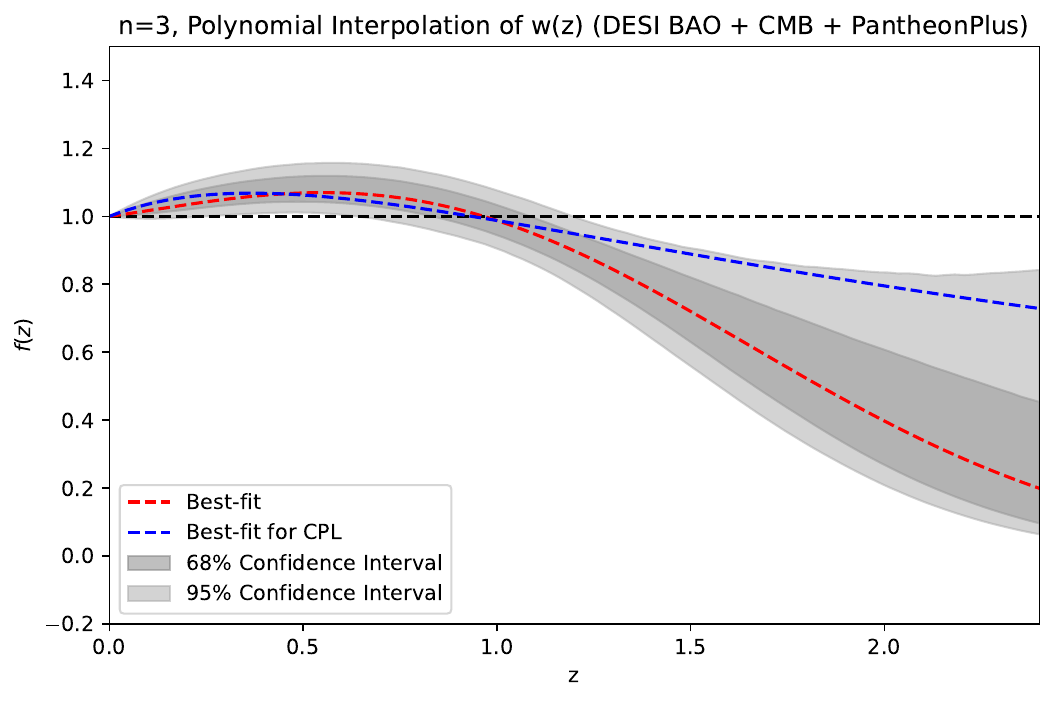}
        \includegraphics[width=0.33\textwidth]{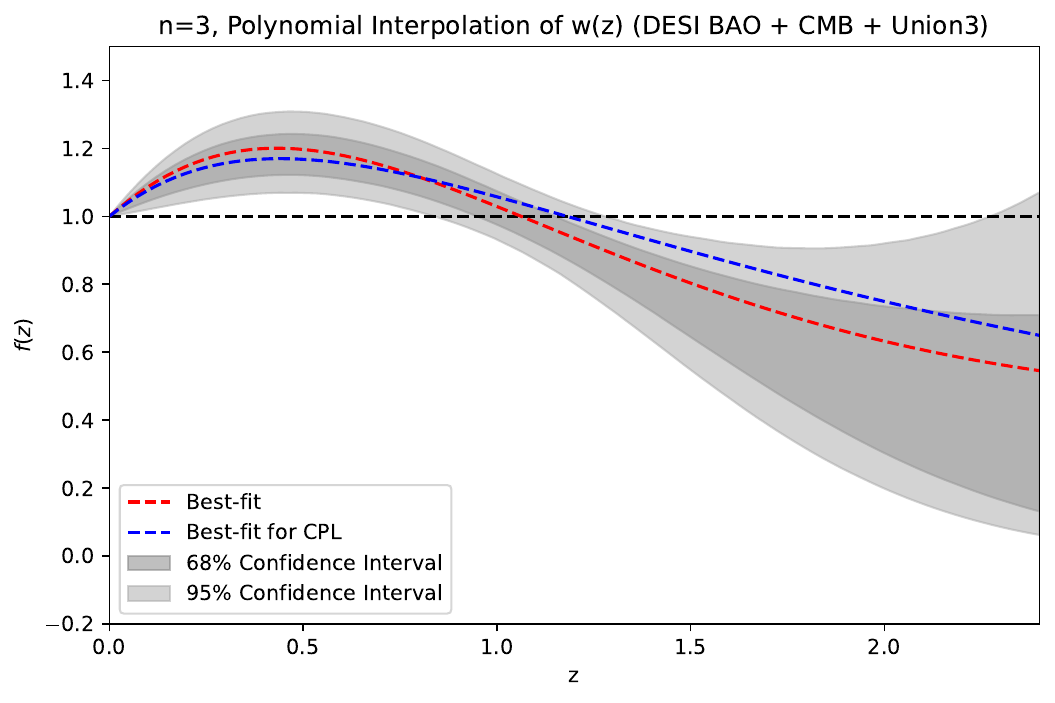}
        \includegraphics[width=0.33\textwidth]{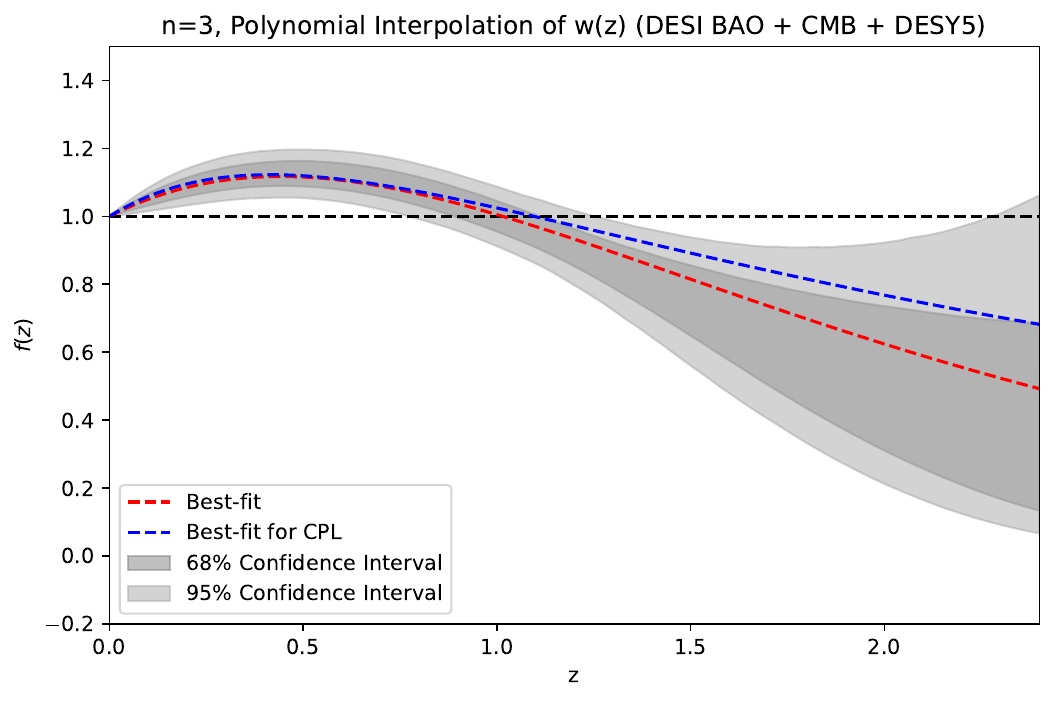}

    \caption{Reconstruction of the DE EoS $w(z)$ (upper panels) and the corresponding DE density $f(z)$ using polynomial interpolation method (n=3). Red dashed lines denote the best-fit of the reconstruction results, while the blue dashed lines corresponds to the best-fit CPL model for comparison. Shaded regions represent the 68\% (dark gray) and 95\% (light gray) confidence intervals. Results presented from left to right columns combine DESI BAO + CMB data with PantheonPlus, Union3, and DESY5 supernovae datasets, respectively.}
    \label{fig.3wpi}
\end{figure*}

\begin{figure*}[htbp]
\small
    \centering
        \includegraphics[width=0.33\textwidth]{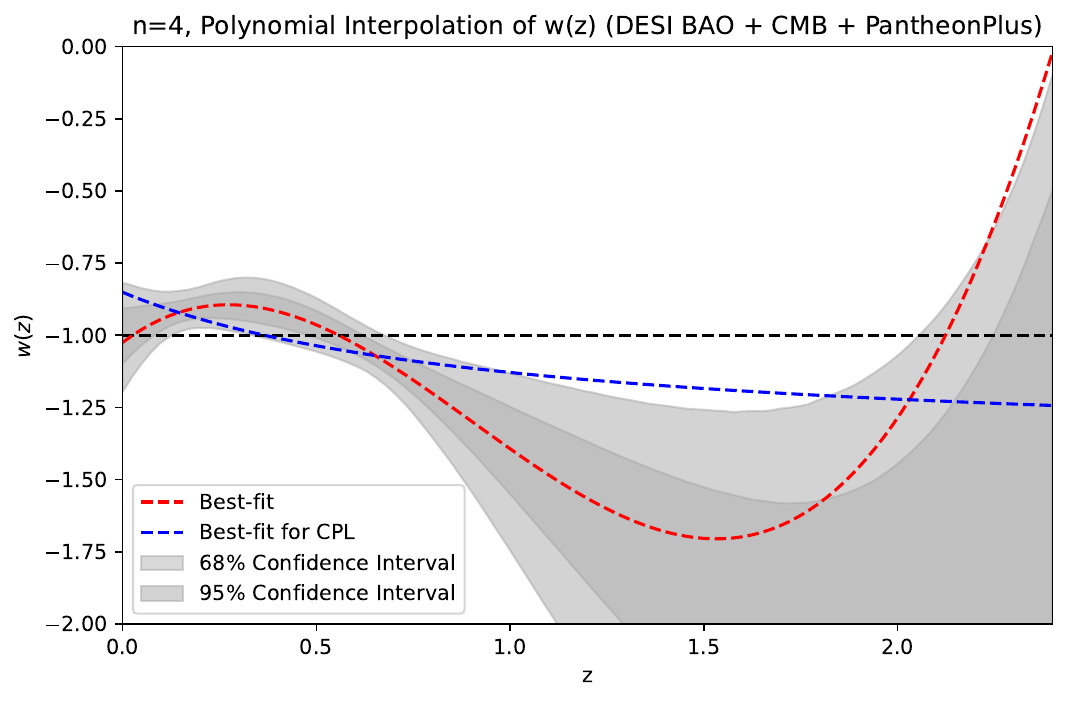}
        \includegraphics[width=0.33\textwidth]{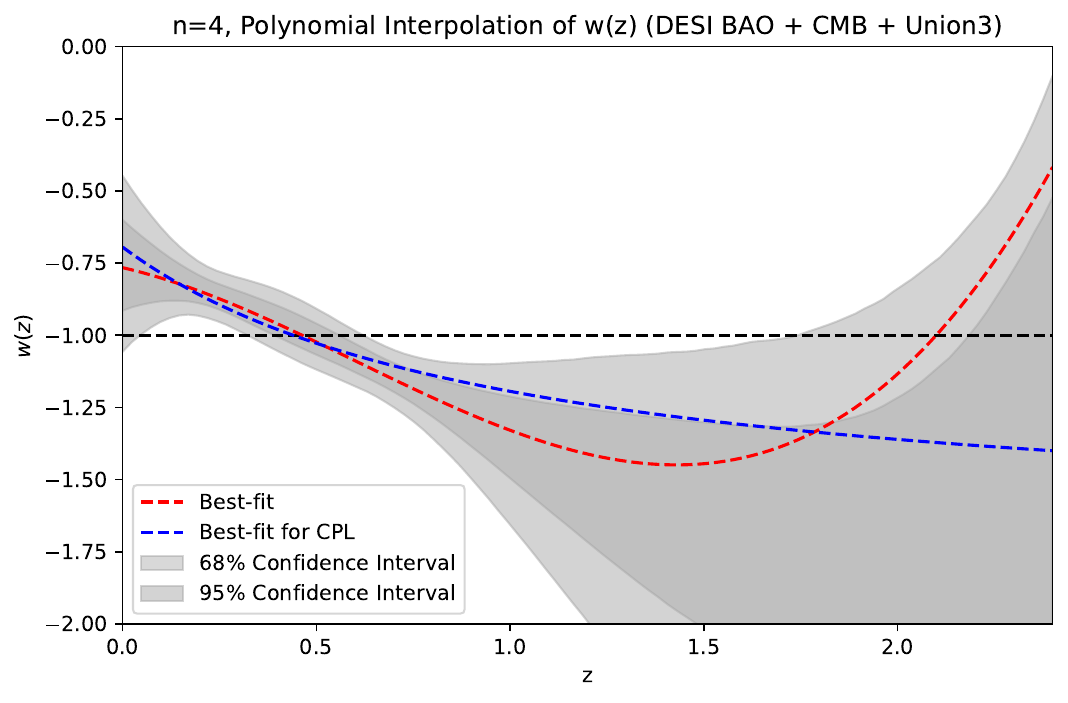}
        \includegraphics[width=0.33\textwidth]{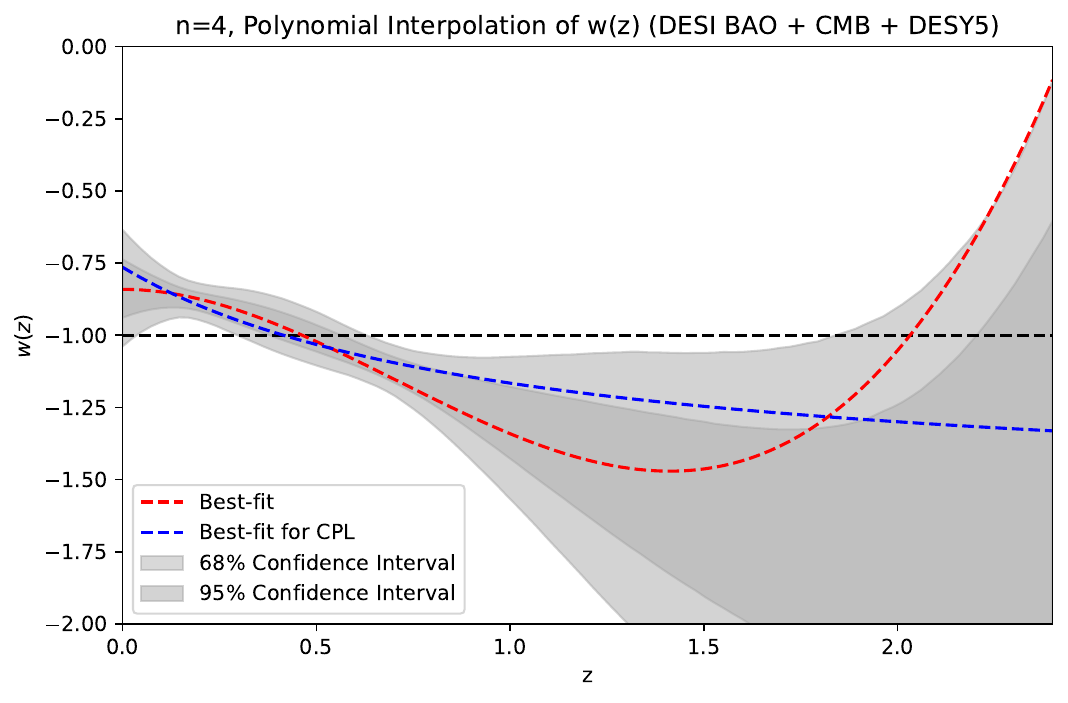}
    
        \includegraphics[width=0.33\textwidth]{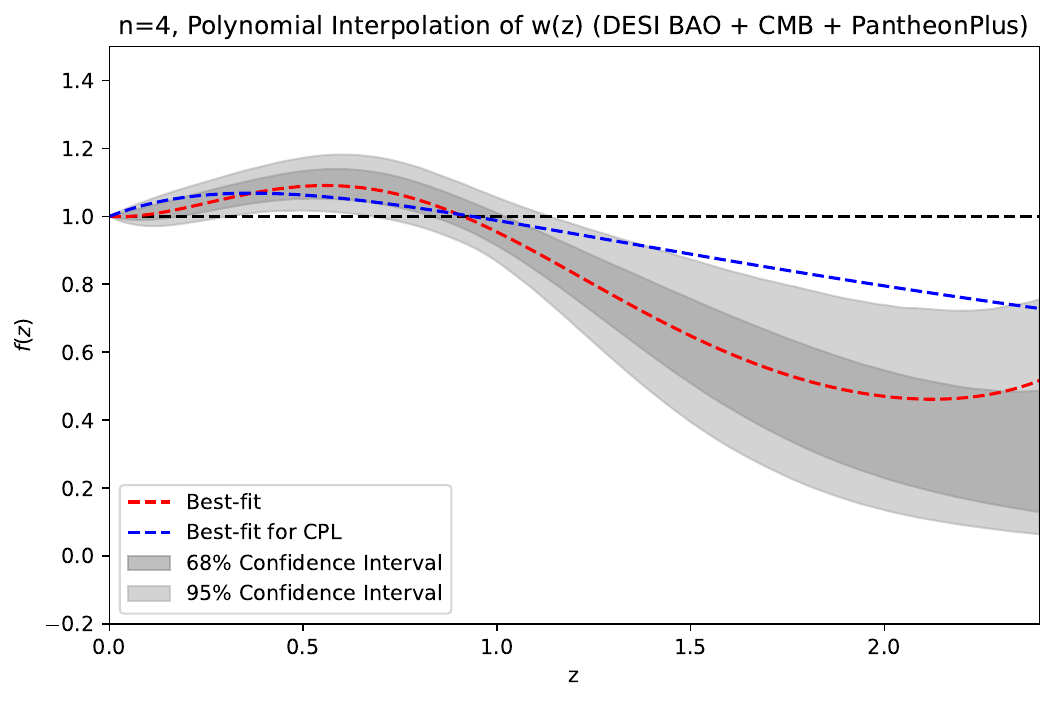}
        \includegraphics[width=0.33\textwidth]{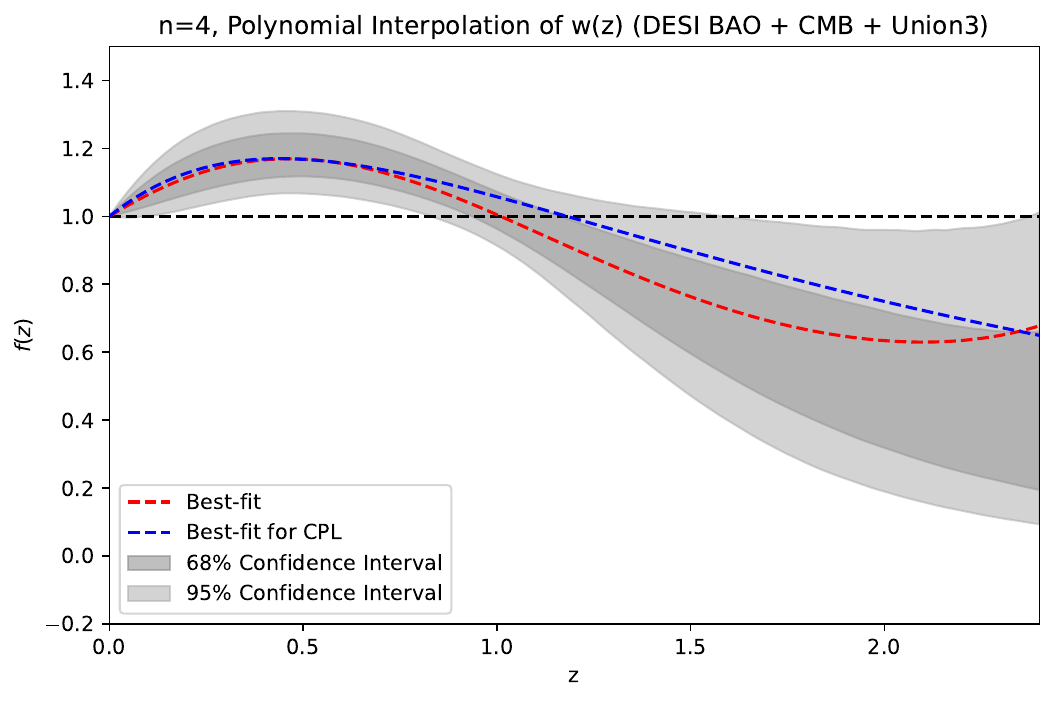}
        \includegraphics[width=0.33\textwidth]{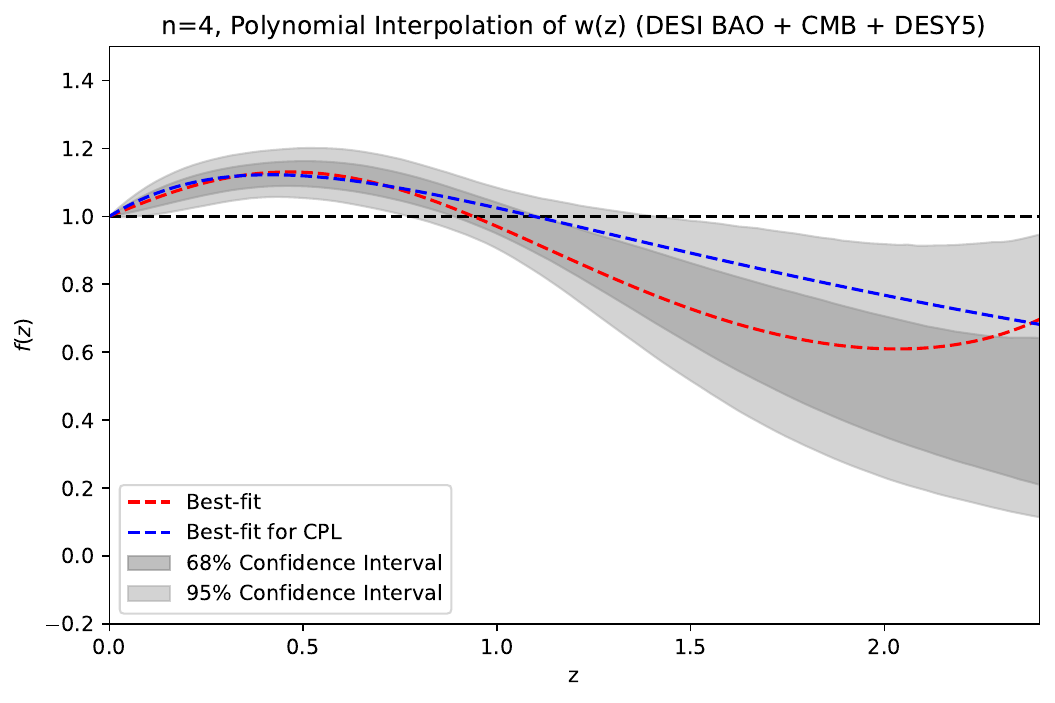}

    \caption{Reconstruction of the DE EoS $w(z)$ (upper panels) and the corresponding DE density $f(z)$ using polynomial interpolation method (n=4). Red dashed lines denote the best-fit of the reconstruction results, while the blue dashed lines corresponds to the best-fit CPL model for comparison. Shaded regions represent the 68\% (dark gray) and 95\% (light gray) confidence intervals. Results presented from left to right columns combine DESI BAO + CMB data with PantheonPlus, Union3, and DESY5 supernovae datasets, respectively.}
    \label{fig.4wpi}
\end{figure*}

\begin{figure*}[htbp]
\small
    \centering
        \includegraphics[width=0.33\textwidth]{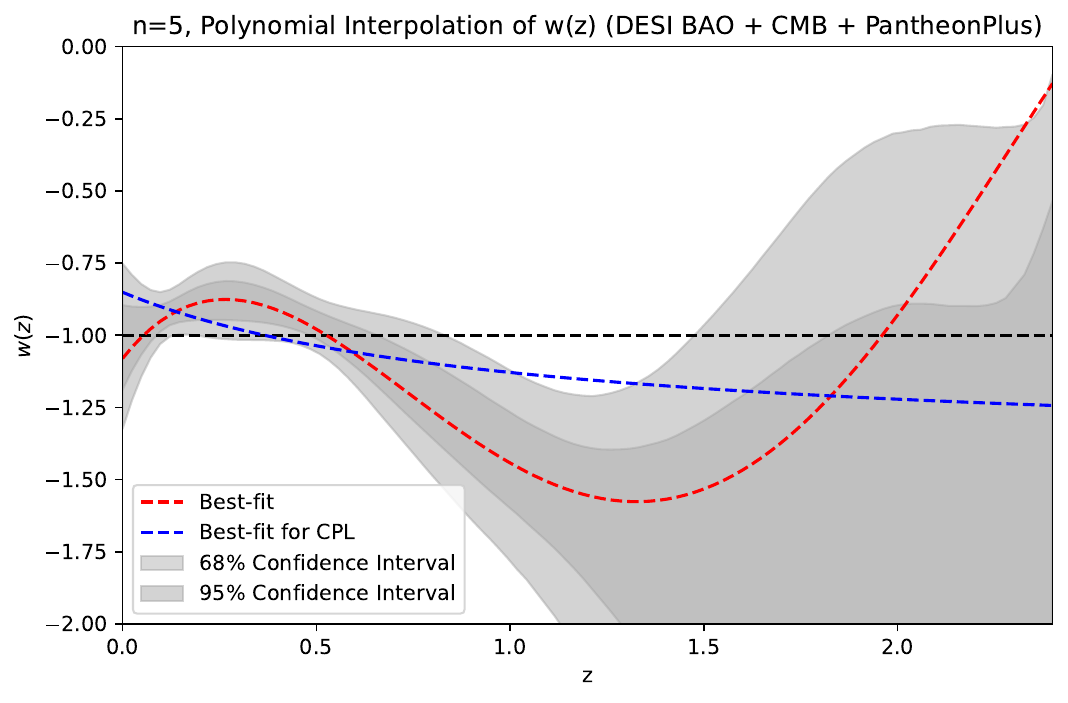}
        \includegraphics[width=0.33\textwidth]{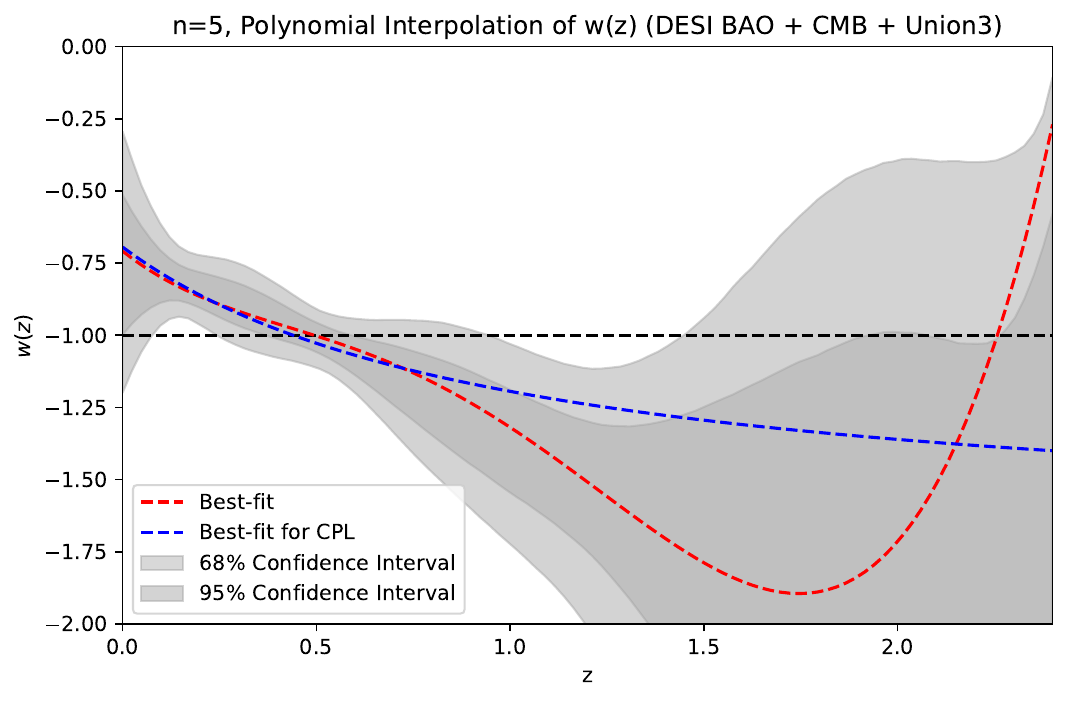}
        \includegraphics[width=0.33\textwidth]{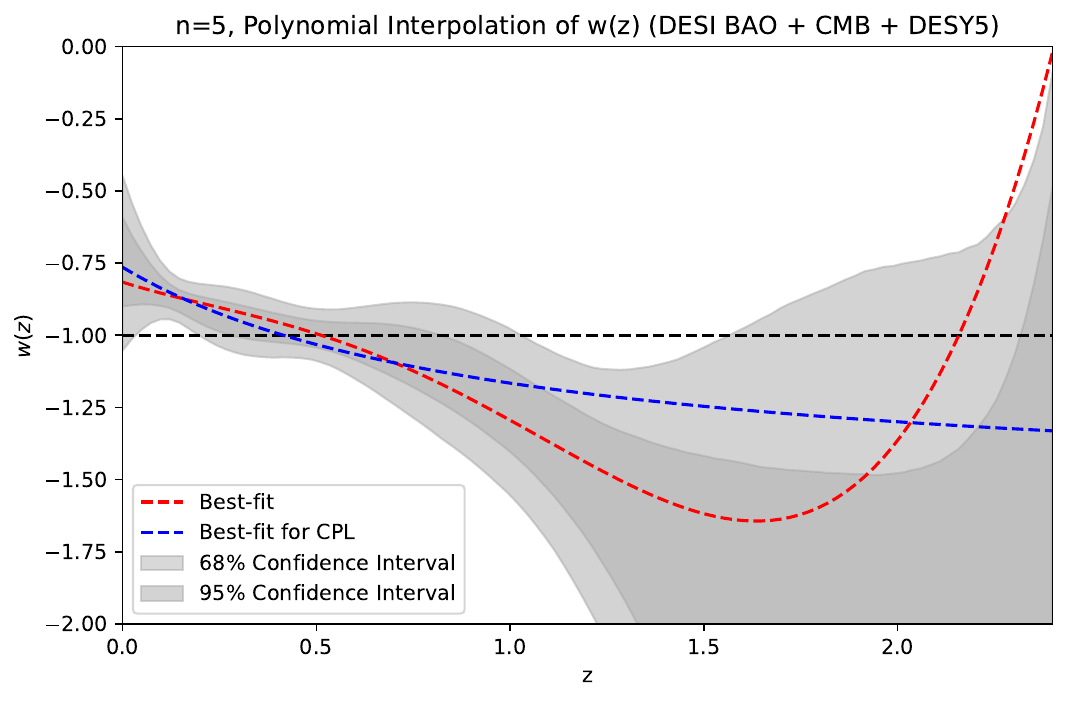}
    
        \includegraphics[width=0.33\textwidth]{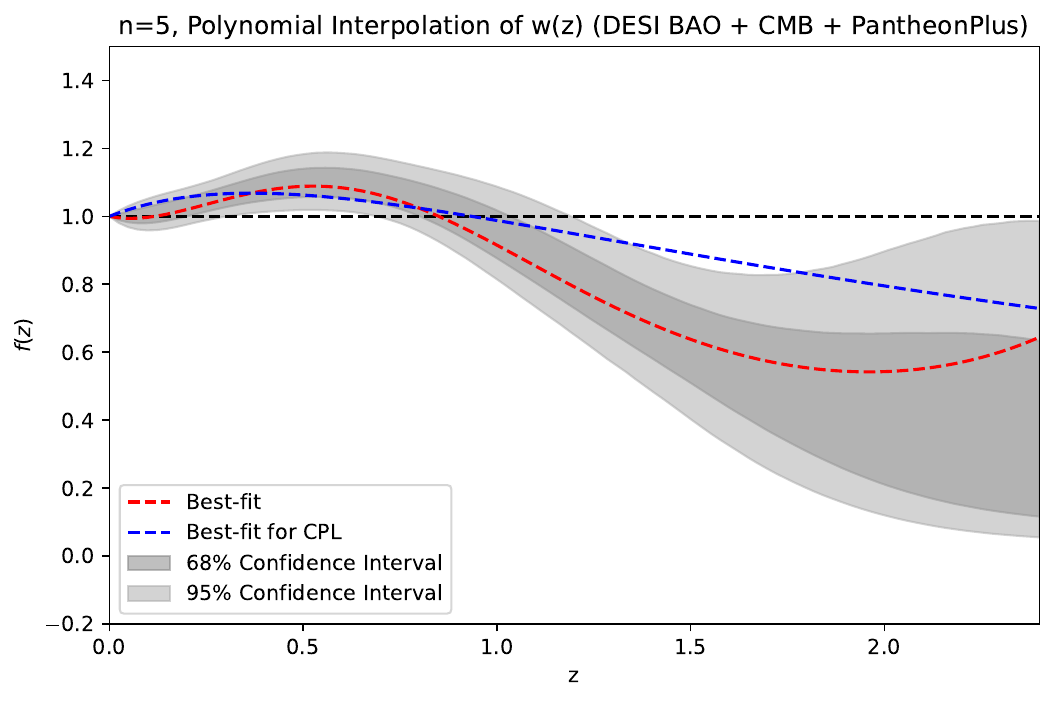}
        \includegraphics[width=0.33\textwidth]{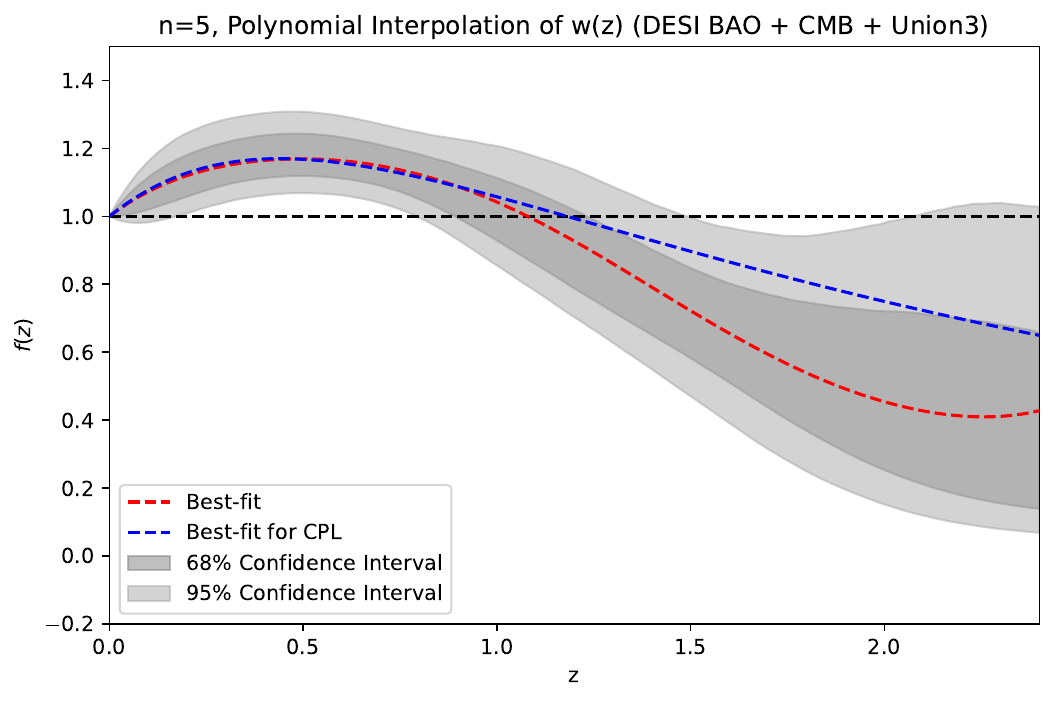}
        \includegraphics[width=0.33\textwidth]{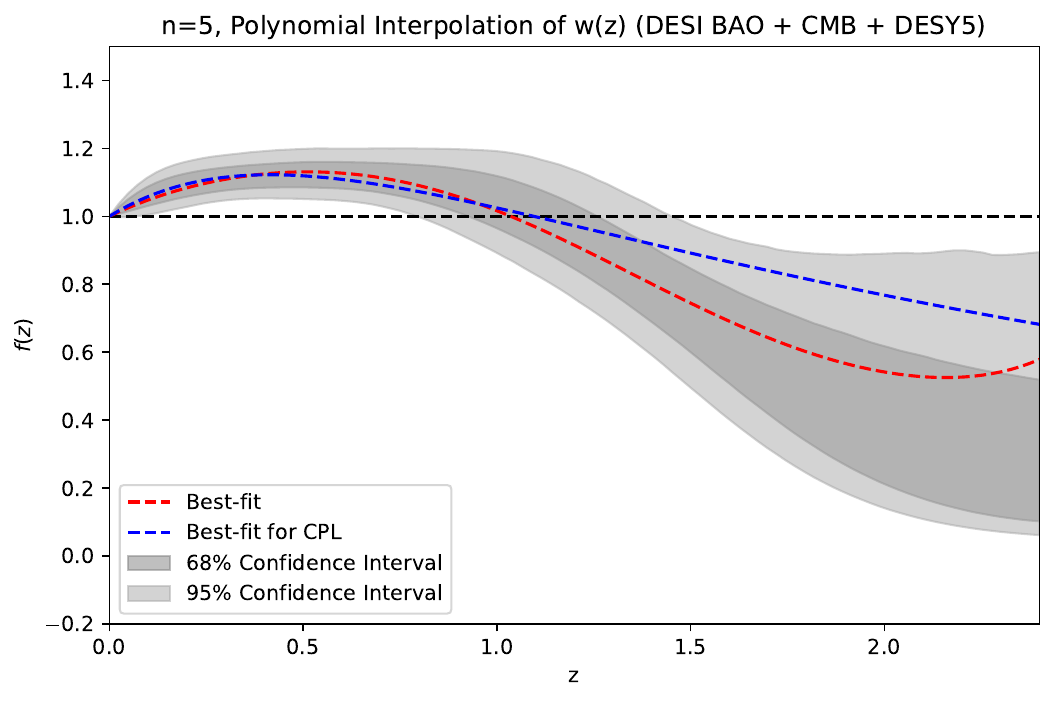}

    \caption{Reconstruction of the DE EoS $w(z)$ (upper panels) and the corresponding DE density $f(z)$ using polynomial interpolation method (n=5). Red dashed lines denote the best-fit of the reconstruction results, while the blue dashed lines corresponds to the best-fit CPL model for comparison. Shaded regions represent the 68\% (dark gray) and 95\% (light gray) confidence intervals. Results presented from left to right columns combine DESI BAO + CMB data with PantheonPlus, Union3, and DESY5 supernovae datasets, respectively.}
    \label{fig.5wpi}
\end{figure*}

For the polynomial interpolation method applied to $w(z)$, the DE EoS is characterized by a global polynomial. 
The reconstruction results are shown in Figs. \ref{fig.3wpi}, \ref{fig.4wpi}, \ref{fig.5wpi}, where the interpolation node numbers are the cases of $n =3, 4$, and $5$, respectively.
For simplicity, we refer to these cases as "3pi-w", "4pi-w", and "5pi-w", respectively.
In these figures, the upper rows represent the reconstruction of the DE EoS $w(z)$, while the lower rows denote the corresponding DE density $f(z)$.
The three columns represent results from combing DESI BAO and CMB with PantheonPlus, Union3, and DESY5, respectively.
The red dashed lines denote the best-fit results from the polynomial interpolation models, while the blue dashed lines represent the CPL best-fit curves under the same data combinations. 

For the 3pi-w case, different data combinations yield different evolution behaviors for $w(z)$. However, in the redshift region $0.5<z<1.5$, these three data combinations all reveal a decreasing trend in $w(z)$. In particular, $w(z)$ obviously exhibits a behavior of crossing the $w=-1$ boundary. It should be mentioned that due to the limited number of data points at high redshift ($z>2.0$), the error bars of $w(z)$ become large, resulting in weak constraints on the last interpolation coefficient. Based on the upper rows of Fig. \ref{fig.3binnedwz} and Fig. \ref{fig.3wpi}, one can see that the results of 3bins-w and 3pi-w have a common feature. The EoS parameter $w(z)$ would significantly decrease in the redshift range $0.5<z<1.5$. As for the corresponding DE density, results are consistent with the lower row of Fig. \ref{fig.3binnedwz}, where the density $f(z)$ first increases and then decreases.

For the 4pi-w case, the results are broadly consistent with those from the 3pi-w case. In the redshift range $0.5<z<1.5$, the decreasing trend of $w(z)$ persists. Additionally, the evolution behavior of $w(z)$ is clear at low redshift, indicating a deviation from the cosmological constant. However, at high redshift, the best-fit of $w(z)$ lies outside the 1$\sigma$ region. This is due to the lack of sufficient data in high redshift, suggesting that the last parameter is poorly constrained. This situation is similar to that observed in the upper row of Fig. \ref{fig.4binnedwz}. The corresponding DE evolution is also consistent with the lower row of Fig. \ref{fig.4binnedwz}, showing a patten of increase followed by decrease.

For the 5pi-w case, the overall trends remain consistent with those of the 4pi-w case. The decreasing trend of $w(z)$ persists in the redshift range $0.5<z<1.5$. Besides, the crossing of $w=-1$ occurs at $z \approx 0.5$. As in the previous cases, the best-fit of $w(z)$ also lies outside the 1$\sigma$ region at high redshift. This is also attributed to the lack of high-redshift data. These results align with the upper row of Fig. \ref{fig.5binnedwz}. The corresponding DE density exhibits the same behavior of increasing at low redshift and then decreasing at high redshift.

In summary, Figs. \ref{fig.3wpi}, \ref{fig.4wpi}, \ref{fig.5wpi} reveal an evolution behavior of DE, which is consistent with the results from redshift binning of $w(z)$. In the redshift range $0.5<z<1.5$, $w(z)$ exhibits a decreasing trend and shows a behavior of crossing the $w=-1$ boundary. specifically, the DE exhibits a quintom-B behavior \cite{Cai:2025mas}, in which the crossing occurs from the below to above. These reconstruction results of $w(z)$ are generally in agreement with Fig.1 of the DESI results \cite{DESI:2024aqx}. 

\subsubsection{Interpolation of density $f(z)$}
For the polynomial interpolation method applied to $f(z)$, the DE density function is characterized by a global Lagrange polynomial.
The reconstruction results are presented in Fig. \ref{fig.binnedfz}, where the upper, middle, and lower rows correspond to interpolation node numbers of $n =3, 4$, and $5$, respectively.
For simplicity, we refer to these cases as "3bins-f", "4bins-f", and "5bins-f", respectively.
The three columns represent the results obtained by combining DESI BAO and CMB with PantheonPlus, Union3, and DESY5, respectively.
The red dashed lines denote the best-fit results from the polynomial interpolation models, while the blue dashed lines represent the CPL best-fit curves.

\begin{figure*}[htbp]
\small
    \centering
        \includegraphics[width=0.33\textwidth]{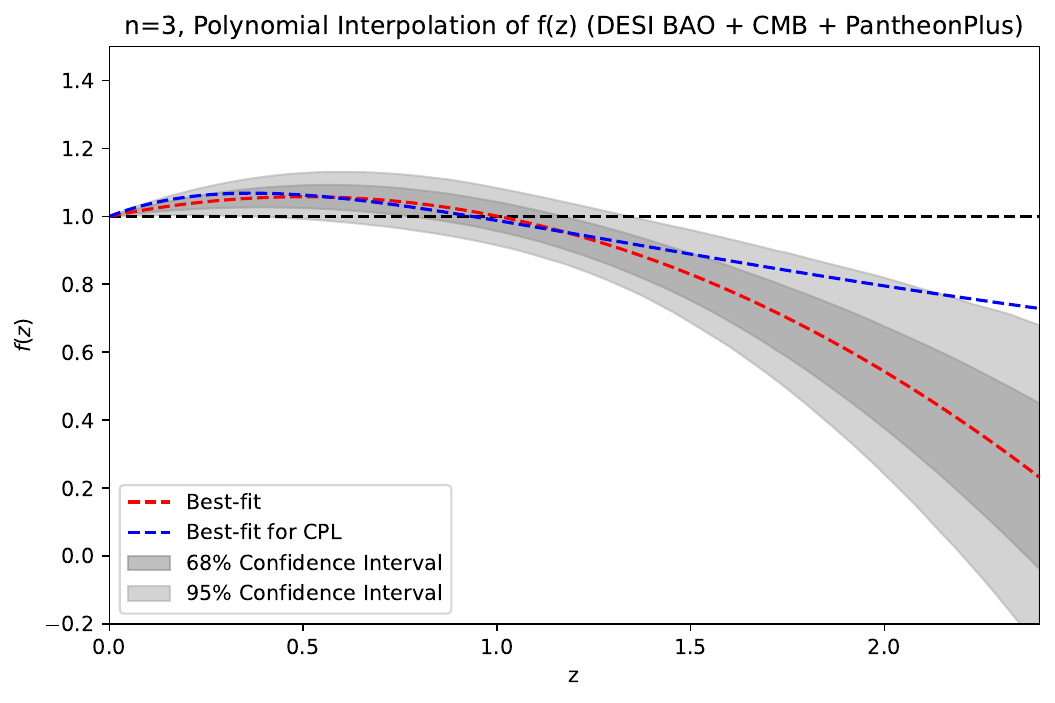}
        \includegraphics[width=0.33\textwidth]{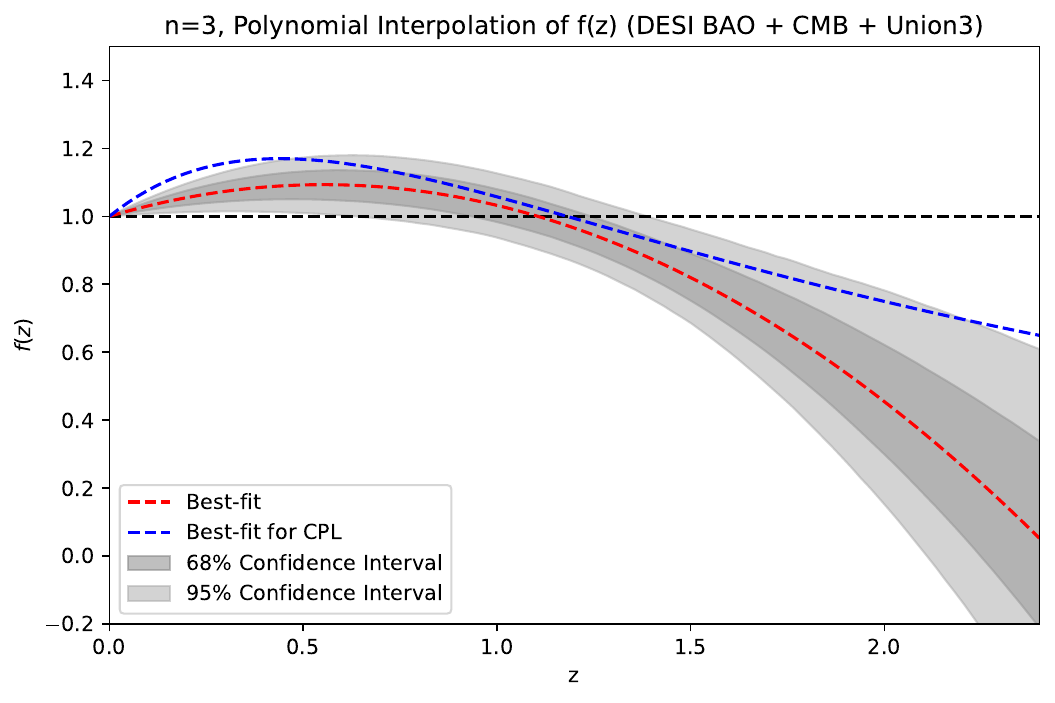}
        \includegraphics[width=0.33\textwidth]{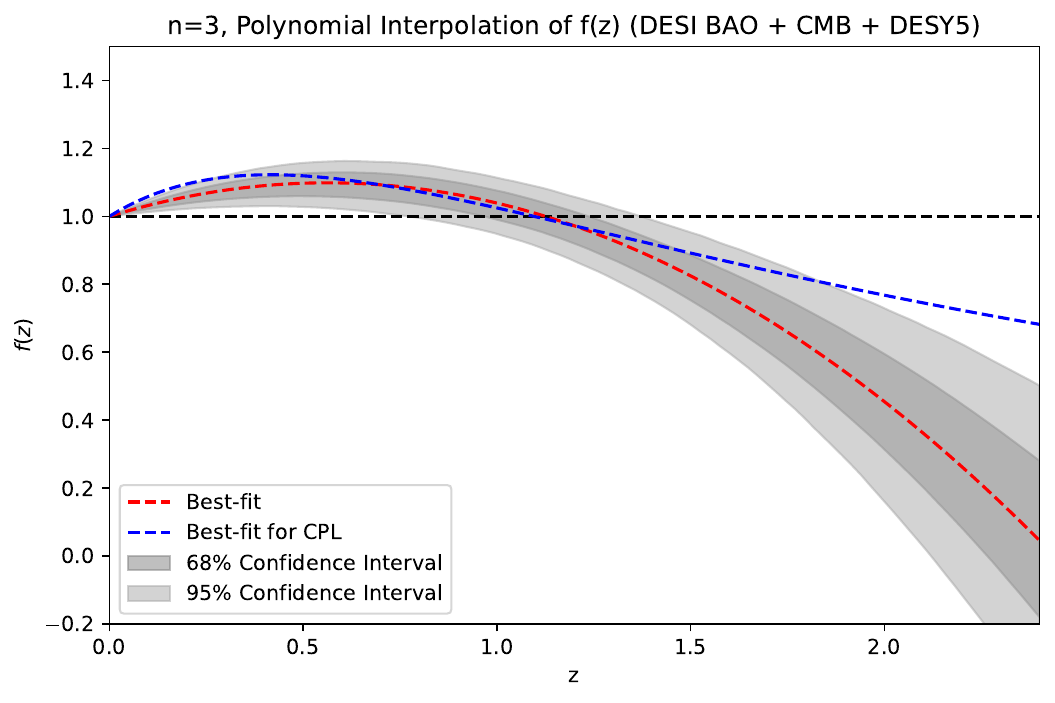}
    
        \includegraphics[width=0.33\textwidth]{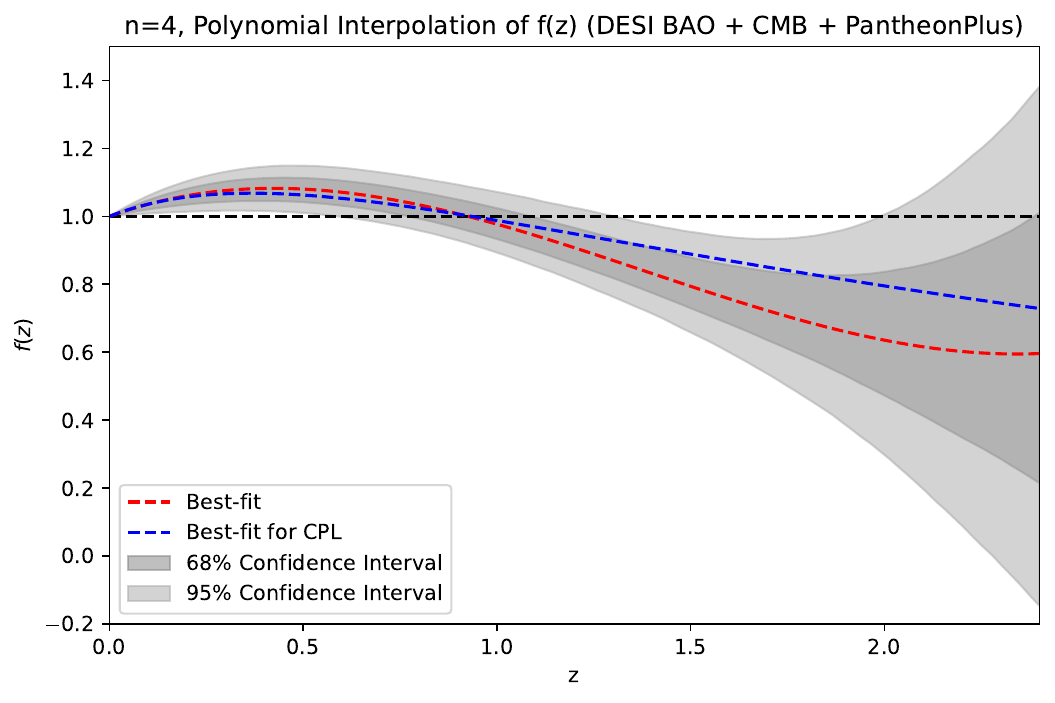}
        \includegraphics[width=0.33\textwidth]{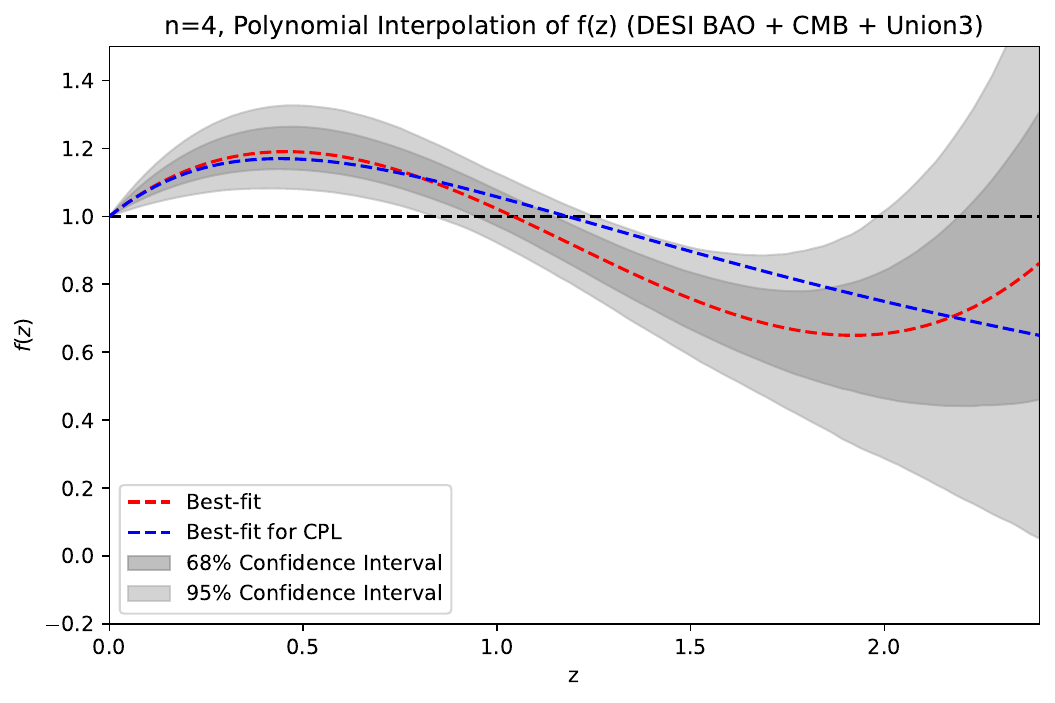}
        \includegraphics[width=0.33\textwidth]{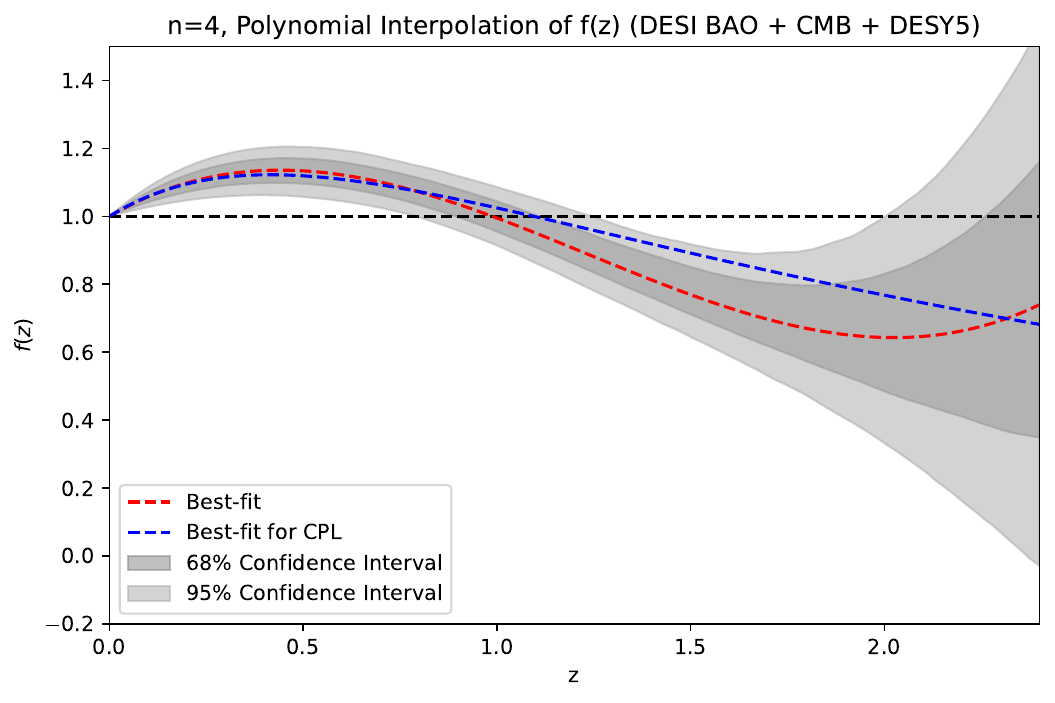}
    
        \includegraphics[width=0.33\textwidth]{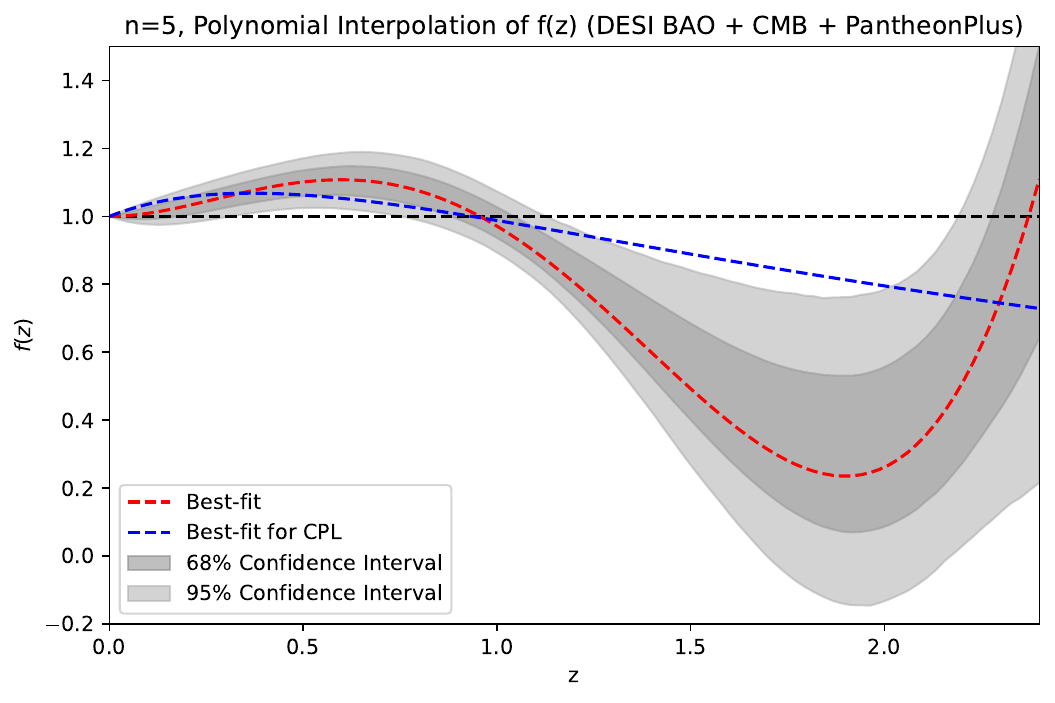}
        \includegraphics[width=0.33\textwidth]{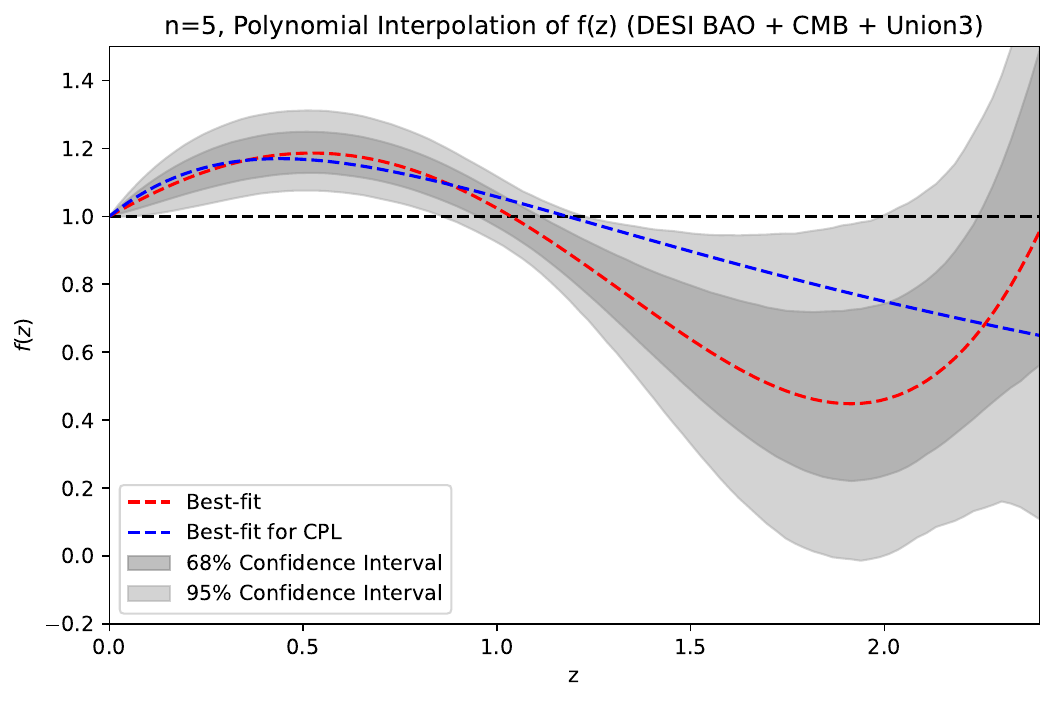}
        \includegraphics[width=0.33\textwidth]{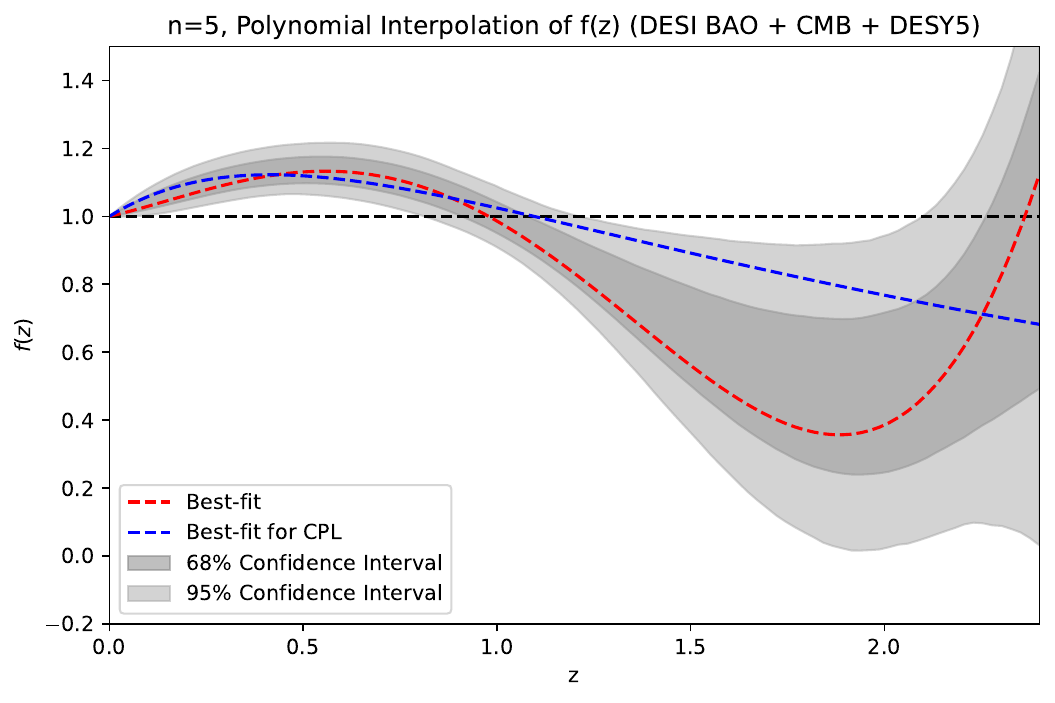}

    \caption{Reconstruction of the DE density $f(z)$ using polynomial interpolation with $n=3, 4$, and $5$ nodes shown in the upper, middle, and lower rows, respectively. Red dashed lines denote the best-fit values within each bin, while the blue dashed lines correspond to the best-fit CPL model for comparison. Shaded regions represent the 68\% (dark gray) and 95\% (light gray) confidence intervals. Results presented from left to right columns combine DESI BAO + CMB data with PantheonPlus, Union3, and DESY5 supernovae datasets, respectively.}
    \label{fig.pifz}
\end{figure*}

For the 3pi-f case, the results of all three data combinations exhibit a common trend. The density function $f(z)$ increases slowly at low redshift and then decreases at high redshift. These results are consistent with the lower row of Fig. \ref{fig.3wpi}.

For the 4pi-f case, the results are basically consistent with those from the 3pi-f case. The density function $f(z)$ increases more obviously at low redshift and reaches a hump at around $z \approx 0.5$, and then rapidly decreases at $z>1$. Due to the limited high-redshift data, the density $f(z)$ rebound at $z>2.0$. 

For the 5pi-f case, the general trend remains consistent with the 4pi-f case. There are two key points are worth mentioning. Firstly, the DE density reaches a maximum around $z \approx 0.5$. Secondly, the rebound at $z>2$ is mainly caused by the lack of high-redshift data. This situation is similar to the 5bins-f cases observed in Fig. \ref{fig.binnedfz}.

In summary, as shown in Fig. \ref{fig.pifz}, we adopt three kinds of interpolation node schemes and three data combinations to probe the evolution of DE density. All the results give a common trend. 
The DE density $f(z)$ reaches a hump near $z \approx 0.5$ and then decrease significantly in the redshift range $1.0<z<2.0$. This conclusion is generally in agreement with the results of DESI \cite{DESI:2024aqx}.

\section{Conclusion}
\label{sec:conclusion}
In this paper, we reconstruct the DE EoS $w(z)$ and the DE density function $f(z)$ using two model-independent approaches (redshift binning method and polynomial interpolation method). In order to ensure the results are model independent, we adopt three different redshift binning schemes (n=3, 4, 5) and three polynomial interpolation schemes with the same number of nodes (n=3, 4, 5). For cosmological data, we incorporate the DESI DR2 BAO measurements, CMB distance priors from Planck 2018 and ACT DR6, and three SN compilations (PantheonPlus, Union3, and DESY5). 

In our analysis, the highest–redshift bin contains very few data points, particularly for larger $n$. Consequently, it contains large uncertainty and fails to provide meaningful cosmological constraints. Therefore, we excluded the highest–redshift bin for larger $n$. The same applies to the polynomial interpolation method. Our main conclusions are as follows: 

1. After incorporating DESI data, there is a trend that DE should evolve with redshift (with deviations from EoS $w=-1$ reaching at least a $2.13\sigma$ confidence level), indicating that current observations favor a dynamical DE.

2. In the redshift range $0.5 < z < 1.5$, the DE EoS $w(z)$ exhibits a decreasing trend and crosses the phantom divide $w=-1$, suggesting quintom-like behavior.

3. The DE density $f(z)$ first increases at low redshift, reaching a hump around $z\approx 0.5$, and then decreases at $0.5 < z < 1.5$, with a rapid decrease at $z>1.0$.

4. For $z > 1.5$, current data are insufficient to place strong constraints on the evolution of DE, resulting in large uncertainties in the DE reconstruction.

Note that two different reconstruction approaches and three different data combinations all give common results. Therefore, these four main conclusions are independent of specific reconstruction models and are insensitive to the choice of SN compilations. 

In our analysis, we only consider two DE reconstruction approaches, i.e. the redshift binning method and the polynomial interpolation method. It would be worthwhile to further study some different DE reconstruction, such as the principle component approach. Moreover, future cosmological observations, including the full five-year data from DESI measurements, data from the Nearby Supernova Factory, and data from Vera C. Rubin observatory, are expected to provide improved precision and broader redshift coverage. These will offer deeper insights into the nature of dark energy and further test its potential deviation from the $\Lambda$CDM model.

\section{Acknowledgments}
We want to thank the anonymous referee, whose valuable suggestions greatly improve the quality of our work. S.Wang is supported by a grant of Guangdong Provincial Department of Science and Technology under No. 2020A1414040009.

\section{Appendix A: Comparison of cosmological results from CMB distance priors and full CMB likelihood}
To ensure the robustness of our results, we repeat our analysis using the $\textit{Planck}$ 2018 full CMB likelihood (TT, TE, EE), in place of the CMB distance priors used in the main body of the paper.
Considering time and computational cost, for the SN data, we only use the Union3 compilation; for the redshift binning method and polynomial interpolation method, we specifically examine four representative cases: 5bins-w, 5bins-f, 5pi-w and 5pi-f.

For the redshift binning method, the mean results are summarized in Table \ref{tab:compare_5bins-wf}. From this table, we see only a slight difference when replacing the CMB distance priors with the full CMB likelihood. As noted in the main text, the high-redshift bins are weakly constrained due to the scarcity of data points, which accounts for the observed discrepancies. Overall, the data combination using CMB distance priors reproduces the full likelihood results to a good approximation.

\begin{table*}
\small
\begin{center}
\resizebox{\textwidth}{!}{
\renewcommand{\arraystretch}{1.5}
\begin{tabular}{llllll}
\hline
$\text{Model/Dataset}$ & $w_1$/CL & $w_2$/CL & $w_3$/CL & $w_4$/CL & $w_5$/CL \\
\hline
\multirow{4}{*}{}
\textbf{5bins-w} & & & & & \\
DESI+Union3+CMB distance priors  &  $-0.859 \pm 0.050$/ $2.82\sigma$ & $-1.161 \pm 0.104$/ $1.54\sigma$ & $-1.630 \pm 0.400$/ $1.57\sigma$ & $-1.563 \pm 0.891$/ $0.63\sigma$ & $-0.702 \pm 1.091$/ $0.27\sigma$ \\
DESI+Unino3+full CMB likelihood  &  $-0.851 \pm 0.052$/ $2.86\sigma$ & $-1.108 \pm 0.114$/ $0.94\sigma$ & $-1.476 \pm 0.398$/ $1.19\sigma$ & $-1.602 \pm 0.901$/ $0.66\sigma$ & $-1.107 \pm 1.112$/ $0.09\sigma$ \\
\hline
$\text{Model/Dataset}$ & $f_1$/CL & $f_2$/CL & $f_3$/CL & $f_4$/CL & $f_5$/CL \\
\hline
\multirow{4}{*}{}
\textbf{5bins-f} & & & & & \\
DESI+Union3+CMB distance priors & -- & $1.061 \pm 0.040$/ $1.52\sigma$ & $0.802 \pm 0.144$/ $1.37\sigma$ & $0.352 \pm 0.348$/ $1.86\sigma$ & $0.796 \pm 0.358$/ $0.57\sigma$ \\
DESI+Unino3+full CMB likelihood  & -- & $1.169 \pm 0.074$/ $2.28\sigma$ & $0.789 \pm 0.147$/ $1.43\sigma$ & $0.459 \pm 0.436$/ $1.24\sigma$ & $0.884 \pm 0.381$/ $0.30\sigma$ \\
\hline
\end{tabular}%
}
\end{center}
\caption{Constraints on $w_i$ or $f_i$ in each redshift bin, derived from DESI+Union3 with CMB distance priors or CMB full likelihood. The upper panel denotes the 5bins-w case while the lower panel denotes the 5bins-f case. The confidence level (CL) indicates the deviation from $w(z)=-1$ (or $f(z)=1$).}
\label{tab:compare_5bins-wf}
\end{table*}

\begin{figure*}[htbp]
\small
    \centering
        \includegraphics[width=0.45\textwidth]{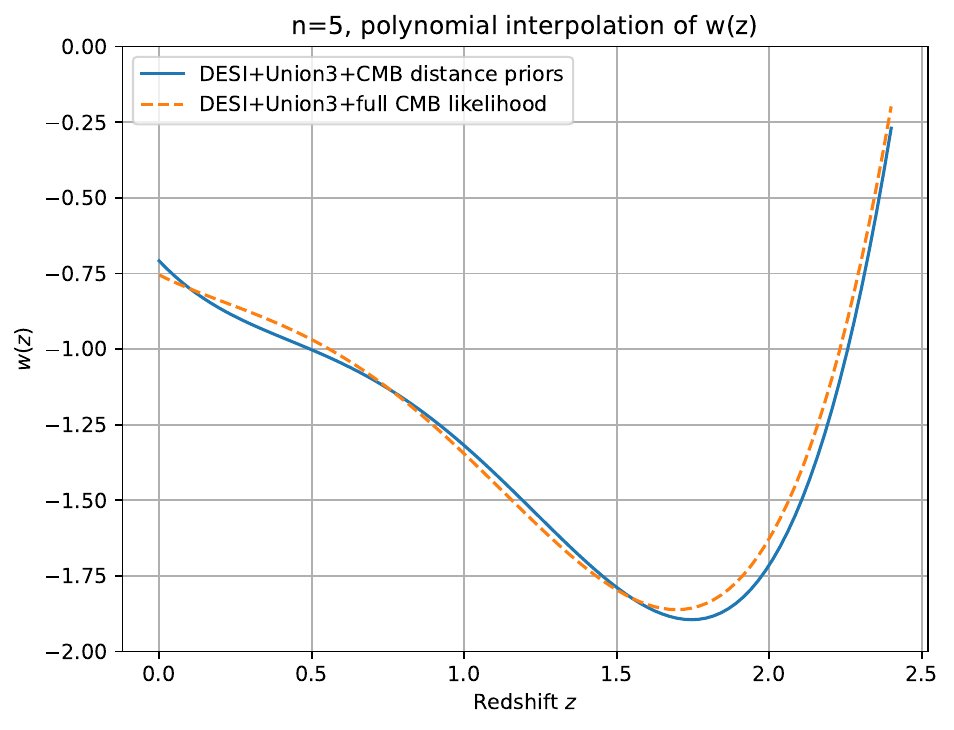}
        \includegraphics[width=0.45\textwidth]{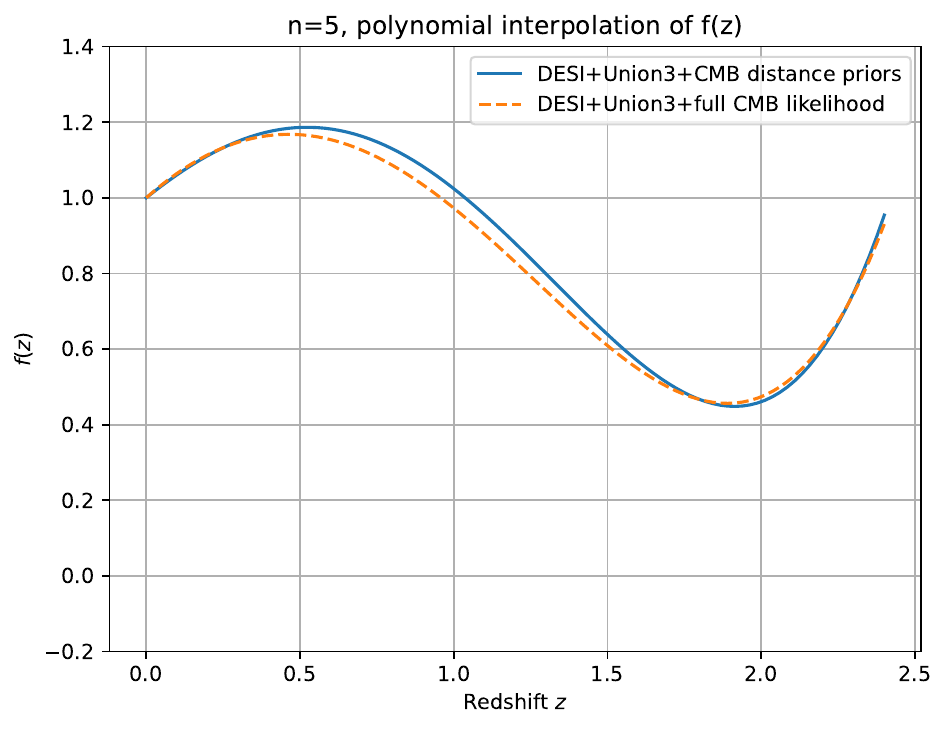}

    \caption{Left panel: Reconstruction results of the DE EoS $w(z)$ using polynomial interpolation method (n=5). Right panel: Reconstruction results of the DE density $f(z)$ using polynomial interpolation method (n=5).  Blue lines denote the best-fit result derived form DESI+Union3+CMB distance priors, while the orange dash lines denote the best-fit result derived from DESI+Union3+full CMB likelihood.}
    \label{fig.compare_5pi}
\end{figure*}

For the polynomial interpolation method, since the values of these interpolation nodes do not provide any physical meaning, we use their best-fit values to plot the reconstructed evolution curves of DE EoS $w(z)$ and density $f(z)$. The corresponding results are shown in Fig. \ref{fig.compare_5pi}. From this figure, the data combination using CMB distance priors reproduces the full likelihood results to a good approximation.

Based on Table.\ref{tab:compare_5bins-wf} and Figure.\ref{fig.compare_5pi}, one can see that the cosmological results of CMB distance priors and full CMB likelihood only have slight differences.

\newcommand{\bibcommenthead}{}
\bibliographystyle{sn-aps}
\bibliography{paper}

\begin{thebibliography}{100}
\providecommand{\url}[1]{{#1}}
\providecommand{\urlprefix}{URL }
\providecommand{\doi}[1]{\url{https://doi.org/#1}}
\bibcommenthead

\bibitem{SupernovaSearchTeam:1998fmf}
A.G. Riess, et~al., {Observational evidence from supernovae for an accelerating universe and a cosmological constant}.
\newblock Astron. J. \textbf{116}, 1009--1038 (1998).
\newblock \doi{10.1086/300499}.
\newblock {\href{https://arxiv.org/abs/astro-ph/9805201}{{arXiv:astro-ph/9805201}}}

\bibitem{SupernovaCosmologyProject:1998vns}
S.~Perlmutter, et~al., {Measurements of $\Omega$ and $\Lambda$ from 42 High Redshift Supernovae}.
\newblock Astrophys. J. \textbf{517}, 565--586 (1999).
\newblock \doi{10.1086/307221}.
\newblock {\href{https://arxiv.org/abs/astro-ph/9812133}{{arXiv:astro-ph/9812133}}}

\bibitem{Poulin:2023lkg}
V.~Poulin, T.L. Smith, T.~Karwal, {The Ups and Downs of Early Dark Energy solutions to the Hubble tension: A review of models, hints and constraints circa 2023}.
\newblock Phys. Dark Univ. \textbf{42}, 101348 (2023).
\newblock \doi{10.1016/j.dark.2023.101348}.
\newblock {\href{https://arxiv.org/abs/2302.09032}{{arXiv:2302.09032}}} {[astro-ph.CO]}

\bibitem{Rong-Gen:2023dcz}
C.~Rong-Gen, L.~Li, W.~Shao-Jiang, {Hubble-constant crisis}.
\newblock Acta Phys. Sin. \textbf{72}(23), 239801 (2023).
\newblock \doi{10.7498/aps.72.20231270}

\bibitem{Efstathiou:2024dvn}
G.~Efstathiou, {Challenges to the \ensuremath{\Lambda}CDM cosmology}.
\newblock Phil. Trans. Roy. Soc. Lond. A \textbf{383}(2290), 20240022 (2025).
\newblock \doi{10.1098/rsta.2024.0022}.
\newblock {\href{https://arxiv.org/abs/2406.12106}{{arXiv:2406.12106}}} {[astro-ph.CO]}

\bibitem{DiValentino:2025sru}
E.~Di~Valentino, et~al., {The CosmoVerse White Paper: Addressing observational tensions in cosmology with systematics and fundamental physics}  (2025).
\newblock {\href{https://arxiv.org/abs/2504.01669}{{arXiv:2504.01669}}} {[astro-ph.CO]}

\bibitem{DESI:2024mwx}
A.G. Adame, et~al., {DESI 2024 VI: Cosmological Constraints from the Measurements of Baryon Acoustic Oscillations}  (2024).
\newblock {\href{https://arxiv.org/abs/2404.03002}{{arXiv:2404.03002}}} {[astro-ph.CO]}

\bibitem{DESI:2025zgx}
M.~Abdul~Karim, et~al., {DESI DR2 Results II: Measurements of Baryon Acoustic Oscillations and Cosmological Constraints}  (2025).
\newblock {\href{https://arxiv.org/abs/2503.14738}{{arXiv:2503.14738}}} {[astro-ph.CO]}

\bibitem{Brout:2022vxf}
D.~Brout, et~al., {The Pantheon+ Analysis: Cosmological Constraints}.
\newblock Astrophys. J. \textbf{938}(2), 110 (2022).
\newblock \doi{10.3847/1538-4357/ac8e04}.
\newblock {\href{https://arxiv.org/abs/2202.04077}{{arXiv:2202.04077}}} {[astro-ph.CO]}

\bibitem{Rubin:2023ovl}
D.~Rubin, et~al., {Union Through UNITY: Cosmology with 2,000 SNe Using a Unified Bayesian Framework}  (2023).
\newblock {\href{https://arxiv.org/abs/2311.12098}{{arXiv:2311.12098}}} {[astro-ph.CO]}

\bibitem{DES:2024jxu}
T.M.C. Abbott, et~al., {The Dark Energy Survey: Cosmology Results with \ensuremath{\sim}1500 New High-redshift Type Ia Supernovae Using the Full 5 yr Data Set}.
\newblock Astrophys. J. Lett. \textbf{973}(1), L14 (2024).
\newblock \doi{10.3847/2041-8213/ad6f9f}.
\newblock {\href{https://arxiv.org/abs/2401.02929}{{arXiv:2401.02929}}} {[astro-ph.CO]}

\bibitem{Chevallier:2000qy}
M.~Chevallier, D.~Polarski, {Accelerating universes with scaling dark matter}.
\newblock Int. J. Mod. Phys. D \textbf{10}, 213--224 (2001).
\newblock \doi{10.1142/S0218271801000822}.
\newblock {\href{https://arxiv.org/abs/gr-qc/0009008}{{arXiv:gr-qc/0009008}}}

\bibitem{Linder:2002et}
E.V. Linder, {Exploring the expansion history of the universe}.
\newblock Phys. Rev. Lett. \textbf{90}, 091301 (2003).
\newblock \doi{10.1103/PhysRevLett.90.091301}.
\newblock {\href{https://arxiv.org/abs/astro-ph/0208512}{{arXiv:astro-ph/0208512}}}

\bibitem{Yin:2024hba}
W.~Yin, {Cosmic clues: DESI, dark energy, and the cosmological constant problem}.
\newblock JHEP \textbf{05}, 327 (2024).
\newblock \doi{10.1007/JHEP05(2024)327}.
\newblock {\href{https://arxiv.org/abs/2404.06444}{{arXiv:2404.06444}}} {[hep-ph]}

\bibitem{Shlivko:2024llw}
D.~Shlivko, P.J. Steinhardt, {Assessing observational constraints on dark energy}.
\newblock Phys. Lett. B \textbf{855}, 138826 (2024).
\newblock \doi{10.1016/j.physletb.2024.138826}.
\newblock {\href{https://arxiv.org/abs/2405.03933}{{arXiv:2405.03933}}} {[astro-ph.CO]}

\bibitem{DESI:2024kob}
K.~Lodha, et~al., {DESI 2024: Constraints on physics-focused aspects of dark energy using DESI DR1 BAO data}.
\newblock Phys. Rev. D \textbf{111}(2), 023532 (2025).
\newblock \doi{10.1103/PhysRevD.111.023532}.
\newblock {\href{https://arxiv.org/abs/2405.13588}{{arXiv:2405.13588}}} {[astro-ph.CO]}

\bibitem{Notari:2024rti}
A.~Notari, M.~Redi, A.~Tesi, {Consistent theories for the DESI dark energy fit}.
\newblock JCAP \textbf{11}, 025 (2024).
\newblock \doi{10.1088/1475-7516/2024/11/025}.
\newblock {\href{https://arxiv.org/abs/2406.08459}{{arXiv:2406.08459}}} {[astro-ph.CO]}

\bibitem{Wolf:2024eph}
W.J. Wolf, C.~Garc\'\i{}a-Garc\'\i{}a, D.J. Bartlett, P.G. Ferreira, {Scant evidence for thawing quintessence}.
\newblock Phys. Rev. D \textbf{110}(8), 083528 (2024).
\newblock \doi{10.1103/PhysRevD.110.083528}.
\newblock {\href{https://arxiv.org/abs/2408.17318}{{arXiv:2408.17318}}} {[astro-ph.CO]}

\bibitem{Shajib:2025tpd}
A.J. Shajib, J.A. Frieman, {Evolving dark energy models: Current and forecast constraints}  (2025).
\newblock {\href{https://arxiv.org/abs/2502.06929}{{arXiv:2502.06929}}} {[astro-ph.CO]}

\bibitem{Anchordoqui:2025fgz}
L.A. Anchordoqui, I.~Antoniadis, D.~Lust, {S-dual Quintessence, the Swampland, and the DESI DR2 Results}  (2025).
\newblock {\href{https://arxiv.org/abs/2503.19428}{{arXiv:2503.19428}}} {[hep-th]}

\bibitem{Gialamas:2025pwv}
I.D. Gialamas, G.~H{\"u}tsi, M.~Raidal, J.~Urrutia, M.~Vasar, H.~Veerm{\"a}e, {Quintessence and phantoms in light of DESI 2025}  (2025).
\newblock {\href{https://arxiv.org/abs/2506.21542}{{arXiv:2506.21542}}} {[astro-ph.CO]}

\bibitem{Li:2024bwr}
J.X. Li, S.~Wang, {A comprehensive numerical study on four categories of holographic dark energy models}.
\newblock JCAP \textbf{07}, 047 (2025).
\newblock \doi{10.1088/1475-7516/2025/07/047}.
\newblock {\href{https://arxiv.org/abs/2412.09064}{{arXiv:2412.09064}}} {[astro-ph.CO]}

\bibitem{Tyagi:2024cqp}
U.K. Tyagi, S.~Haridasu, S.~Basak, {Holographic and gravity-thermodynamic approaches in entropic cosmology: Bayesian assessment using late-time data}.
\newblock Phys. Rev. D \textbf{110}(6), 063503 (2024).
\newblock \doi{10.1103/PhysRevD.110.063503}.
\newblock {\href{https://arxiv.org/abs/2406.07446}{{arXiv:2406.07446}}} {[astro-ph.CO]}

\bibitem{Astashenok:2024jje}
A.V. Astashenok, A.S. Tepliakov, {General constraints on Tsallis holographic dark energy from observational data}.
\newblock Phys. Dark Univ. \textbf{47}, 101747 (2025).
\newblock \doi{10.1016/j.dark.2024.101747}.
\newblock {\href{https://arxiv.org/abs/2410.00597}{{arXiv:2410.00597}}} {[astro-ph.CO]}

\bibitem{Li:2024qus}
T.N. Li, Y.H. Li, G.H. Du, P.J. Wu, L.~Feng, J.F. Zhang, X.~Zhang, {Revisiting holographic dark energy after DESI 2024}.
\newblock Eur. Phys. J. C \textbf{85}(6), 608 (2025).
\newblock \doi{10.1140/epjc/s10052-025-14279-7}.
\newblock {\href{https://arxiv.org/abs/2411.08639}{{arXiv:2411.08639}}} {[astro-ph.CO]}

\bibitem{Han:2024sxm}
T.~Han, Z.~Li, J.F. Zhang, X.~Zhang, {Revisiting Holographic Dark Energy from the Perspective of Multi-Messenger Gravitational Wave Astronomy: Future Joint Observations with Short Gamma-Ray Bursts}.
\newblock Universe \textbf{11}(3), 85 (2025).
\newblock \doi{10.3390/universe11030085}.
\newblock {\href{https://arxiv.org/abs/2412.06873}{{arXiv:2412.06873}}} {[astro-ph.CO]}

\bibitem{Giare:2024smz}
W.~Giar\`e, M.A. Sabogal, R.C. Nunes, E.~Di~Valentino, {Interacting Dark Energy after DESI Baryon Acoustic Oscillation Measurements}.
\newblock Phys. Rev. Lett. \textbf{133}(25), 251003 (2024).
\newblock \doi{10.1103/PhysRevLett.133.251003}.
\newblock {\href{https://arxiv.org/abs/2404.15232}{{arXiv:2404.15232}}} {[astro-ph.CO]}

\bibitem{Montani:2024pou}
G.~Montani, N.~Carlevaro, L.A. Escamilla, E.~Di~Valentino, {Kinetic model for dark energy\textemdash{}dark matter interaction: Scenario for the hubble tension}.
\newblock Phys. Dark Univ. \textbf{48}, 101848 (2025).
\newblock \doi{10.1016/j.dark.2025.101848}.
\newblock {\href{https://arxiv.org/abs/2404.15977}{{arXiv:2404.15977}}} {[gr-qc]}

\bibitem{Li:2024qso}
T.N. Li, P.J. Wu, G.H. Du, S.J. Jin, H.L. Li, J.F. Zhang, X.~Zhang, {Constraints on Interacting Dark Energy Models from the DESI Baryon Acoustic Oscillation and DES Supernovae Data}.
\newblock Astrophys. J. \textbf{976}(1), 1 (2024).
\newblock \doi{10.3847/1538-4357/ad87f0}.
\newblock {\href{https://arxiv.org/abs/2407.14934}{{arXiv:2407.14934}}} {[astro-ph.CO]}

\bibitem{Li:2025owk}
T.N. Li, G.H. Du, Y.H. Li, P.J. Wu, S.J. Jin, J.F. Zhang, X.~Zhang, {Probing the sign-changeable interaction between dark energy and dark matter with DESI baryon acoustic oscillations and DES supernovae data}  (2025).
\newblock {\href{https://arxiv.org/abs/2501.07361}{{arXiv:2501.07361}}} {[astro-ph.CO]}

\bibitem{Yang:2025vnm}
Y.~Yang, Y.~Wang, X.~Dai, {Cosmological constraints on two vacuum decay models}.
\newblock Eur. Phys. J. C \textbf{85}(3), 224 (2025).
\newblock \doi{10.1140/epjc/s10052-025-13990-9}.
\newblock {\href{https://arxiv.org/abs/2502.17792}{{arXiv:2502.17792}}} {[astro-ph.CO]}

\bibitem{Zhai:2025hfi}
Y.~Zhai, M.~de~Cesare, C.~van~de Bruck, E.~Di~Valentino, E.~Wilson-Ewing, {A low-redshift preference for an interacting dark energy model}  (2025).
\newblock {\href{https://arxiv.org/abs/2503.15659}{{arXiv:2503.15659}}} {[astro-ph.CO]}

\bibitem{Pan:2025qwy}
S.~Pan, S.~Paul, E.N. Saridakis, W.~Yang, {Interacting dark energy after DESI DR2: a challenge for $\Lambda$CDM paradigm?}  (2025).
\newblock {\href{https://arxiv.org/abs/2504.00994}{{arXiv:2504.00994}}} {[astro-ph.CO]}

\bibitem{Yang:2025boq}
Y.~Yang, X.~Dai, Y.~Wang, {New cosmological constraints on the evolution of dark matter energy density}  (2025).
\newblock {\href{https://arxiv.org/abs/2505.09879}{{arXiv:2505.09879}}} {[astro-ph.CO]}

\bibitem{vanderWesthuizen:2025iam}
M.~van~der Westhuizen, D.~Figueruelo, R.~Thubisi, S.~Sahlu, A.~Abebe, A.~Paliathanasis, {Compartmentalization in the Dark Sector of the Universe after DESI DR2 BAO data}  (2025).
\newblock {\href{https://arxiv.org/abs/2505.23306}{{arXiv:2505.23306}}} {[astro-ph.CO]}

\bibitem{Chaussidon:2025npr}
E.~Chaussidon, et~al., {Early time solution as an alternative to the late time evolving dark energy with DESI DR2 BAO}  (2025).
\newblock {\href{https://arxiv.org/abs/2503.24343}{{arXiv:2503.24343}}} {[astro-ph.CO]}

\bibitem{Qu:2024lpx}
F.J. Qu, K.M. Surrao, B.~Bolliet, J.C. Hill, B.D. Sherwin, H.T. Jense, {Accelerated inference on accelerated cosmic expansion: New constraints on axion-like early dark energy with DESI BAO and ACT DR6 CMB lensing}  (2024).
\newblock {\href{https://arxiv.org/abs/2404.16805}{{arXiv:2404.16805}}} {[astro-ph.CO]}

\bibitem{Seto:2024cgo}
O.~Seto, Y.~Toda, {DESI constraints on the varying electron mass model and axionlike early dark energy}.
\newblock Phys. Rev. D \textbf{110}(8), 083501 (2024).
\newblock \doi{10.1103/PhysRevD.110.083501}.
\newblock {\href{https://arxiv.org/abs/2405.11869}{{arXiv:2405.11869}}} {[astro-ph.CO]}

\bibitem{Wang:2024rus}
J.~Wang, Z.~Huang, Y.~Yao, J.~Liu, L.~Huang, Y.~Su, {A PAge-like Unified Dark Fluid model}.
\newblock JCAP \textbf{09}, 053 (2024).
\newblock \doi{10.1088/1475-7516/2024/09/053}.
\newblock {\href{https://arxiv.org/abs/2405.05798}{{arXiv:2405.05798}}} {[astro-ph.CO]}

\bibitem{RoyChoudhury:2024wri}
S.~Roy~Choudhury, T.~Okumura, {Updated Cosmological Constraints in Extended Parameter Space with Planck PR4, DESI Baryon Acoustic Oscillations, and Supernovae: Dynamical Dark Energy, Neutrino Masses, Lensing Anomaly, and the Hubble Tension}.
\newblock Astrophys. J. Lett. \textbf{976}(1), L11 (2024).
\newblock \doi{10.3847/2041-8213/ad8c26}.
\newblock {\href{https://arxiv.org/abs/2409.13022}{{arXiv:2409.13022}}} {[astro-ph.CO]}

\bibitem{Odintsov:2025kyw}
S.D. Odintsov, V.K. Oikonomou, G.S. Sharov, {Einstein-Gauss-Bonnet cosmology confronted with observations}.
\newblock JHEAp \textbf{47}, 100398 (2025).
\newblock \doi{10.1016/j.jheap.2025.100398}.
\newblock {\href{https://arxiv.org/abs/2503.17946}{{arXiv:2503.17946}}} {[gr-qc]}

\bibitem{Scherer:2025esj}
M.~Scherer, M.A. Sabogal, R.C. Nunes, A.~De~Felice, {Challenging $\Lambda$CDM: 5$\sigma$ Evidence for a Dynamical Dark Energy Late-Time Transition}  (2025).
\newblock {\href{https://arxiv.org/abs/2504.20664}{{arXiv:2504.20664}}} {[astro-ph.CO]}

\bibitem{RoyChoudhury:2025dhe}
S.~Roy~Choudhury, {Cosmology in Extended Parameter Space with DESI Data Release 2 Baryon Acoustic Oscillations: A 2$\sigma$+ Detection of Nonzero Neutrino Masses with an Update on Dynamical Dark Energy and Lensing Anomaly}.
\newblock Astrophys. J. Lett. \textbf{986}, L31 (2025).
\newblock \doi{10.3847/2041-8213/ade1cc}.
\newblock {\href{https://arxiv.org/abs/2504.15340}{{arXiv:2504.15340}}} {[astro-ph.CO]}

\bibitem{Paliathanasis:2025hjw}
A.~Paliathanasis, {Testing non-coincident f(Q)-gravity with DESI DR2 BAO and GRBs}.
\newblock Phys. Dark Univ. \textbf{49}, 101993 (2025).
\newblock \doi{10.1016/j.dark.2025.101993}.
\newblock {\href{https://arxiv.org/abs/2504.11132}{{arXiv:2504.11132}}} {[gr-qc]}

\bibitem{Paliathanasis:2025xxm}
A.~Paliathanasis, {Observational Constraints on Scalar Field--Matter Interaction in Weyl Integrable Spacetime}  (2025).
\newblock {\href{https://arxiv.org/abs/2506.16223}{{arXiv:2506.16223}}} {[gr-qc]}

\bibitem{Odintsov:2025jfq}
S.D. Odintsov, V.K. Oikonomou, G.S. Sharov, {Dynamical Dark Energy from $F(R)$ Gravity Models Unifying Inflation with Dark Energy: Confronting the Latest Observational Data}  (2025).
\newblock {\href{https://arxiv.org/abs/2506.02245}{{arXiv:2506.02245}}} {[gr-qc]}

\bibitem{Yoo:2012ug}
J.~Yoo, Y.~Watanabe, {Theoretical Models of Dark Energy}.
\newblock Int. J. Mod. Phys. D \textbf{21}, 1230002 (2012).
\newblock \doi{10.1142/S0218271812300029}.
\newblock {\href{https://arxiv.org/abs/1212.4726}{{arXiv:1212.4726}}} {[astro-ph.CO]}

\bibitem{Arun:2017uaw}
K.~Arun, S.B. Gudennavar, C.~Sivaram, {Dark matter, dark energy, and alternate models: A review}.
\newblock Adv. Space Res. \textbf{60}, 166--186 (2017).
\newblock \doi{10.1016/j.asr.2017.03.043}.
\newblock {\href{https://arxiv.org/abs/1704.06155}{{arXiv:1704.06155}}} {[physics.gen-ph]}

\bibitem{Bahamonde:2017ize}
S.~Bahamonde, C.G. B\"ohmer, S.~Carloni, E.J. Copeland, W.~Fang, N.~Tamanini, {Dynamical systems applied to cosmology: dark energy and modified gravity}.
\newblock Phys. Rept. \textbf{775-777}, 1--122 (2018).
\newblock \doi{10.1016/j.physrep.2018.09.001}.
\newblock {\href{https://arxiv.org/abs/1712.03107}{{arXiv:1712.03107}}} {[gr-qc]}

\bibitem{Wang:2016och}
S.~Wang, Y.~Wang, M.~Li, {Holographic Dark Energy}.
\newblock Phys. Rept. \textbf{696}, 1--57 (2017).
\newblock \doi{10.1016/j.physrep.2017.06.003}.
\newblock {\href{https://arxiv.org/abs/1612.00345}{{arXiv:1612.00345}}} {[astro-ph.CO]}

\bibitem{Wang:2023gov}
S.~Wang, M.~Li, {Theoretical aspects of holographic dark energy}.
\newblock Commun. Theor. Phys. \textbf{75}(11), 117401 (2023).
\newblock \doi{10.1088/1572-9494/acf27c}

\bibitem{Cai:2025mas}
Y.~Cai, X.~Ren, T.~Qiu, M.~Li, X.~Zhang, {The Quintom theory of dark energy after DESI DR2}  (2025).
\newblock {\href{https://arxiv.org/abs/2505.24732}{{arXiv:2505.24732}}} {[astro-ph.CO]}

\bibitem{Huterer:2002hy}
D.~Huterer, G.~Starkman, {Parameterization of dark-energy properties: A Principal-component approach}.
\newblock Phys. Rev. Lett. \textbf{90}, 031301 (2003).
\newblock \doi{10.1103/PhysRevLett.90.031301}.
\newblock {\href{https://arxiv.org/abs/astro-ph/0207517}{{arXiv:astro-ph/0207517}}}

\bibitem{Shafieloo:2005nd}
A.~Shafieloo, U.~Alam, V.~Sahni, A.A. Starobinsky, {Smoothing Supernova Data to Reconstruct the Expansion History of the Universe and its Age}.
\newblock Mon. Not. Roy. Astron. Soc. \textbf{366}, 1081--1095 (2006).
\newblock \doi{10.1111/j.1365-2966.2005.09911.x}.
\newblock {\href{https://arxiv.org/abs/astro-ph/0505329}{{arXiv:astro-ph/0505329}}}

\bibitem{Holsclaw:2010sk}
T.~Holsclaw, U.~Alam, B.~Sanso, H.~Lee, K.~Heitmann, S.~Habib, D.~Higdon, {Nonparametric Dark Energy Reconstruction from Supernova Data}.
\newblock Phys. Rev. Lett. \textbf{105}, 241302 (2010).
\newblock \doi{10.1103/PhysRevLett.105.241302}.
\newblock {\href{https://arxiv.org/abs/1011.3079}{{arXiv:1011.3079}}} {[astro-ph.CO]}

\bibitem{Clarkson:2010bm}
C.~Clarkson, C.~Zunckel, {Direct reconstruction of dark energy}.
\newblock Phys. Rev. Lett. \textbf{104}, 211301 (2010).
\newblock \doi{10.1103/PhysRevLett.104.211301}.
\newblock {\href{https://arxiv.org/abs/1002.5004}{{arXiv:1002.5004}}} {[astro-ph.CO]}

\bibitem{Shafieloo:2012ht}
A.~Shafieloo, A.G. Kim, E.V. Linder, {Gaussian Process Cosmography}.
\newblock Phys. Rev. D \textbf{85}, 123530 (2012).
\newblock \doi{10.1103/PhysRevD.85.123530}.
\newblock {\href{https://arxiv.org/abs/1204.2272}{{arXiv:1204.2272}}} {[astro-ph.CO]}

\bibitem{Zhao:2012aw}
G.B. Zhao, R.G. Crittenden, L.~Pogosian, X.~Zhang, {Examining the evidence for dynamical dark energy}.
\newblock Phys. Rev. Lett. \textbf{109}, 171301 (2012).
\newblock \doi{10.1103/PhysRevLett.109.171301}.
\newblock {\href{https://arxiv.org/abs/1207.3804}{{arXiv:1207.3804}}} {[astro-ph.CO]}

\bibitem{Nesseris:2012tt}
S.~Nesseris, J.~Garcia-Bellido, {A new perspective on Dark Energy modeling via Genetic Algorithms}.
\newblock JCAP \textbf{11}, 033 (2012).
\newblock \doi{10.1088/1475-7516/2012/11/033}.
\newblock {\href{https://arxiv.org/abs/1205.0364}{{arXiv:1205.0364}}} {[astro-ph.CO]}

\bibitem{Shafieloo:2012yh}
A.~Shafieloo, {Crossing Statistic: Reconstructing the Expansion History of the Universe}.
\newblock JCAP \textbf{08}, 002 (2012).
\newblock \doi{10.1088/1475-7516/2012/08/002}.
\newblock {\href{https://arxiv.org/abs/1204.1109}{{arXiv:1204.1109}}} {[astro-ph.CO]}

\bibitem{Zhao:2017cud}
G.B. Zhao, et~al., {Dynamical dark energy in light of the latest observations}.
\newblock Nature Astron. \textbf{1}(9), 627--632 (2017).
\newblock \doi{10.1038/s41550-017-0216-z}.
\newblock {\href{https://arxiv.org/abs/1701.08165}{{arXiv:1701.08165}}} {[astro-ph.CO]}

\bibitem{Wang:2018fng}
Y.~Wang, L.~Pogosian, G.B. Zhao, A.~Zucca, {Evolution of dark energy reconstructed from the latest observations}.
\newblock Astrophys. J. Lett. \textbf{869}, L8 (2018).
\newblock \doi{10.3847/2041-8213/aaf238}.
\newblock {\href{https://arxiv.org/abs/1807.03772}{{arXiv:1807.03772}}} {[astro-ph.CO]}

\bibitem{LHuillier:2019imn}
B.~L'Huillier, A.~Shafieloo, D.~Polarski, A.A. Starobinsky, {Defying the laws of Gravity I: Model-independent reconstruction of the Universe expansion from growth data}.
\newblock Mon. Not. Roy. Astron. Soc. \textbf{494}(1), 819--826 (2020).
\newblock \doi{10.1093/mnras/staa633}.
\newblock {\href{https://arxiv.org/abs/1906.05991}{{arXiv:1906.05991}}} {[astro-ph.CO]}

\bibitem{Raveri:2021dbu}
M.~Raveri, L.~Pogosian, M.~Martinelli, K.~Koyama, A.~Silvestri, G.B. Zhao, {Principal reconstructed modes of dark energy and gravity}.
\newblock JCAP \textbf{02}, 061 (2023).
\newblock \doi{10.1088/1475-7516/2023/02/061}.
\newblock {\href{https://arxiv.org/abs/2107.12990}{{arXiv:2107.12990}}} {[astro-ph.CO]}

\bibitem{DESI:2024aqx}
R.~Calderon, et~al., {DESI 2024: reconstructing dark energy using crossing statistics with DESI DR1 BAO data}.
\newblock JCAP \textbf{10}, 048 (2024).
\newblock \doi{10.1088/1475-7516/2024/10/048}.
\newblock {\href{https://arxiv.org/abs/2405.04216}{{arXiv:2405.04216}}} {[astro-ph.CO]}

\bibitem{Pang:2024qyh}
Y.H. Pang, X.~Zhang, Q.G. Huang, {Constraints on Redshift-Binned Dark Energy using DESI BAO Data}  (2024).
\newblock {\href{https://arxiv.org/abs/2408.14787}{{arXiv:2408.14787}}} {[astro-ph.CO]}

\bibitem{Bansal:2025ipo}
P.~Bansal, D.~Huterer, {Expansion-history preferences of DESI and external data}  (2025).
\newblock {\href{https://arxiv.org/abs/2502.07185}{{arXiv:2502.07185}}} {[astro-ph.CO]}

\bibitem{Reboucas:2024smm}
J.a. Rebou\c{c}as, D.H.F. de~Souza, K.~Zhong, V.~Miranda, R.~Rosenfeld, {Investigating late-time dark energy and massive neutrinos in light of DESI Y1 BAO}.
\newblock JCAP \textbf{02}, 024 (2025).
\newblock \doi{10.1088/1475-7516/2025/02/024}.
\newblock {\href{https://arxiv.org/abs/2408.14628}{{arXiv:2408.14628}}} {[astro-ph.CO]}

\bibitem{Orchard:2024bve}
L.~Orchard, V.H. C\'ardenas, {Probing dark energy evolution post-DESI 2024}.
\newblock Phys. Dark Univ. \textbf{46}, 101678 (2024).
\newblock \doi{10.1016/j.dark.2024.101678}.
\newblock {\href{https://arxiv.org/abs/2407.05579}{{arXiv:2407.05579}}} {[astro-ph.CO]}

\bibitem{Ormondroyd:2025exu}
A.N. Ormondroyd, W.J. Handley, M.P. Hobson, A.N. Lasenby, {Nonparametric reconstructions of dynamical dark energy via flexknots}  (2025).
\newblock {\href{https://arxiv.org/abs/2503.08658}{{arXiv:2503.08658}}} {[astro-ph.CO]}

\bibitem{Ormondroyd:2025iaf}
A.N. Ormondroyd, W.J. Handley, M.P. Hobson, A.N. Lasenby, {Comparison of dynamical dark energy with \ensuremath{\Lambda}CDM in light of DESI DR2}  (2025).
\newblock {\href{https://arxiv.org/abs/2503.17342}{{arXiv:2503.17342}}} {[astro-ph.CO]}

\bibitem{Berti:2025phi}
M.~Berti, E.~Bellini, C.~Bonvin, M.~Kunz, M.~Viel, M.~Zumalacarregui, {Reconstructing the dark energy density in light of DESI BAO observations}  (2025).
\newblock {\href{https://arxiv.org/abs/2503.13198}{{arXiv:2503.13198}}} {[astro-ph.CO]}

\bibitem{Luongo:2024fww}
O.~Luongo, M.~Muccino, {Model-independent cosmographic constraints from DESI 2024}.
\newblock Astron. Astrophys. \textbf{690}, A40 (2024).
\newblock \doi{10.1051/0004-6361/202450512}.
\newblock {\href{https://arxiv.org/abs/2404.07070}{{arXiv:2404.07070}}} {[astro-ph.CO]}

\bibitem{Mukherjee:2024ryz}
P.~Mukherjee, A.A. Sen, {Model-independent cosmological inference post DESI DR1 BAO measurements}.
\newblock Phys. Rev. D \textbf{110}(12), 123502 (2024).
\newblock \doi{10.1103/PhysRevD.110.123502}.
\newblock {\href{https://arxiv.org/abs/2405.19178}{{arXiv:2405.19178}}} {[astro-ph.CO]}

\bibitem{Jiang:2024xnu}
J.Q. Jiang, D.~Pedrotti, S.S. da~Costa, S.~Vagnozzi, {Nonparametric late-time expansion history reconstruction and implications for the Hubble tension in light of recent DESI and type Ia supernovae data}.
\newblock Phys. Rev. D \textbf{110}(12), 123519 (2024).
\newblock \doi{10.1103/PhysRevD.110.123519}.
\newblock {\href{https://arxiv.org/abs/2408.02365}{{arXiv:2408.02365}}} {[astro-ph.CO]}

\bibitem{Liu:2024yib}
T.~Liu, S.~Wang, H.~Wu, S.~Cao, J.~Wang, {Newest Measurements of Cosmic Curvature with BOSS/eBOSS and DESI DR1 Baryon Acoustic Oscillation Observations}.
\newblock Astrophys. J. Lett. \textbf{981}(2), L24 (2025).
\newblock \doi{10.3847/2041-8213/adb7de}.
\newblock {\href{https://arxiv.org/abs/2411.14154}{{arXiv:2411.14154}}} {[astro-ph.CO]}

\bibitem{Yang:2025kgc}
Y.~Yang, Q.~Wang, C.~Li, P.~Yuan, X.~Ren, E.N. Saridakis, Y.F. Cai, {Gaussian-process reconstructions and model building of quintom dark energy from latest cosmological observations}  (2025).
\newblock {\href{https://arxiv.org/abs/2501.18336}{{arXiv:2501.18336}}} {[astro-ph.CO]}

\bibitem{Paliathanasis:2025cuc}
A.~Paliathanasis, {Observational constraints on dark energy models with {\ensuremath{\Lambda}} as an equilibrium point}.
\newblock Phys. Dark Univ. \textbf{48}, 101956 (2025).
\newblock \doi{10.1016/j.dark.2025.101956}.
\newblock {\href{https://arxiv.org/abs/2502.16221}{{arXiv:2502.16221}}} {[astro-ph.CO]}

\bibitem{Gao:2025ozb}
S.~Gao, Q.~Gao, Y.~Gong, X.~Lu, {Null tests with Gaussian Process}  (2025).
\newblock {\href{https://arxiv.org/abs/2503.15943}{{arXiv:2503.15943}}} {[astro-ph.CO]}

\bibitem{Gu:2025xie}
G.~Gu, et~al., {Dynamical Dark Energy in light of the DESI DR2 Baryonic Acoustic Oscillations Measurements}  (2025).
\newblock {\href{https://arxiv.org/abs/2504.06118}{{arXiv:2504.06118}}} {[astro-ph.CO]}

\bibitem{Cheng:2025cmb}
H.~Cheng, Z.~Yin, E.~Di~Valentino, D.J.E. Marsh, L.~Visinelli, {Constraining exotic high-$z$ reionization histories with Gaussian processes and the cosmic microwave background}  (2025).
\newblock {\href{https://arxiv.org/abs/2506.19096}{{arXiv:2506.19096}}} {[astro-ph.CO]}

\bibitem{Li:2025ula}
Y.H. Li, X.~Zhang, {Cosmic Sign-Reversal: Non-Parametric Reconstruction of Interacting Dark Energy with DESI DR2}  (2025).
\newblock {\href{https://arxiv.org/abs/2506.18477}{{arXiv:2506.18477}}} {[astro-ph.CO]}

\bibitem{Gonzalez-Fuentes:2025lei}
A.~Gonz{\'a}lez-Fuentes, A.~G{\'o}mez-Valent, {Reconstruction of dark energy and late-time cosmic expansion using the Weighted Function Regression method}  (2025).
\newblock {\href{https://arxiv.org/abs/2506.11758}{{arXiv:2506.11758}}} {[astro-ph.CO]}

\bibitem{Zhao:2007ew}
G.B. Zhao, D.~Huterer, X.~Zhang, {High-resolution temporal constraints on the dynamics of dark energy}.
\newblock Phys. Rev. D \textbf{77}, 121302 (2008).
\newblock \doi{10.1103/PhysRevD.77.121302}.
\newblock {\href{https://arxiv.org/abs/0712.2277}{{arXiv:0712.2277}}} {[astro-ph]}

\bibitem{Serra:2009yp}
P.~Serra, A.~Cooray, D.E. Holz, A.~Melchiorri, S.~Pandolfi, D.~Sarkar, {No Evidence for Dark Energy Dynamics from a Global Analysis of Cosmological Data}.
\newblock Phys. Rev. D \textbf{80}, 121302 (2009).
\newblock \doi{10.1103/PhysRevD.80.121302}.
\newblock {\href{https://arxiv.org/abs/0908.3186}{{arXiv:0908.3186}}} {[astro-ph.CO]}

\bibitem{Zhao:2009ti}
G.B. Zhao, X.m. Zhang, {Probing Dark Energy Dynamics from Current and Future Cosmological Observations}.
\newblock Phys. Rev. D \textbf{81}, 043518 (2010).
\newblock \doi{10.1103/PhysRevD.81.043518}.
\newblock {\href{https://arxiv.org/abs/0908.1568}{{arXiv:0908.1568}}} {[astro-ph.CO]}

\bibitem{AlbertoVazquez:2012ofj}
J.~Alberto~Vazquez, M.~Bridges, M.P. Hobson, A.N. Lasenby, {Reconstruction of the Dark Energy equation of state}.
\newblock JCAP \textbf{09}, 020 (2012).
\newblock \doi{10.1088/1475-7516/2012/09/020}.
\newblock {\href{https://arxiv.org/abs/1205.0847}{{arXiv:1205.0847}}} {[astro-ph.CO]}

\bibitem{Wang:2015wga}
Y.~Wang, G.B. Zhao, D.~Wands, L.~Pogosian, R.G. Crittenden, {Reconstruction of the dark matter\textendash{}vacuum energy interaction}.
\newblock Phys. Rev. D \textbf{92}, 103005 (2015).
\newblock \doi{10.1103/PhysRevD.92.103005}.
\newblock {\href{https://arxiv.org/abs/1505.01373}{{arXiv:1505.01373}}} {[astro-ph.CO]}

\bibitem{Wang:2025vfb}
Y.~Wang, K.~Freese, {Model-Independent Dark Energy Measurements from DESI DR2 and Planck 2015 Data}  (2025).
\newblock {\href{https://arxiv.org/abs/2505.17415}{{arXiv:2505.17415}}} {[astro-ph.CO]}

\bibitem{Huterer:2004ch}
D.~Huterer, A.~Cooray, {Uncorrelated estimates of dark energy evolution}.
\newblock Phys. Rev. D \textbf{71}, 023506 (2005).
\newblock \doi{10.1103/PhysRevD.71.023506}.
\newblock {\href{https://arxiv.org/abs/astro-ph/0404062}{{arXiv:astro-ph/0404062}}}

\bibitem{Qi:2008zk}
S.~Qi, F.Y. Wang, T.~Lu, {Constraining the evolution of dark energy with type Ia supernovae and gamma-ray bursts}.
\newblock Astron. Astrophys. \textbf{483}, 49 (2008).
\newblock \doi{10.1051/0004-6361:20079329}.
\newblock {\href{https://arxiv.org/abs/0803.4304}{{arXiv:0803.4304}}} {[astro-ph]}

\bibitem{Gong:2009ye}
Y.~Gong, R.G. Cai, Y.~Chen, Z.H. Zhu, {Observational constraint on dynamical evolution of dark energy}.
\newblock JCAP \textbf{01}, 019 (2010).
\newblock \doi{10.1088/1475-7516/2010/01/019}.
\newblock {\href{https://arxiv.org/abs/0909.0596}{{arXiv:0909.0596}}} {[astro-ph.CO]}

\bibitem{Cai:2010qp}
R.G. Cai, Q.~Su, H.B. Zhang, {Probing the dynamical behavior of dark energy}.
\newblock JCAP \textbf{04}, 012 (2010).
\newblock \doi{10.1088/1475-7516/2010/04/012}.
\newblock {\href{https://arxiv.org/abs/1001.2207}{{arXiv:1001.2207}}} {[astro-ph.CO]}

\bibitem{Lazkoz:2012eh}
R.~Lazkoz, V.~Salzano, I.~Sendra, {Revisiting a model-independent dark energy reconstruction method}.
\newblock Eur. Phys. J. C \textbf{72}, 2130 (2012).
\newblock \doi{10.1140/epjc/s10052-012-2130-y}.
\newblock {\href{https://arxiv.org/abs/1202.4689}{{arXiv:1202.4689}}} {[astro-ph.CO]}

\bibitem{Liu:2015mkm}
Z.E. Liu, H.F. Qin, J.~Zhang, T.J. Zhang, H.R. Yu, {Reconstructing equation of state of dark energy with principal component analysis}.
\newblock Phys. Dark Univ. \textbf{26}, 100379 (2019).
\newblock \doi{10.1016/j.dark.2019.100379}.
\newblock {\href{https://arxiv.org/abs/1501.02971}{{arXiv:1501.02971}}} {[astro-ph.CO]}

\bibitem{Zheng:2017ulu}
W.~Zheng, H.~Li, {Constraints on parameterized dark energy properties from new observations with principal component analysis}.
\newblock Astropart. Phys. \textbf{86}, 1--10 (2017).
\newblock \doi{10.1016/j.astropartphys.2016.10.005}

\bibitem{Dai:2018zwv}
J.P. Dai, Y.~Yang, J.Q. Xia, {Reconstruction of the Dark Energy Equation of State from the Latest Observations}.
\newblock Astrophys. J. \textbf{857}(1), 9 (2018).
\newblock \doi{10.3847/1538-4357/aab49a}

\bibitem{Luo:2018yvq}
X.~Luo, S.~Wang, S.~Wen, {Probing Cosmic Acceleration Using Model-independent Parameterizations and Three Kinds of Supernova Statistics Techniques}.
\newblock Astrophys. J. \textbf{873}(1), 47 (2019).
\newblock \doi{10.3847/1538-4357/ab0416}.
\newblock {\href{https://arxiv.org/abs/1812.10542}{{arXiv:1812.10542}}} {[astro-ph.CO]}

\bibitem{Huang:2009rf}
Q.G. Huang, M.~Li, X.D. Li, S.~Wang, {Fitting the constitution type Ia supernova data with the redshift-binned parametrization method}.
\newblock Phys. Rev. D \textbf{80}, 083515 (2009).
\newblock \doi{10.1103/PhysRevD.80.083515}.
\newblock {\href{https://arxiv.org/abs/0905.0797}{{arXiv:0905.0797}}} {[astro-ph.CO]}

\bibitem{Wang:2010vj}
S.~Wang, X.D. Li, M.~Li, {Exploring the Latest Union2 SNIa Dataset by Using Model-Independent Parametrization Methods}.
\newblock Phys. Rev. D \textbf{83}, 023010 (2011).
\newblock \doi{10.1103/PhysRevD.83.023010}.
\newblock {\href{https://arxiv.org/abs/1009.5837}{{arXiv:1009.5837}}} {[astro-ph.CO]}

\bibitem{Escamilla:2021uoj}
L.A. Escamilla, J.A. Vazquez, {Model selection applied to reconstructions of the Dark Energy}.
\newblock Eur. Phys. J. C \textbf{83}(3), 251 (2023).
\newblock \doi{10.1140/epjc/s10052-023-11404-2}.
\newblock {\href{https://arxiv.org/abs/2111.10457}{{arXiv:2111.10457}}} {[astro-ph.CO]}

\bibitem{Wang:2009gt}
Y.~Wang, {Clarifying Forecasts of Dark Energy Constraints from Baryon Acoustic Oscillations}.
\newblock Mod. Phys. Lett. A \textbf{25}, 3093--3113 (2010).
\newblock \doi{10.1142/S0217732310034316}.
\newblock {\href{https://arxiv.org/abs/0904.2218}{{arXiv:0904.2218}}} {[astro-ph.CO]}

\bibitem{Sullivan:2007pd}
S.~Sullivan, A.~Cooray, D.E. Holz, {Narrowing Constraints with Type Ia Supernovae: Converging on a Cosmological Constant}.
\newblock JCAP \textbf{09}, 004 (2007).
\newblock \doi{10.1088/1475-7516/2007/09/004}.
\newblock {\href{https://arxiv.org/abs/0706.3730}{{arXiv:0706.3730}}} {[astro-ph]}

\bibitem{Maor:2000jy}
I.~Maor, R.~Brustein, P.J. Steinhardt, {Limitations in using luminosity distance to determine the equation of state of the universe}.
\newblock Phys. Rev. Lett. \textbf{86}, 6 (2001).
\newblock \doi{10.1103/PhysRevLett.86.6}.
\newblock [Erratum: Phys.Rev.Lett. 87, 049901 (2001)].
\newblock {\href{https://arxiv.org/abs/astro-ph/0007297}{{arXiv:astro-ph/0007297}}}

\bibitem{Wang:2001ht}
Y.~Wang, P.M. Garnavich, {Measuring time dependence of dark energy density from type Ia supernova data}.
\newblock Astrophys. J. \textbf{552}, 445 (2001).
\newblock \doi{10.1086/320552}.
\newblock {\href{https://arxiv.org/abs/astro-ph/0101040}{{arXiv:astro-ph/0101040}}}

\bibitem{Wang:2001da}
Y.~Wang, G.~Lovelace, {Unbiased estimate of dark energy density from type IA supernova data}.
\newblock Astrophys. J. Lett. \textbf{562}, L115--L120 (2001).
\newblock \doi{10.1086/338142}.
\newblock {\href{https://arxiv.org/abs/astro-ph/0109233}{{arXiv:astro-ph/0109233}}}

\bibitem{Cardenas:2014jya}
V.H. Cardenas, {Exploring hints for dark energy density evolution in light of recent data}.
\newblock Phys. Lett. B \textbf{750}, 128--134 (2015).
\newblock \doi{10.1016/j.physletb.2015.08.064}.
\newblock {\href{https://arxiv.org/abs/1405.5116}{{arXiv:1405.5116}}} {[astro-ph.CO]}

\bibitem{Grandon:2021nls}
D.~Grandon, V.H. Cardenas, {Studies on dark energy evolution}.
\newblock Class. Quant. Grav. \textbf{38}(14), 145008 (2021).
\newblock \doi{10.1088/1361-6382/ac0357}.
\newblock {\href{https://arxiv.org/abs/2107.04876}{{arXiv:2107.04876}}} {[astro-ph.CO]}

\bibitem{Bernardo:2021cxi}
R.C. Bernardo, D.~Grand\'on, J.~Said~Levi, V.H. C\'ardenas, {Parametric and nonparametric methods hint dark energy evolution}.
\newblock Phys. Dark Univ. \textbf{36}, 101017 (2022).
\newblock \doi{10.1016/j.dark.2022.101017}.
\newblock {\href{https://arxiv.org/abs/2111.08289}{{arXiv:2111.08289}}} {[astro-ph.CO]}

\bibitem{gezerlis2020numerical}
A.~Gezerlis, \emph{Numerical methods in physics with Python}, vol.~1 (Cambridge University Press Cambridge, 2020)

\bibitem{Brieden:2022heh}
S.~Brieden, H.~Gil-Mar\'\i{}n, L.~Verde, {A tale of two (or more) h's}.
\newblock JCAP \textbf{04}, 023 (2023).
\newblock \doi{10.1088/1475-7516/2023/04/023}.
\newblock {\href{https://arxiv.org/abs/2212.04522}{{arXiv:2212.04522}}} {[astro-ph.CO]}

\bibitem{Schoneberg:2022ggi}
N.~Sch\"oneberg, L.~Verde, H.~Gil-Mar\'\i{}n, S.~Brieden, {BAO+BBN revisited \textemdash{} growing the Hubble tension with a 0.7 km/s/Mpc constraint}.
\newblock JCAP \textbf{11}, 039 (2022).
\newblock \doi{10.1088/1475-7516/2022/11/039}.
\newblock {\href{https://arxiv.org/abs/2209.14330}{{arXiv:2209.14330}}} {[astro-ph.CO]}

\bibitem{Planck:2018vyg}
N.~Aghanim, et~al., {Planck 2018 results. VI. Cosmological parameters}.
\newblock Astron. Astrophys. \textbf{641}, A6 (2020).
\newblock \doi{10.1051/0004-6361/201833910}.
\newblock [Erratum: Astron.Astrophys. 652, C4 (2021)].
\newblock {\href{https://arxiv.org/abs/1807.06209}{{arXiv:1807.06209}}} {[astro-ph.CO]}

\bibitem{Planck:2019nip}
N.~Aghanim, et~al., {Planck 2018 results. V. CMB power spectra and likelihoods}.
\newblock Astron. Astrophys. \textbf{641}, A5 (2020).
\newblock \doi{10.1051/0004-6361/201936386}.
\newblock {\href{https://arxiv.org/abs/1907.12875}{{arXiv:1907.12875}}} {[astro-ph.CO]}

\bibitem{ACT:2023dou}
F.J. Qu, et~al., {The Atacama Cosmology Telescope: A Measurement of the DR6 CMB Lensing Power Spectrum and Its Implications for Structure Growth}.
\newblock Astrophys. J. \textbf{962}(2), 112 (2024).
\newblock \doi{10.3847/1538-4357/acfe06}.
\newblock {\href{https://arxiv.org/abs/2304.05202}{{arXiv:2304.05202}}} {[astro-ph.CO]}

\bibitem{ACT:2023kun}
M.S. Madhavacheril, et~al., {The Atacama Cosmology Telescope: DR6 Gravitational Lensing Map and Cosmological Parameters}.
\newblock Astrophys. J. \textbf{962}(2), 113 (2024).
\newblock \doi{10.3847/1538-4357/acff5f}.
\newblock {\href{https://arxiv.org/abs/2304.05203}{{arXiv:2304.05203}}} {[astro-ph.CO]}

\bibitem{Efstathiou:1998xx}
G.~Efstathiou, J.R. Bond, {Cosmic confusion: Degeneracies among cosmological parameters derived from measurements of microwave background anisotropies}.
\newblock Mon. Not. Roy. Astron. Soc. \textbf{304}, 75--97 (1999).
\newblock \doi{10.1046/j.1365-8711.1999.02274.x}.
\newblock {\href{https://arxiv.org/abs/astro-ph/9807103}{{arXiv:astro-ph/9807103}}}

\bibitem{Wang:2006ts}
Y.~Wang, P.~Mukherjee, {Robust dark energy constraints from supernovae, galaxy clustering, and three-year wilkinson microwave anisotropy probe observations}.
\newblock Astrophys. J. \textbf{650}, 1--6 (2006).
\newblock \doi{10.1086/507091}.
\newblock {\href{https://arxiv.org/abs/astro-ph/0604051}{{arXiv:astro-ph/0604051}}}

\bibitem{Wang:2007mza}
Y.~Wang, P.~Mukherjee, {Observational Constraints on Dark Energy and Cosmic Curvature}.
\newblock Phys. Rev. D \textbf{76}, 103533 (2007).
\newblock \doi{10.1103/PhysRevD.76.103533}.
\newblock {\href{https://arxiv.org/abs/astro-ph/0703780}{{arXiv:astro-ph/0703780}}}

\bibitem{Hu:1995en}
W.~Hu, N.~Sugiyama, {Small scale cosmological perturbations: An Analytic approach}.
\newblock Astrophys. J. \textbf{471}, 542--570 (1996).
\newblock \doi{10.1086/177989}.
\newblock {\href{https://arxiv.org/abs/astro-ph/9510117}{{arXiv:astro-ph/9510117}}}

\bibitem{Torrado:2020dgo}
J.~Torrado, A.~Lewis, {Cobaya: Code for Bayesian Analysis of hierarchical physical models}.
\newblock JCAP \textbf{05}, 057 (2021).
\newblock \doi{10.1088/1475-7516/2021/05/057}.
\newblock {\href{https://arxiv.org/abs/2005.05290}{{arXiv:2005.05290}}} {[astro-ph.IM]}

\bibitem{Lewis:2013hha}
A.~Lewis, {Efficient sampling of fast and slow cosmological parameters}.
\newblock Phys. Rev. D \textbf{87}(10), 103529 (2013).
\newblock \doi{10.1103/PhysRevD.87.103529}.
\newblock {\href{https://arxiv.org/abs/1304.4473}{{arXiv:1304.4473}}} {[astro-ph.CO]}

\bibitem{Lewis:2002ah}
A.~Lewis, S.~Bridle, {Cosmological parameters from CMB and other data: A Monte Carlo approach}.
\newblock Phys. Rev. D \textbf{66}, 103511 (2002).
\newblock \doi{10.1103/PhysRevD.66.103511}.
\newblock {\href{https://arxiv.org/abs/astro-ph/0205436}{{arXiv:astro-ph/0205436}}}

\bibitem{Gelman:1992zz}
A.~Gelman, D.B. Rubin, {Inference from Iterative Simulation Using Multiple Sequences}.
\newblock Statist. Sci. \textbf{7}, 457--472 (1992).
\newblock \doi{10.1214/ss/1177011136}

\bibitem{Lewis:2019xzd}
A.~Lewis, {GetDist: a Python package for analysing Monte Carlo samples}  (2019).
\newblock {\href{https://arxiv.org/abs/1910.13970}{{arXiv:1910.13970}}} {[astro-ph.IM]}

\bibitem{DESI:2025fii}
K.~Lodha, et~al., {Extended Dark Energy analysis using DESI DR2 BAO measurements}  (2025).
\newblock {\href{https://arxiv.org/abs/2503.14743}{{arXiv:2503.14743}}} {[astro-ph.CO]}

\end{thebibliography}

\end{document}